\documentclass[reprint,aps,pra,showkeys,showpacs]{revtex4-2}  
\usepackage{epsfig}
\usepackage{xcolor}   
\usepackage{amsmath}
\definecolor{RED}{rgb}{1., 0., 0.}

\begin{document}

\title{Modulational instability and discrete quantum droplets in a deep quasi-one-dimensional optical lattice}

\author{Sherzod R. Otajonov$^{1, 2}$}
\author{Bakhram A. Umarov$^1$}
\author{Fatkhulla Kh. Abdullaev$^{1, 2}$}

\address{$^1$Physical-Technical Institute of the Uzbek Academy of Sciences,
Chingiz Aytmatov Str. 2-B, Tashkent, 100084, Uzbekistan}

\address{$^2$National University of Uzbekistan, Department of Theoretical Physics, 100174, Tashkent, Uzbekistan}

\begin{abstract}
We study the properties of modulational instability and discrete breathers arising in a quasi-one-dimensional discrete Gross-Pitaevskii equation with Lee-Huang-Yang corrections. Conditions for modulation instability and instability regions of nonlinear plane waves are determined in parameter space. We analytically investigate the existence of different quantum droplet solutions, including intersite, onsite, front-like, flat-top and dark localized modes, using the Page method and variational approach. Their stability is checked using linear stability analyses and numerical simulations. The analytical predictions corroborated with the numerical simulations. 
\end{abstract}

\maketitle

\section{Introduction}
\label{sec:intro} 
In recent years, significant progress has been made in exploring the behaviour of quantum matter waves, specifically through the use of binary Bose-Einstein condensates (BECs). A major breakthrough in this field has been the discovery of ultra-dilute superfluids that form quantum droplets (QDs) representing a new quantum state of matter~\cite{Petrov2015, Luo2021, Barbut2019}. These QDs are the result of the interplay between mean-field interactions and Lee-Huang-Yang corrections, which are induced by quantum fluctuations~\cite{LHY}. In three dimensions, Petrov first demonstrated the creation of QDs in binary BECs by modelling the system with the modified Gross-Pitaevskii equation (GPE). The LHY correction in three-dimensional (3D) to the condensate energy is $E_{\mathrm{LHY}} \sim n^{5/2} $, and the corresponding correction to the Gross-Pitaevskii equation is $\sim |\psi|^3 \psi$, where $n$ is the BEC density and $\psi$ is the condensate wavefunction. This equation includes both the MF self-attractive cubic term and the repulsive Lee-Huang-Yang quartic term. In both three-dimensional and two-dimensional (2D) geometries, quantum fluctuations can help to stabilise a binary condensate against collapse driven by cross-attraction between its components. In each component, the cross-attraction is slightly greater than self-repulsion, but the residual attraction is counteracted by the quantum fluctuations. Subsequent experiments have confirmed the existence of QDs in dipolar~\cite{Barbut2016}, binary homonuclear~\cite{Cheiney2018} and heteronuclear~\cite{D'Errico2019} BECs.
One interesting property of quantum fluctuations is that the dimensional reduction from three dimension to two or one dimension in a BEC, due to the action of a tightly confining potential, induces different nonlinear terms in the Gross-Pitaevskii equation for different dimensional settings~\cite{Petrov2016}. For instance, in the one-dimensional (1D) limit, the Lee-Huang-Yang term has a negative sign, $E_{\mathrm{LHY}} \sim n^{3/2} $, and the beyond-mean-field correction in the Gross-Pitaevskii equation is quadratic in the condensate wavefunction, i.e., $\sim -|\psi| \psi$, rather than quartic as in the 3D case. In all space geometries, stationary and dynamical properties of QDs are investigated in Refs.~\cite{Otajonov2019, Otajonov2020, Otajonov2022_1, Otajonov2024} by using the Lagrangian formalism. 

For the elongated Bose-Bose mixtures with the transverse confinement $l_{\perp} \sim$ a few $\mu m$, using the effective 1D and 2D GPE with repulsive LHY term is relevant~\cite{Debnath2021, Debnath2022, Liu2023}.  
The formation of discrete quantum droplets in BECs has been studied in various theoretical models, including the existence and properties of discrete vortex quantum droplets with topological charges of up to $S=5$ in a binary BEC loaded in a deep 2D optical lattice.  Investigation of semidiscrete quantum droplets and vortices in a quasi-one-dimensional geometry is reported in reference~\cite{Zhang2019}. The findings indicate that these semidiscrete vortex quantum droplets can also be stabilized up to at least S = 5. In reference~\cite{Zhao2021}, the emergence and stability of discrete quantum droplets in one-dimensional optical lattices are investigated, along with an analysis of their mobility and collisions. In the papers~\cite{Katsimiga2023, Debnath2023, Khan2024, Debnath2023a, Sakkaf2024} the interaction of quantum droplets in 1D Bose-Bose mixtures is studied.

Modulational instability is a phenomenon that occurs when a perturbation of a nonlinear plane wave becomes unstable, leading to the formation of bright localized structures~\cite{Benjamin1967}. This instability is present in both continuous and discrete nonlinear systems~\cite{Kivshar1992}. In discrete systems, there are self-localized states known as nonlinear localized modes or intrinsic localized modes, which arise from the interplay between nonlinear effects and lattice coupling. These localized states are also referred to as discrete breathers. Discrete breathers have been studied in various fields of physics, including solid-state physics, biophysics, nonlinear optics~\cite{Sukhorukov2003, Kevrekidis2009}, photonic-crystal waveguides~\cite{Mingaleev2000}, and Bose-Einstein condensates in optical lattices~\cite{Trombettoni2001, Abdullaev2001, Morsch2006, Flach2008}. 

In this work, we investigate the modulational instability and discrete breather solutions in elongated BEC loaded in a deep optical lattice in the presence of quantum fluctuations. The paper is structured as follows. The model is introduced in Sec.~\ref{sec:model} for the description of a elongated BEC. In Sec.~\ref{sec:MI} modulational instability and instability regions of nonlinear plane waves are discussed. In Sec.~\ref{sec:Page}, we study the different types of discrete breather solutions of the system. The stability of these discrete breathers is also investigated in Sec.~\ref{sec:Page}.
The quasi-continuous limit is explored using the variational approach (VA) in Section~\ref{sec:Variational}. 
The results of the last three sections were also complemented with numerical simulations.
Finally, Sec.~\ref{sec:conc} concludes the paper.
\section{The Model}
\label{sec:model}
The dynamics of a two-component Bose-Einstein condensate in three-dimensional geometries in the beyond mean-field approach is described by the Gross-Pitaevskii equation. 
This equation includes the LHY-induced quartic repulsive term as indicated by references~\cite{Petrov2015, Otajonov2022_1}. 
\begin{eqnarray}
& i \hbar \cfrac{\partial \Psi}{\partial T} + \cfrac{\hbar^2}{2 m_0} \nabla^2 \Psi -U(x,y,z) \Psi + \cfrac{4 \pi \hbar^2 \delta a }{m_0} |\Psi|^2 \Psi 
\nonumber \\
& - \cfrac{256 \sqrt{2 \pi} \hbar^2 a^{5/2}}{3 m_0} |\Psi|^3 \Psi=0\,,
\label{dimGPE3D}
\end{eqnarray}
where $\Psi=\Psi(x,y,z,T)$ is the condensate wave function, $|\Psi|^2$ represents the density of the condensate, $\nabla^2$ is the Laplacian operator, which accounts for the spatial derivatives in all three dimensions, $T$ is time, $m_0$ is atomic mass, $U(x,y,z)=m_0 \omega_{\perp}^2 / 2\,(x^2+y^2)^2+V(z)$ is the external potential acting on the condensate, $\delta a=-a+a_{12}$ is residual scattering length, where $a_{11}=a_{22}=a$ and $a_{12}$ are the intra- and inter-species scattering lengths, respectively.

The quasi-1D regime of a system is characterized by the condition $\xi = g \tilde{n}/\hbar \omega_{\perp} \ll 1,$ where $g \tilde{n}$ represents the mean-field energy, $\tilde{n}$ is the density of the BEC and $g=4 \pi \hbar^2 a/m_0$ is coupling constant, and $\hbar \omega_{\perp}$ is the transverse confinement energy. References~\cite{Edler2017, Zin2018} analyze the LHY correction term, showing that this regime is realized in a Bose-Bose mixture for $\xi \leq 0.0004$. This condition corresponds to extremely narrow traps with a transverse confinement length of $ l_{\perp} = \sqrt{\hbar/m_0 \omega_{\perp}} \approx 20-30 \text{ nm}$. In this regime, the LHY correction forms an attractive interaction, with $E_{\mathrm{LHY}} \sim -|\Psi|^3.$ However, in typical experiments where $l_{\perp}$ is on the order of a few micrometres, the LHY correction is repulsive. For $\xi \geq 0.3$, the correction follows the 3D form, scaling as $\sim |\Psi|^5 $. Consequently, the elongated condensate factorization technique can be applied: 
\begin{equation}
\Psi(x,y,z,t)=R(x,y) \Phi (z,T).
\label{factor}
\end{equation}
The transverse distribution $R(x,y)=\exp[-(x^2+y^2)/2 l_{\perp}^2]/\sqrt{\pi} l_{\perp}$ is commonly approximated by a Gaussian function. The effect of interactions on this Gaussian distribution can be described by treating $l_{\perp}$ as a variational parameter~\cite{Salasnich1, Blackie1, Blackie2, Bloch_discr}. In the first approximation, $ l_{\perp}$ is typically taken as its unperturbed value, $ l_{\perp}^{(0)} =l_{\perp},$ as studied in both 1D and 2D cases~\cite{Debnath2021, Liu2023, Malomed1, Malomed2, China1, Edmonds2020}. Solving the nonlinear variational problem for $l_{\perp}$ reveals that, for $\xi \approx 0.3$, the deviation from $l_{\perp}^{(0)}$ is less than one percent, see Appendix~\ref{appenA} for details. This justifies the use of Eq.~(\ref{factor}) with $l_{\perp} = l_{\perp}^{(0)}$ in this case~\cite{Palo2023}. However, analyzing the case $\xi \ge 1$ in the discrete system required a separate investigation (see Ref.~\cite{Bloch_discr}). 
 By substituting Eq.(\ref{factor}) into Eq.(\ref{dimGPE3D}) and subsequently multiplying both sides of this equation by $R(x,y)$, we can then integrate the transverse variables. This integration leads us to the one-dimensional Gross-Pitaevskii equation:
\begin{eqnarray}
& i \hbar \cfrac{\partial \Phi}{\partial T} + \cfrac{\hbar^2}{2 m_0} \cfrac{\partial^2 \Phi}{\partial z^2} -U(z) \Phi + \cfrac{2 \hbar^2 \delta a }{m_0 l_{\perp}^2} |\Phi|^2 \Phi 
\nonumber \\
& - \cfrac{512 \sqrt{2} \hbar^2 a^{5/2}}{15 \pi m_0 l_{\perp}^3} |\Phi|^3 \Phi=0\,,
\label{dimGPE1D}
\end{eqnarray}
By introducing the following rescalings $t=T/t_S$, $z=z/z_s$, and $\psi=\Phi/\psi_s$, we can rewrite the Eq.(\ref{dimGPE1D}) in dimensionless form:
\begin{equation}
i \cfrac{\partial \psi}{\partial t}+ \cfrac{1}{2} \cfrac{\partial^2 \psi}{\partial z^2}-V(z) \psi + \tilde{\gamma} |\psi|^2 \psi- \tilde{\delta} |\psi |^3\psi=0\,,
\label{dlessGPE1D}
\end{equation}
where $V(z)=U(z) t_s / \hbar \psi_s$ is the rescaled external potential, and scale parameters are defined as:
$$t_s=\cfrac{2^{16}\, m_0 \tilde{\gamma}^3 a^5}{225\, \pi^2 \hbar \, \tilde{\delta}^2 \delta a^3 }\,, \quad
z_s=\cfrac{256}{15 \pi} \left( \cfrac{\tilde{\gamma}^3 a^5}{\tilde{\delta}^2 \delta a^3} \right)^{1/2}\,,$$  
$$\psi_s=\cfrac{15 \pi \delta a \, l_{\perp} \tilde{\delta} }{256 \sqrt{2}\, a^{5/2} \tilde{\gamma} } \, .$$
Equation~(\ref{dlessGPE1D}) can be normalized so that the coefficients satisfy $|\tilde{\gamma}|= |\tilde{\delta}| = 1$. However, we retain these notations in the dimensionless equation to simplify the analysis of different scenarios, including cases where either $\tilde{\gamma}$ or $\tilde{\delta}$ is zero.

The model we aim to construct involves the discretization of the continuous Gross-Pitaevskii equation using Wannier functions. When a BEC is loaded in a periodic optical lattice with a depth that is sufficiently deep, causing the barrier between adjacent sites is much higher than the chemical potential, and the energy of the system is limited to the lowest band, it can be modelled using the tight-binding approximation~\cite{Alfimov,Lewenstein2012}.
In this approach, the wave function of the BEC is expanded in terms of a set of orthonormal localized Wannier functions, which represent the wave function of a single particle localized around a lattice site. Using this approach, the Hamiltonian of the BEC in the optical lattice can be expressed as a sum of single-site Hamiltonians, which describe the energy of a particle localized at each lattice site, as well as hopping terms that describe the tunnelling of particles between neighbouring lattice sites. In the context of BEC, this system can be modelled by the discrete Gross-Pitaevskii equation.
Let us assume the wave function can be written as:
\begin{equation}
\psi(z, t) = \sum_{n} \psi_n(t) w(z - na)
\label{WannierFunc}
\end{equation}
where $\psi_n(t)$ are time-dependent complex coefficients and $w(z - na)$ represents the Wannier function centered at lattice site $n$, with $a$ being the lattice spacing. The external potential $V(z)=V(z+L)$ is periodic function with period $L$. 

Using the standard discretization technique, we derive a discrete equation that captures the essential dynamics of the system on a lattice~\cite{Trombettoni2001, Abdullaev2001, Lewenstein2012}:
\begin{equation}
i \psi_{n,t}+ \kappa (\psi_{n+1}+\psi_{n-1}-2 \psi_n) + \gamma |\psi_n|^2 \psi_n-\delta |\psi_n|^3\psi_n=0\,,
\label{dnlseQ1D}
\end{equation}
where $\kappa$ is the hopping rate between the lattice sites, $\gamma = \tilde{\gamma} \int |w(z)|^4 \, dz$ and $\delta = \tilde{\delta} \int |w(z)|^5 \, dz$. In the dimensionless Eq.(\ref{dnlseQ1D}), the parameters $\gamma$ and $\delta$ represent the strengths of the two-body interactions and quantum fluctuations, respectively.
The key steps of discretization involve multiplying both sides of the Eq.(\ref{dlessGPE1D}) by $w^*(z-m a)$ and integrating over $z$ then simplifying the time derivative using orthogonality conditions for Wannier functions$\int_{-\infty}^{\infty} w^*(z - m a)\, w(z - n a) \, dz = \delta_{nm}$, evaluating the linear terms to obtain onsite and hopping contributions $\kappa=\int w^*(z-ma)(\frac{1}{2} \frac{\partial^2}{\partial x^2} -V)w(z-(m \pm1)a)\,dz$, approximating the nonlinear terms by assuming the Wannier functions are highly localized, and adjusting the energy reference to simplify the final expression. Then, we replace the subindex $m$ with $n$ for notation purposes, resulting in Eq. (\ref{dnlseQ1D}).

The transformation $\psi_n \rightarrow \psi_n \exp(-2 i \kappa t)$ reduces Eq.~(\ref{dnlseQ1D}) to the following form~\cite{Kevrekidis2009}: 
\begin{equation}
i \psi_{n,t}+ \kappa (\psi_{n+1}+\psi_{n-1}) + \gamma |\psi_n|^2 \psi_n-\delta |\psi_n|^3\psi_n=0\,,
\label{moddnlseQ1D}
\end{equation}
In the next sections, we will deal with this equation. 

\section{Modulational Instability}
\label{sec:MI}

In this section, we examine the stability of stationary plane-wave solutions, as described by Eq.~(\ref{moddnlseQ1D}), when subjected to small amplitude modulations.
\begin{equation}
\psi_n=A \exp[i(q n +k t)]
\label{pwQ1D}
\end{equation}
By using Eq.~(\ref{pwQ1D}) we get following nonlinear dispersion relation:
\begin{equation}
k=2\kappa\, \mathrm{cos}(q)+\gamma A^2-\delta A^3
\label{kk}
\end{equation}
The stability of the solution can be analyzed by evaluating its sensitivity to small initial perturbations. To study the initial stage of the evolution, linear stability analysis can be applied, which involves seeking a solution in the form:
\begin{equation}
\psi_n=(A+\xi_n) \exp[i(q n +k t)],
\label{pertPW}
\end{equation}
where we assume a small perturbation $\xi_n \ll A$ of the form:
\begin{equation}
\xi_n=\varphi_1 \exp[i(Q n +K t)]+\varphi_2 \exp[-i(Q n +K t)]
\label{xik}
\end{equation}
where $Q$ is perturbation wave number. By inserting Eq.~(\ref{pertPW}) into Eq.~(\ref{moddnlseQ1D}) and after linearization, it results in an eigenvalue problem for the wave vector $K$ of the perturbation.
\begin{equation}
  \left| \begin{matrix}
   -K+f^+  & A^2 \Lambda \\
   A^2 \Lambda & K+f^-  \\
   \end{matrix} \right|=0 \, ,
\label{det}
\end{equation}
where $f^{\pm} \equiv 2 \kappa \,[\mathrm{cos}(q \pm Q)-\mathrm{cos}(q)]+A^2 \Lambda $, and $\Lambda \equiv \gamma -\cfrac{3}{2}\, \delta A$.

\begin{figure}[htbp]
  \centerline{ \includegraphics[width=4.55cm]{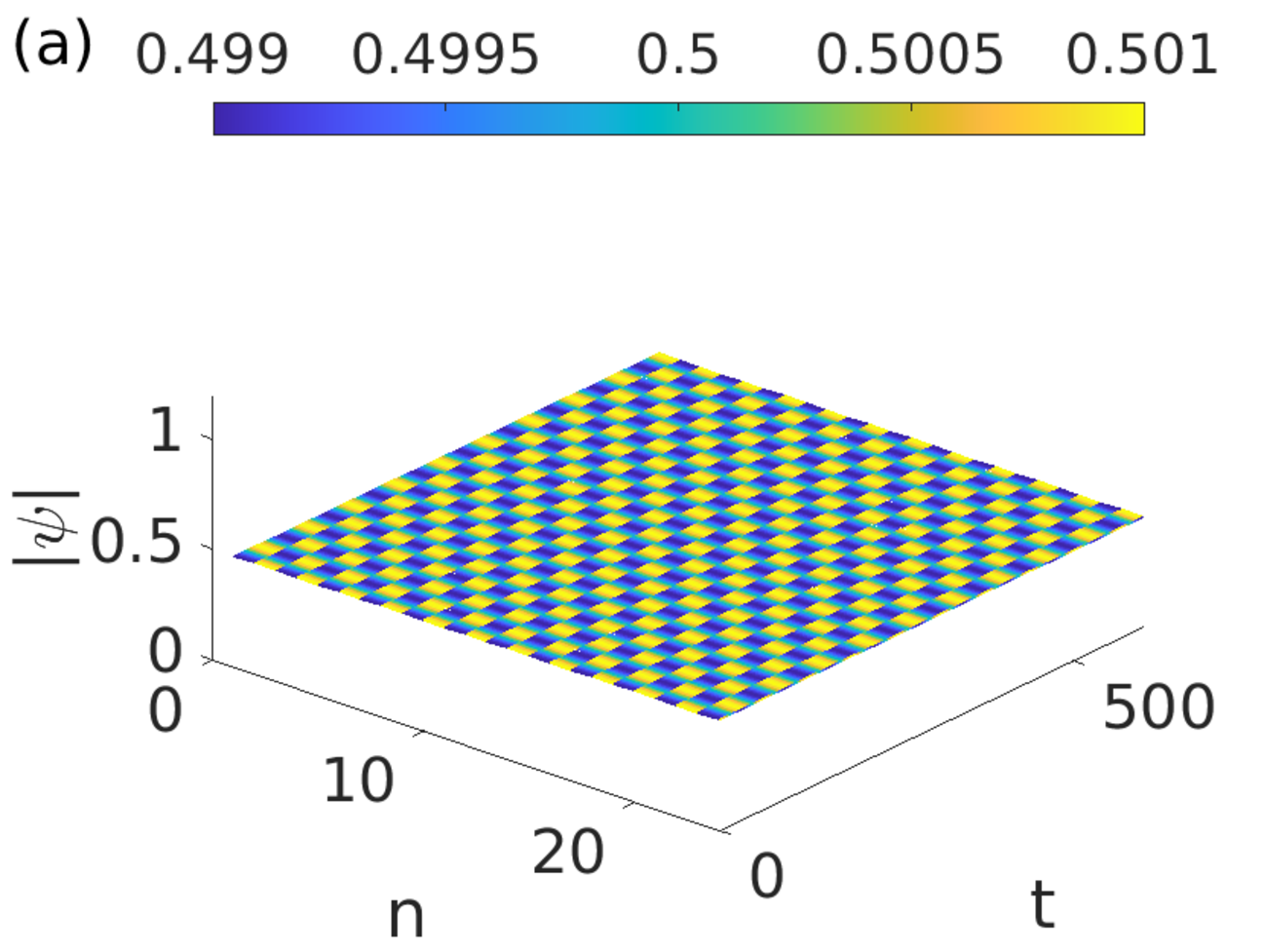} 
  \includegraphics[width=4.5cm]{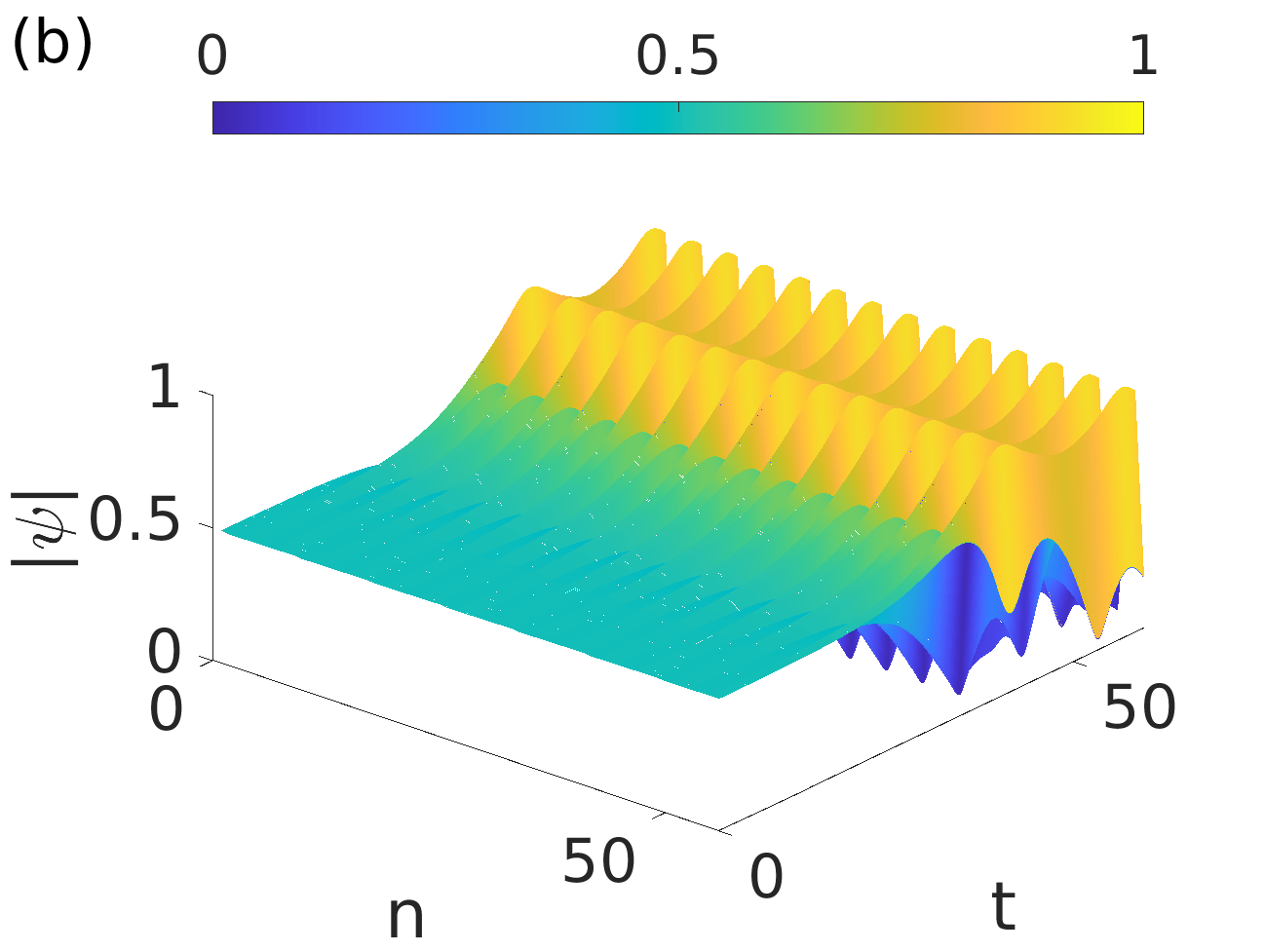}}
\caption{Evaluation of slightly perturbed plane wave solutions. (a) For $Q=\pi>Q_{\mathrm{cr}}$. (b) For $Q=\pi/2<Q_{\mathrm{cr}}$. Other parameters are $A=0.5$, $\kappa=0.1$, $q=\pi$, $\gamma=0$ and $\delta=1$.}
\label{fig-1ab}
\end{figure}

The eigenvalue problem (\ref{det}) leads quadratic equation on $K$ 
\begin{equation}
K^2-(f^+-f^-)K+A^4 \Lambda^2 -f^+ f^-=0
\label{kveq}
\end{equation}
and its solution is
\begin{equation}
K^{\pm}=\cfrac{1}{2}\,[\,f^+-f^- \pm \sqrt{(f^++f^-)^2-4 A^4 \Lambda^2 }\,].
\label{K12}
\end{equation}
The imaginary part will determine the gain spectrum, which represents the growth rate of the perturbations.
\begin{equation}
G=|\mathrm{Im}(K^{\pm})|=\cfrac{1}{2}\,\left|\mathrm{Im}[\,f^+-f^- \pm \sqrt{(f^++f^-)^2-4 A^4 \Lambda^2 }\,]\right|.
\label{gain}
\end{equation}
The maximum value of the wavenumber in the MI gain spectrum 
\begin{equation}
Q_{\mathrm{max}}^\pm=\pm \mathrm{\cos^{-1}}[\cos(2q)-\cfrac{A^2 \Lambda}{2\kappa} \cos(q)]
\label{Qmax}
\end{equation}
is associated with the most unstable mode. This mode experiences the highest amplification or growth rate among all possible perturbations. The values of $Q_{\mathrm{max}}$ is determined from the extremum of Eq.~(\ref{gain}).
The critical value of the wave number 
\begin{equation}
Q_{\mathrm{cr}}^\pm=\pm \mathrm{\cos^{-1}}[\cos(2q)-\cfrac{A^2 \Lambda}{\kappa} \cos(q)]
\label{Qcr}
\end{equation}
corresponds to the threshold value where the MI gain spectrum transitions from unstable behaviour to stability.
We mention that the Eq.~(\ref{Qcr}) proves to be valuable when the gain spectrum exhibits a ``butterfly" shape.
In the case where the wavenumber Q exceeds the critical value, where the growth rate of MI becomes insignificant, the plane-wave solutions remain unchanged, as illustrated in  Fig.~\ref{fig-1ab} (a). Conversely, when $Q$ is smaller than the critical value, the amplitude of the slightly perturbed plane wave experiences exponential growth, as depicted in Fig.~\ref{fig-1ab} (b). It is important to note that the linear stability analysis solely provides the condition for the onset of MI and does not provide insight into the further evolution of the wave field. The nonlinear evolution stage of the wave pattern shown in Fig.~\ref{fig-1ab} (b) goes beyond the scope of linear theory.

To facilitate our analysis, we examine the staggered and unstaggered cases independently. In both cases, we have $f^+=f^-$. First, let us consider the staggered case, where adjacent elements are out of phase, $q=\pi$. In this case, Eq.(\ref{gain}) takes the form:
\begin{equation}
G=2\,\left| \mathrm{Im} \left[ 2 \kappa\,\mathrm{sin}^2\left({Q \over 2}\right) \left( 2 \kappa\,\mathrm{sin}^2\left({Q \over 2}\right)+ A^2 \Lambda \right) \right]^{1/2} \right|.
\label{gainQ=pi}
\end{equation}
Now, let us turn our attention to the unstaggered case, where adjacent elements are in phase, $q=0$. 
In this case, the growth rate has the form:
\begin{equation}
G=2\,\left| \mathrm{Im} \left[ 2 \kappa\,\mathrm{sin}^2\left({Q \over 2}\right) \left( 2 \kappa\,\mathrm{sin}^2\left({Q \over 2}\right)- A^2 \Lambda \right) \right]^{1/2} \right|.
\label{gainQ=pi}
\end{equation}

When a synthetic gauge field is introduced between the lattices, the hopping acquires a complex phase factor: $\kappa \rightarrow \kappa e^{i \theta}$, where $\theta$ is the gauge phase associated with the hopping process. If we choose $\theta = \pi$, then: $\kappa e^{i \theta} = -\kappa$. Thus, the hopping rate switches from positive to negative, altering the system’s dynamics~\cite{Celi2014, Zhao2021}. We reveal an inherent symmetry within the system, where a transition between the staggered and unstaggered configurations can be achieved simply by changing the sign of the hopping rate. For example: the case $ q = \pi, \kappa > 0$ is equivalent to the case $q = 0$, $\kappa < 0$. Similarly, the case $q = \pi$, $\kappa < 0$ is equivalent to $q = 0 , \kappa > 0$. 

\begin{figure}[htbp]
  \centerline{ \includegraphics[width=4.55cm]{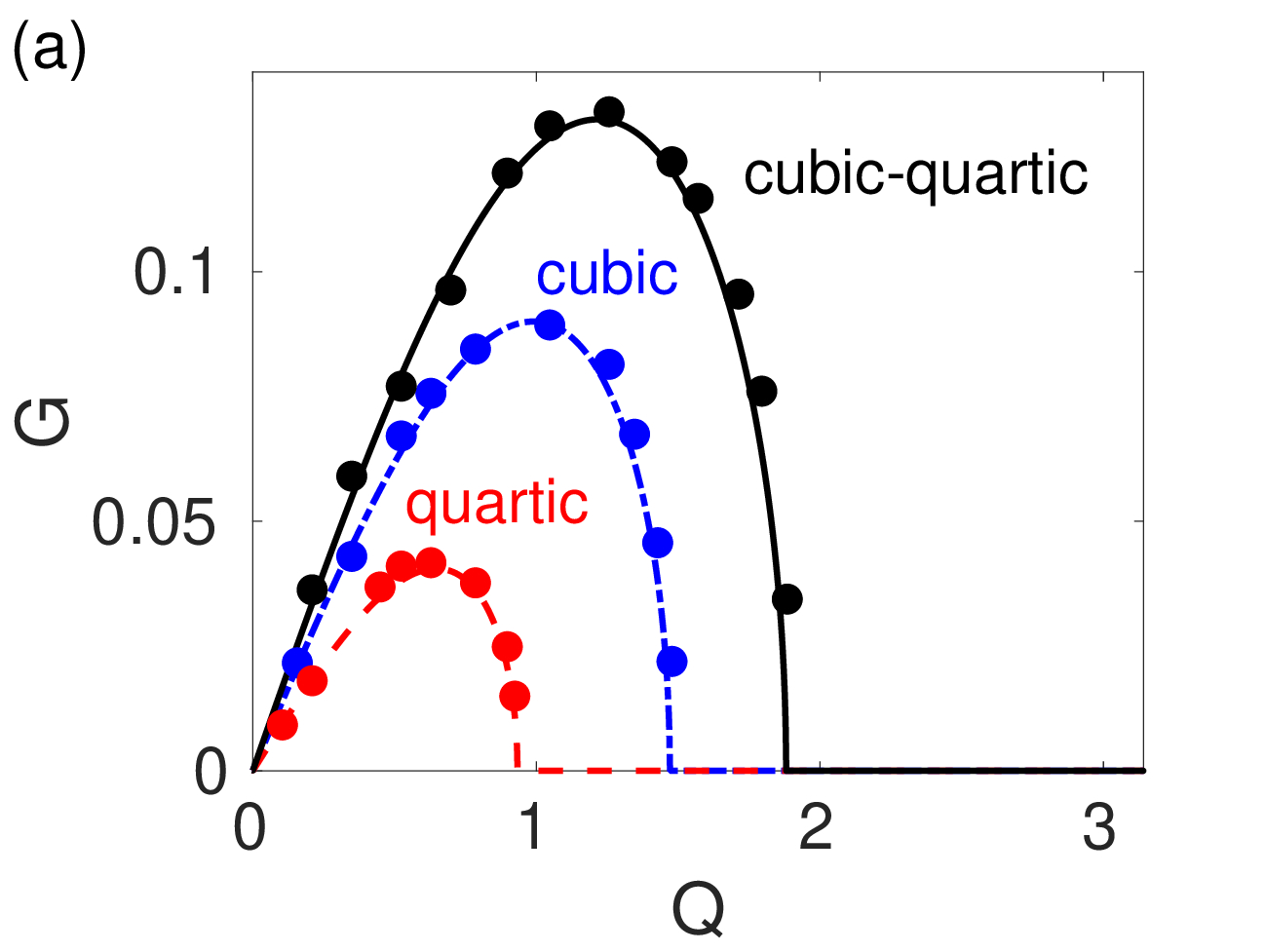} 
  \includegraphics[width=4.5cm]{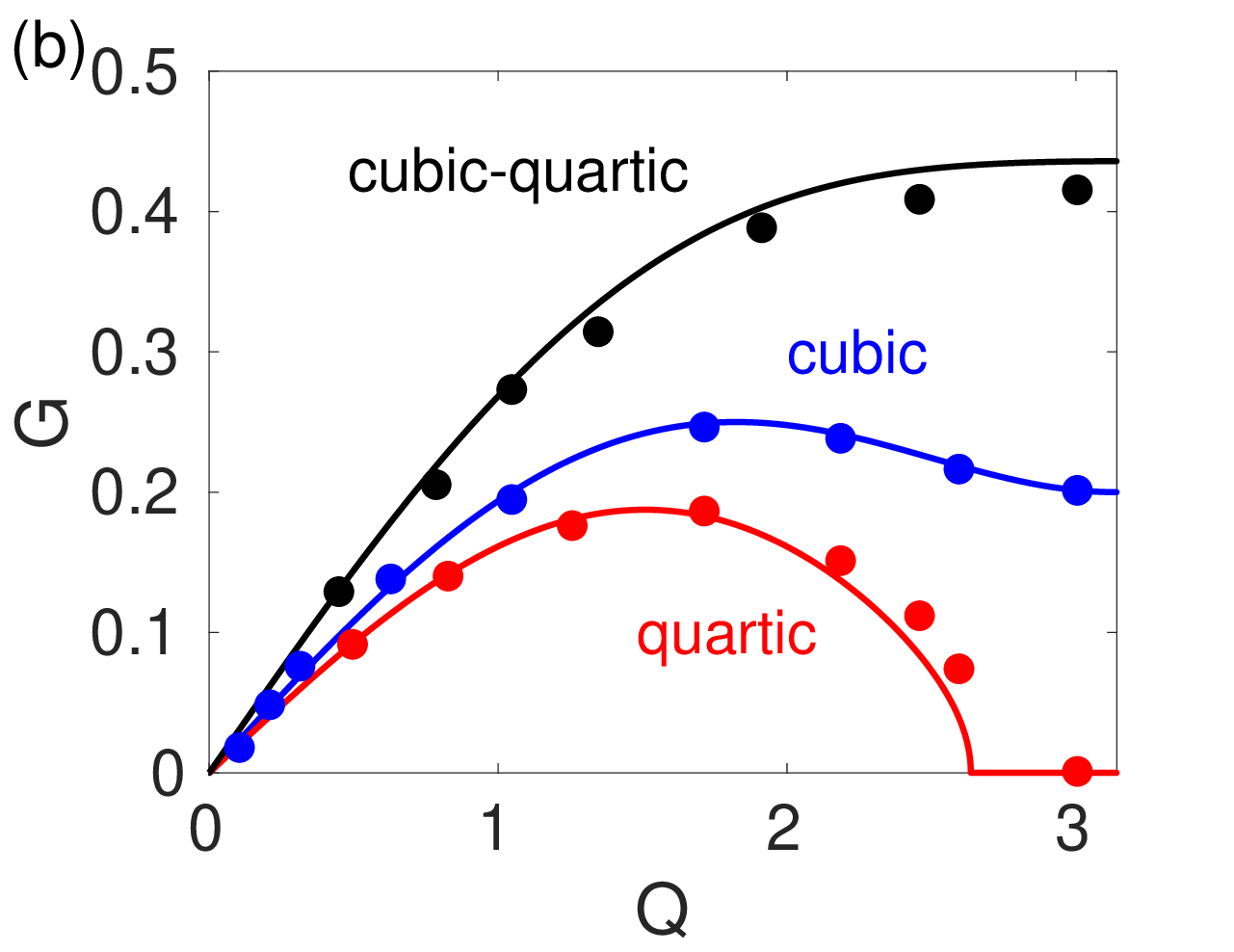}}
\caption{Typical gain spectrum of MI. Lines are found from Eq.~(\ref{gainQ=pi}) and points are found from direct numerical simulations of Eq.~(\ref{moddnlseQ1D}). (a) For $A=0.3$ and the bottom to top curves is for the different values of $(\gamma, \delta)$ that correspond to $(0,1)$, $(-1,0)$ and $(-1,1)$, respectively. (b) The same plot as in the (a) panel, but for $A=0.5$. Other parameters are $\kappa=0.1$, $q=\pi$.}
\label{fig-2ab}
\end{figure}

Figure \ref{fig-2ab} shows the individual impacts of two-body interactions (cubic nonlinear term) and quantum fluctuations (quartic nonlinear term), as well as their collective effect, on the growth rate of MI. 
Due to the even symmetry of the growth rates of MI as described by Eqs.~(\ref{gain})-(\ref{gainQ=pi}) with respect to the perturbation wave number $Q$, we confine our analysis to positive values of this parameter. For negative values of $Q$, the curves exhibit symmetry with respect to the gain axis.

To show the role of specific nonlinearities, we compare the cases of purely cubic, purely quartic, and combined cubic-quartic nonlinearities for staggered states with residual repulsive mean-field interactions, as shown in Fig.~\ref{fig-2ab}~(a) and (b). All lines in these figures are obtained from theoretical predictions, see Eq.~(\ref{gainQ=pi}). 
It can be seen from these figures that the quantum fluctuations cause a notable reduction in the instability domain and a decrease in the value of the growth rate, in contrast to the impact of two-body interactions, when using identical parameters. Additionally, the combined influence of these two interaction terms induces instability in regions where their individual contributions are zero, as seen in the top lines of Fig.~\ref{fig-2ab}~(a). 

Regardless of the specific combination of nonlinear terms whether one is present alone or both coexist, the gain spectrum can take one of two characteristic forms. These forms emerge from slight variations in the amplitude of the plane wave. In the first case, the gain spectrum exhibits a butterfly-like shape, where instability arises for wave numbers $Q$ smaller than the critical value $Q_{\mathrm{cr}}$. In the second case, no specific critical wave number exists, and modulation instability can occur for all values of $Q$ within the interval $0<Q \leq \pi$, as shown in the top and middle lines of Fig.~\ref{fig-2ab}~(b). Notably, $Q = \pi$ corresponds to the edge of the first Brillouin zone.
Points in Fig.~\ref{fig-2ab} are obtained from numerical simulations of Eq.~(\ref{moddnlseQ1D}). To compute the growth rates for different wave numbers $Q$, we used Eq.~(\ref{pertPW}) as the initial condition, with $\varphi_1 = \varphi_2 = 10^{-3}$ representing the perturbation strength. Over time, the amplitude of the perturbed plane waves exhibits exponential growth, and its logarithmic dependence on time closely follows a linear trend. By fitting this curve to a linear function $y = a x + b$, we extract the slope $a$, which corresponds to the exponential growth rate (or gain), as shown by the points in Fig.~\ref{fig-2ab}~(a) and (b). The strong agreement between theoretical predictions and numerical results confirms the accuracy and reliability of the developed method.

\begin{figure}[htbp]
  \centerline{ \includegraphics[width=4.5cm]{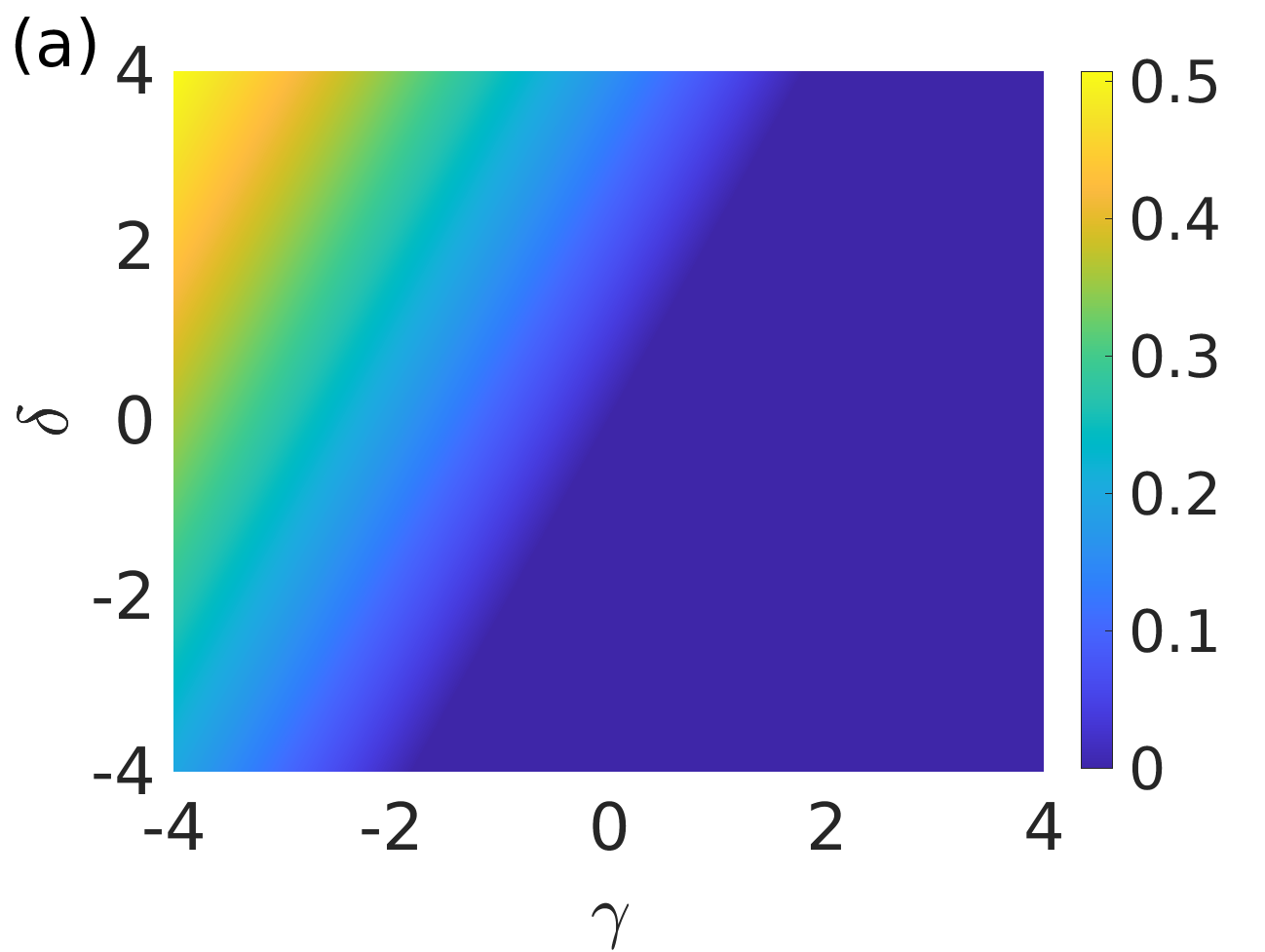} 
       \includegraphics[width=4.5cm]{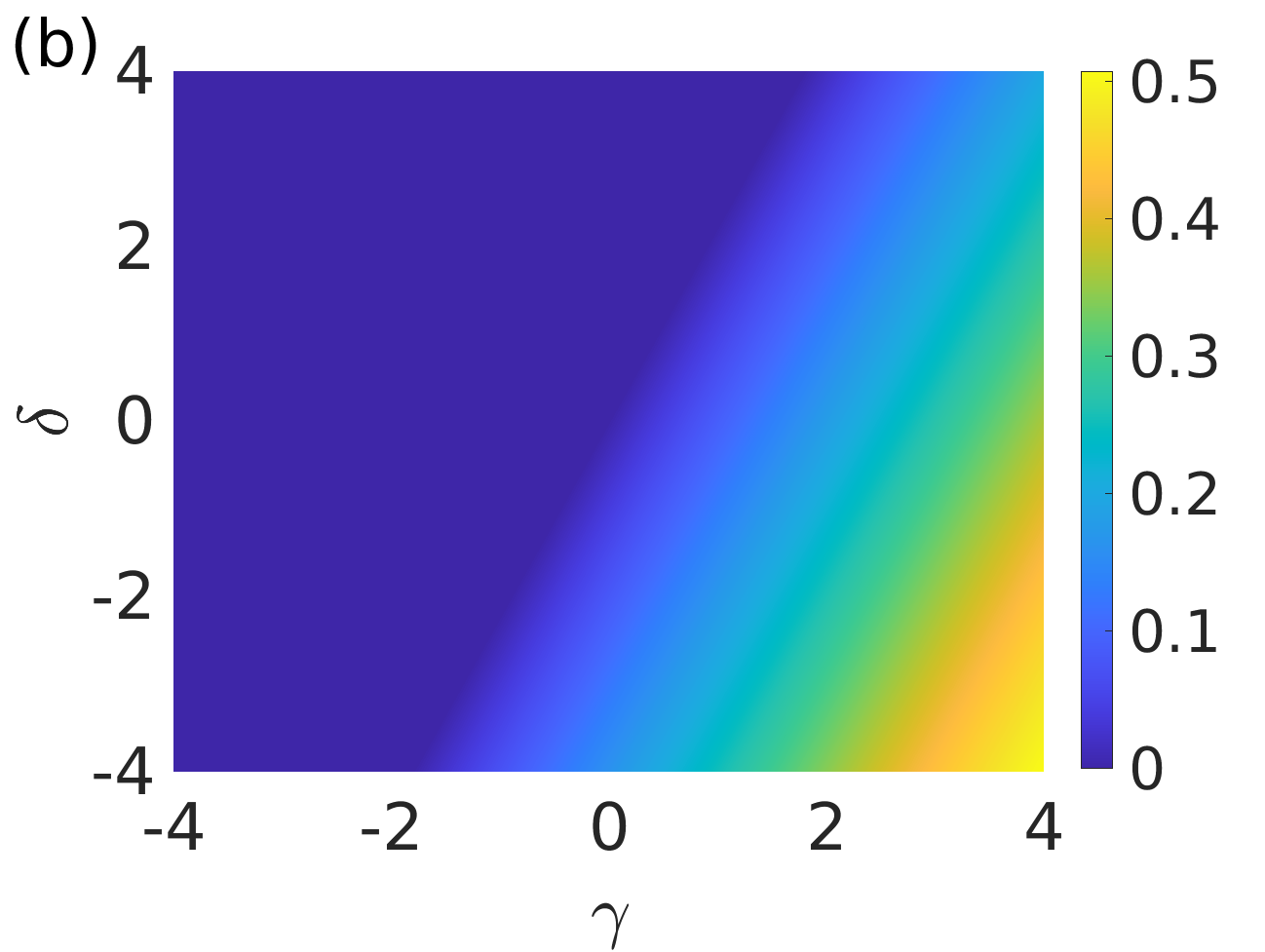}}
\caption{Modulational instability regions in the $(\gamma,\delta)$ plane for different signs of hopping rate. (a) $\kappa=0.1$. (b) $\kappa=-0.1$. In both figures color bar represents maximum values of gain. Other parameters are $A=0.3$ and $q=\pi$.}
\label{fig-3ab}
\end{figure}

Figure \ref{fig-3ab} displays the domains of instability, representing the maximum MI growth rate within the range of $Q \in [0, \pi]$, for various combinations of the $\gamma$ and $\delta$ parameters.
We analyzed the different ranges of the initial parameters and found that the system is always unstable for $\gamma<0$ and $\delta>0$ in contrast to $\gamma>0$ and $\delta<0$ cases where the system is always stable, see Fig.~\ref{fig-3ab}(a). In other cases, the system can be both stable and unstable depending on the parameters. When the sign of the hopping rate changes, the system undergoes a drastic shift in behaviour, leading to a complete reversal. This means that previously stable regions become unstable, while previously unstable regions become stable, see Fig.~\ref{fig-3ab}(b). 

\begin{figure}[htbp]
  \centerline{ \includegraphics[width=4.5cm]{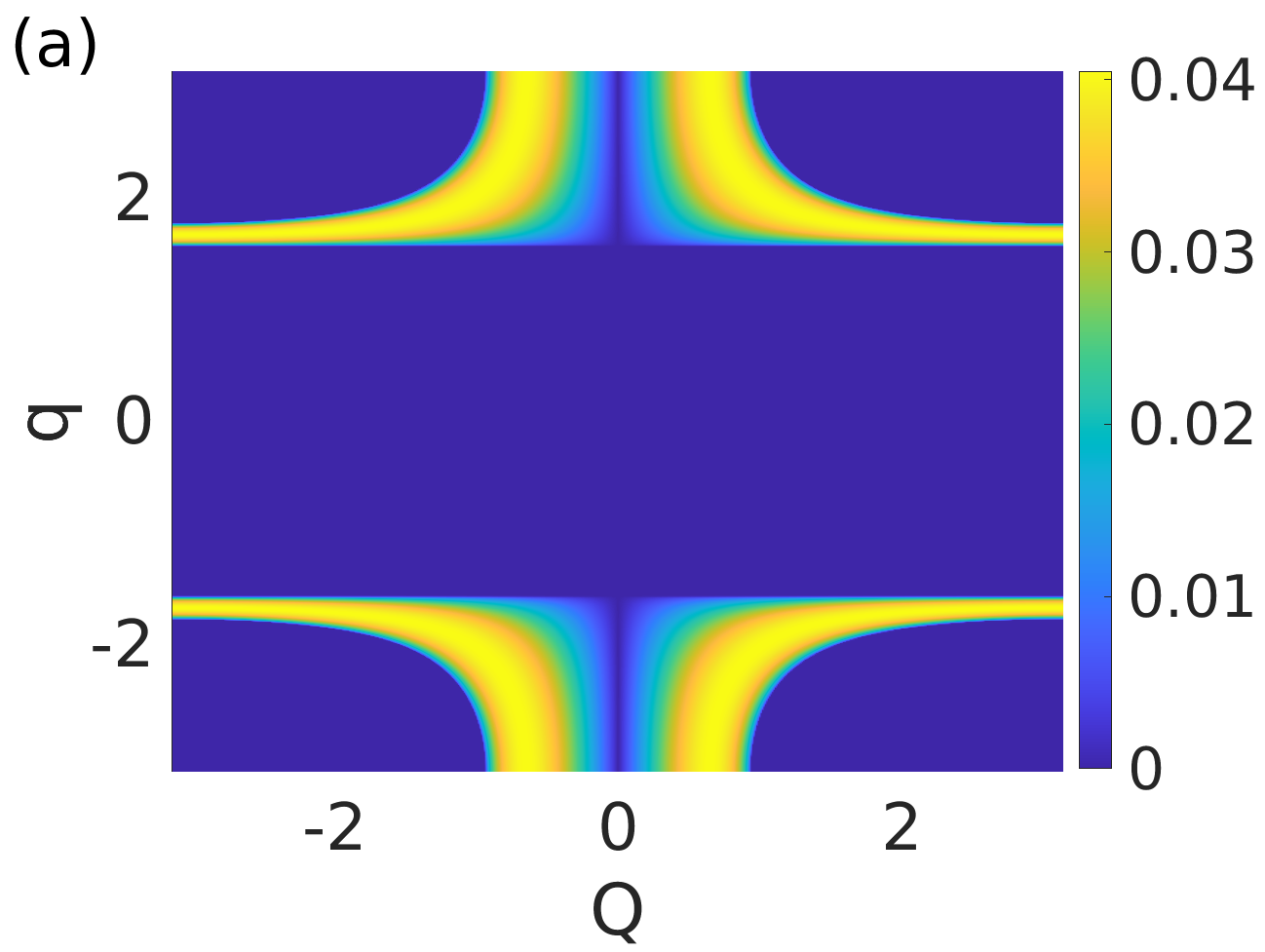} 
     \includegraphics[width=4.5cm]{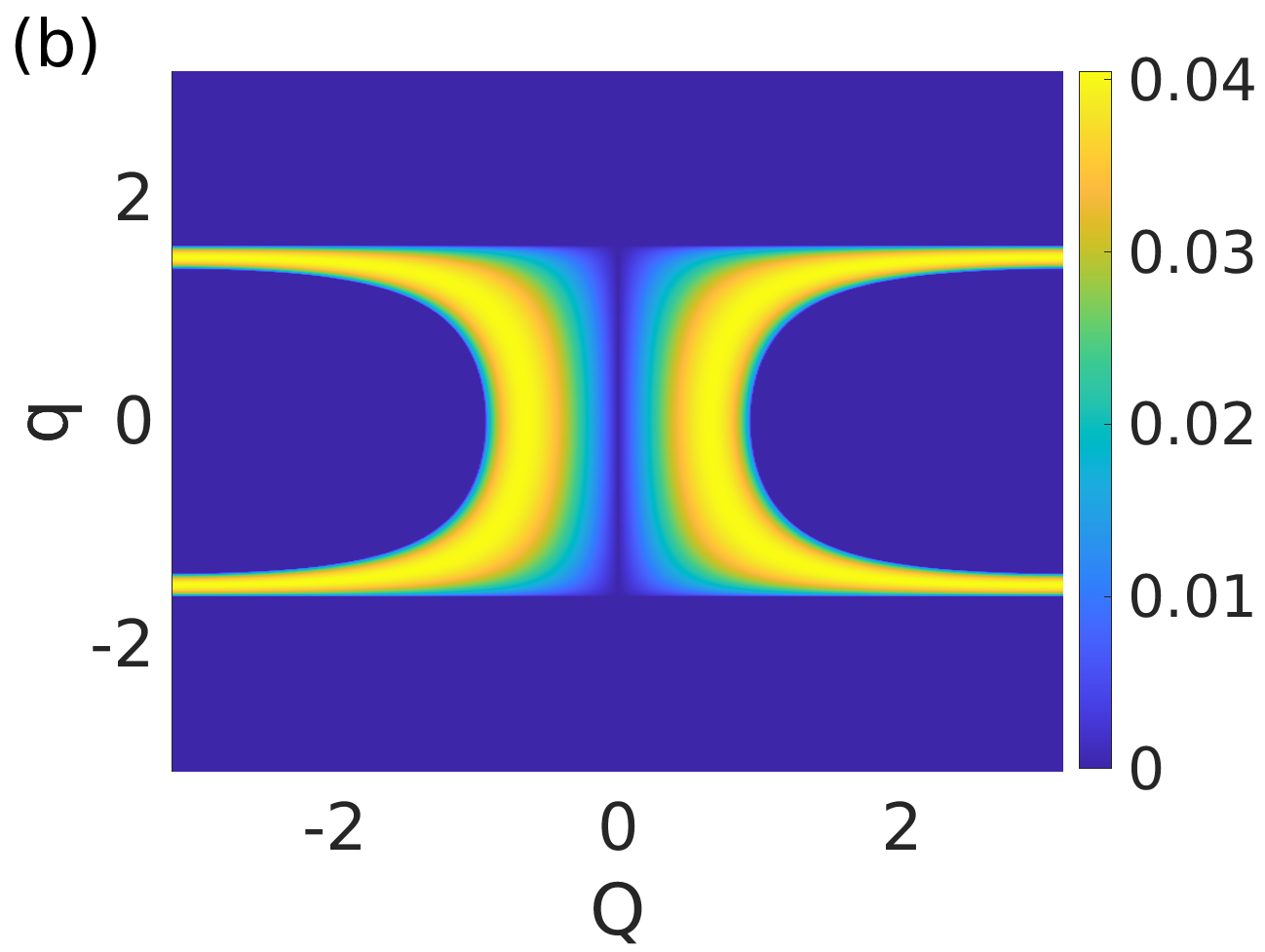}}
\caption{Modulational instability regions in the $(Q,q)$ plane. (a) $\kappa=0.1$. (b) $\kappa=-0.1$. In both figures color bar represents maximum values of gain. Other parameters are $A=0.3$, $\gamma=0$ and $\delta=1$.}
\label{fig-4ab}
\end{figure}

Figure~\ref{fig-4ab} depicts the instability domain in the $(Q,q)$ plane for different signs of the hopping rate. We analyzed multiple sets of initial parameters and observed that when $\kappa>0$, the system remains stable within the $-\pi/2 \le q \le \pi/2$ range, see Fig.~\ref{fig-4ab} (a). However, beyond this range, instability can occur depending on the initial parameters. On the other hand, for the $\kappa<0$ case, instability occurs within $-\pi/2 < q < \pi/2$, while the system remains stable beyond this range, see Fig.~\ref{fig-4ab} (b).

To validate the predictions obtained through linear stability analysis, we performed numerical simulations of the discrete governing GPE by introducing a slight perturbation ($10^{-3}$) to the initial condition, see Fig.~\ref{fig-2ab}. Equation~(\ref{moddnlseQ1D}) describes a set of coupled ordinary differential equations (ODEs) for $n$ complex variables $\psi_n$. By separating the real and imaginary parts of this equation, we obtain a system of $2n$ coupled real ODEs. The initial condition is also split into real and imaginary parts to provide appropriate initial conditions for each solution component. To solve this system of equations, we used MATLAB's ode45 solver with periodic boundary conditions. Relative and absolute tolerance of $10^{-12}$ apply to the real and imaginary parts of the wavefunction. In the nonlinear stage of MI, instability leads to the formation of bright localized modes, see Fig.~\ref{fig-1ab} (b). In the next sections, we consider the properties of different localized modes known as discrete breathers. 

\section{Strongly localized discrete breathers}
\label{sec:Page}
Discrete breather solutions of Eq.~(\ref{moddnlseQ1D}) can be obtained by employing the steady-state ansatz 
\begin{equation}
\psi_n=e_n \exp(-i \mu t),
\label{StatEQ}
\end{equation}
where $\mu$ is chemical potential. Real stationary lattice ﬁeld of an intrinsic localized mode $e_n$ satisfies the following infinite algebraic lattice equation:
\begin{equation}
\mu\, e_n + \kappa \,(e_{n+1}+e_{n-1})+\gamma\, |e_n|^2 e_n -\delta \,|e_n|^3 e_n=0
\label{discStatEQ}
\end{equation}
Through approximating solutions for this infinitely discrete system, we aim to reduce it to a finite set of equations, enabling the identification of localized solutions with different topologies. We employ the Page method \cite{Page} to initiate the process with bright solutions and progress to explore other solution types.

In this section, we conducted a linear stability analysis on discrete quantum droplets by introducing slight perturbations to the stationary plane wave solution~(\ref{StatEQ}):
\begin{equation}
\psi_n=(e_n+f_n \exp(-i \lambda t)+g_n^* \exp (i \lambda^* t)) \exp(-i \mu t),
\label{pertQD}
\end{equation}
where $f_n$ and $g_n$ represent the perturbation eigenmodes, and the asterisk signifies the complex conjugate. 
Inserting Eq.~(\ref{pertQD}) into Eq.~(\ref{StatEQ}) and keeping only linear terms of the perturbation yield the following eigenvalue problem in matrix representation:

\begin{equation}
\lambda 
\begin{pmatrix}
f_n \\
g_n
\end{pmatrix} 
= 
\begin{pmatrix}
\hat{L}_{1} & \hat{L}_{2} \\
-\hat{L}_{2}^* & -\hat{L}_{1}
\end{pmatrix}
\begin{pmatrix}
f_n \\
g_n
\end{pmatrix}
\label{LinEignVal}
\end{equation}
where 
$\hat{L}_{1}=-\kappa (\delta_{n,n+1}+\delta_{n,n-1}) - 2 \gamma |e_n|^2 + \cfrac{5 \delta}{2} |e_n|^3-\mu$,
$\hat{L}_{2}=-\gamma e_n^2 +\cfrac{3 \delta}{2} |e_n| e_n^2$.

We numerically solve the linear eigenvalue problem derived from Eq.~(\ref{LinEignVal}). The presence of an imaginary part in the spectrum $\lambda$ indicates the instability of the solutions and the growth of perturbations in the linear approximation. However, the total energy of the system remains conserved. 
The characteristic values of the eigenvalues are qualitatively determined by the strength of the nonlinear terms in the Gross-Pitaevskii equation and the structure of the solutions. In our case, these eigenvalues can only be calculated numerically.
Also, the numerical results for strongly localized discrete modes are presented in the next sections.
\subsection{Even bright quantum droplets}
\label{sec:Even}
The even (intersite) mode is characterized by the amplitudes, denoted as 
\begin{equation}
e_n=A(...,0,\alpha_3,\alpha_2,1,s,s\alpha_2,s\alpha_3,0,...)\, ,
\label{evenMode}
\end{equation}
with $|n|=1,2,3,...$. Modifying the signs of $s$ and $\alpha$ gives rise to unique topologies of even modes~\cite{Darmanyan1998JETP}. We focus specifically on the strongly localized symmetric modes $s = \pm 1$. The strong localization requires $|\alpha_3|\ll |\alpha_2| \ll 1$, $\alpha_n \approx 0$ for $n>3$. 

By substituting the Eq.~(\ref{evenMode}) into Eq.~(\ref{discStatEQ}), we found the following dispersion relation: 
\begin{eqnarray}
\mu= -\gamma A^2 + \delta A^3 - \kappa s + \cfrac{\kappa^2}{-\gamma A^2 +\delta A^3}, 
\label{discMU}
\end{eqnarray}
and small amplitudes 
\begin{eqnarray}
& \alpha_2=-\cfrac{\kappa}{-\gamma A^2 +\delta A^3} - s \left(\cfrac{\kappa}{-\gamma A^2 +\delta A^3} \right)^2,
\nonumber \\
& \alpha_3= \left(\cfrac{\kappa}{-\gamma A^2 +\delta A^3} \right)^2
\label{discALPHA}
\end{eqnarray}
of the solution. For the reasons of symmetry, the subscript $n=0$ has been omitted. 

\begin{figure}[htbp]
  \centerline{ \includegraphics[width=4.55cm]{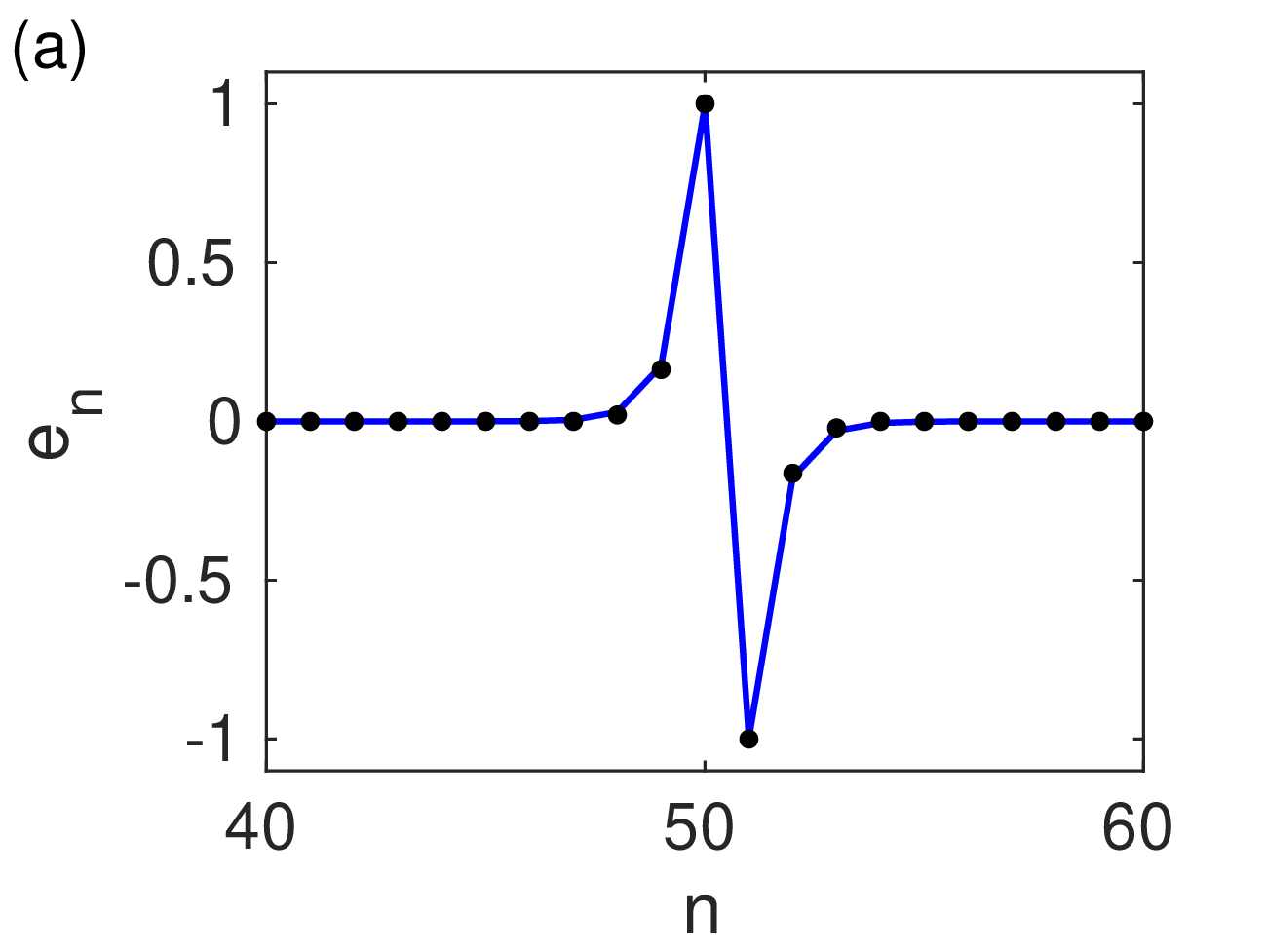} 
  \includegraphics[width=4.5cm]{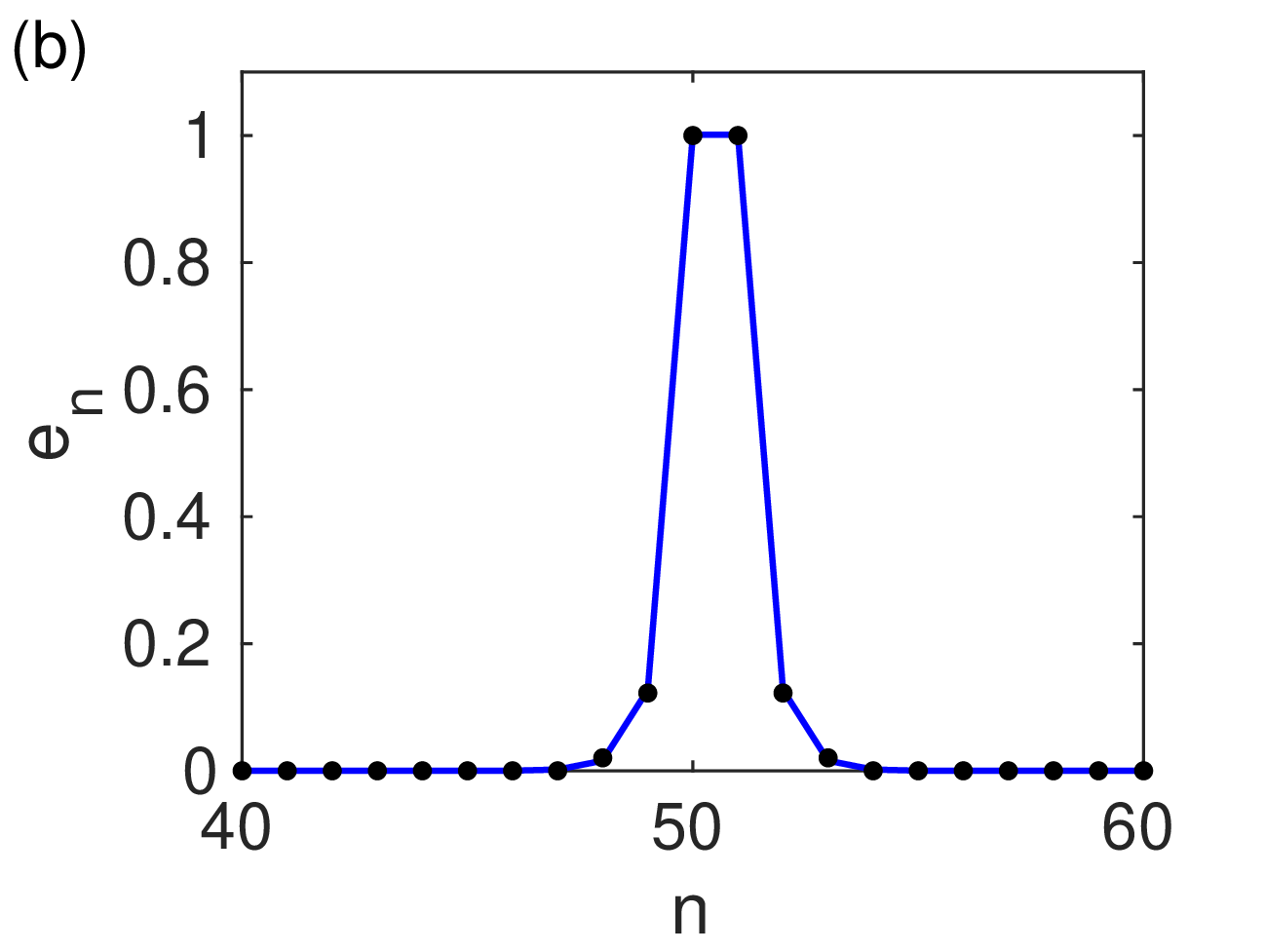}}
\caption{Even strongly localized modes (a) s=-1. (b) s=1. The solid line represents the solution of Eq.~(\ref{discStatEQ}) by the Newton method, and the points represents the approximate solution by the Page method given by Eq.~(\ref{discMU}). Other parameters are $A=1$, $\kappa=0.1$, $\gamma=1$ and $\delta=0.3$. }
\label{fig-5}
\end{figure}

The Fig.~\ref{fig-5}, show the results derived from the numerical solution of Eq.~(\ref{discStatEQ}) by Newton method, as the initial guess we pick the approximate solution given by Page method, see Eqs.~(\ref{discMU}) and (\ref{discALPHA}). 
Fig.~\ref{fig-5} (a) corresponds to the antisymmetric even localized mode, while the Fig.~\ref{fig-5} (b) represents results for symmetric even localized mode.   

One can see that exact solutions are quite close to approximate ones, so one may conclude that the Page method works well in these cases.

\begin{figure}[htbp]
  \centerline{ \includegraphics[width=4.55cm]{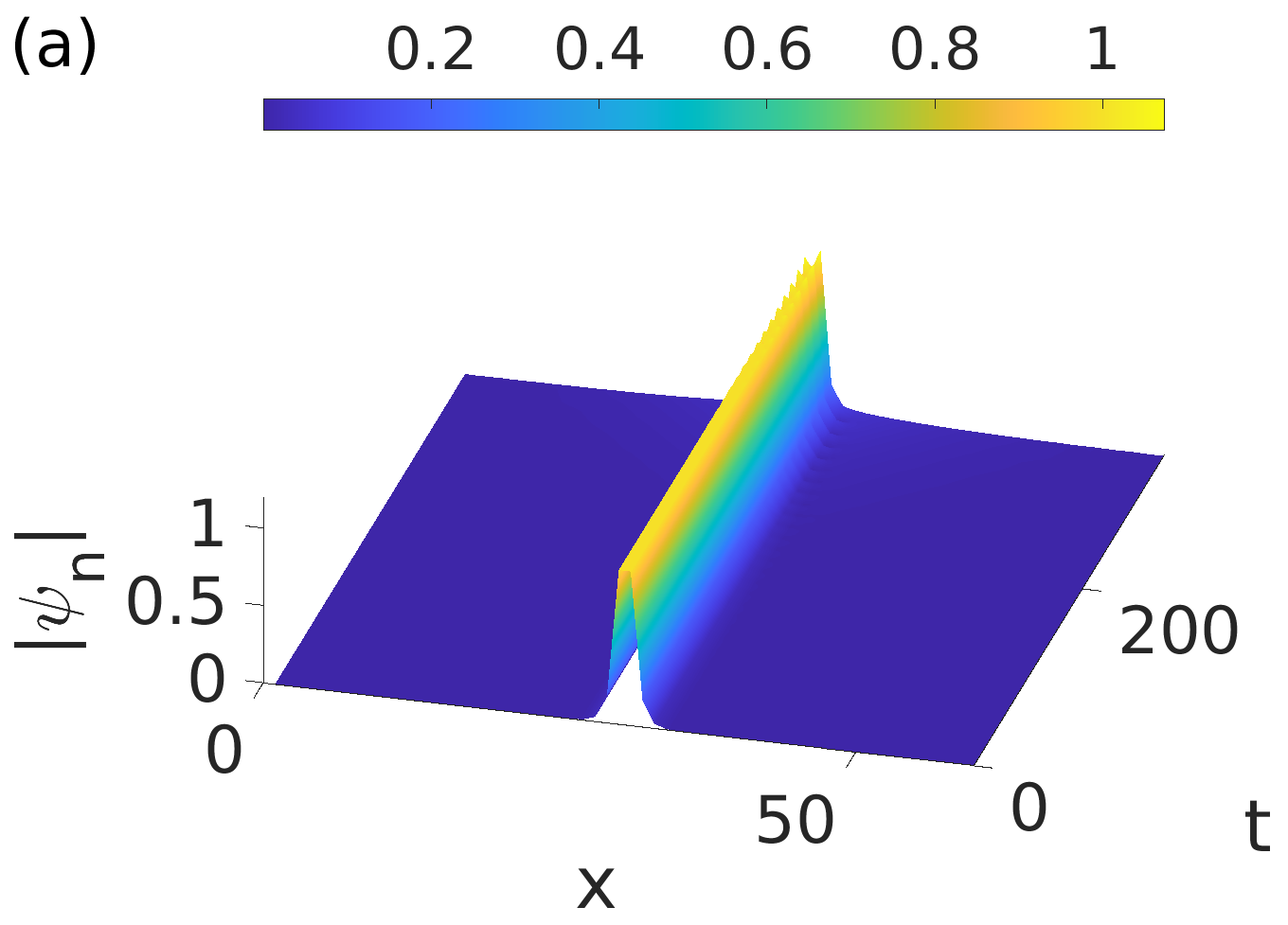} 
  \includegraphics[width=4.5cm]{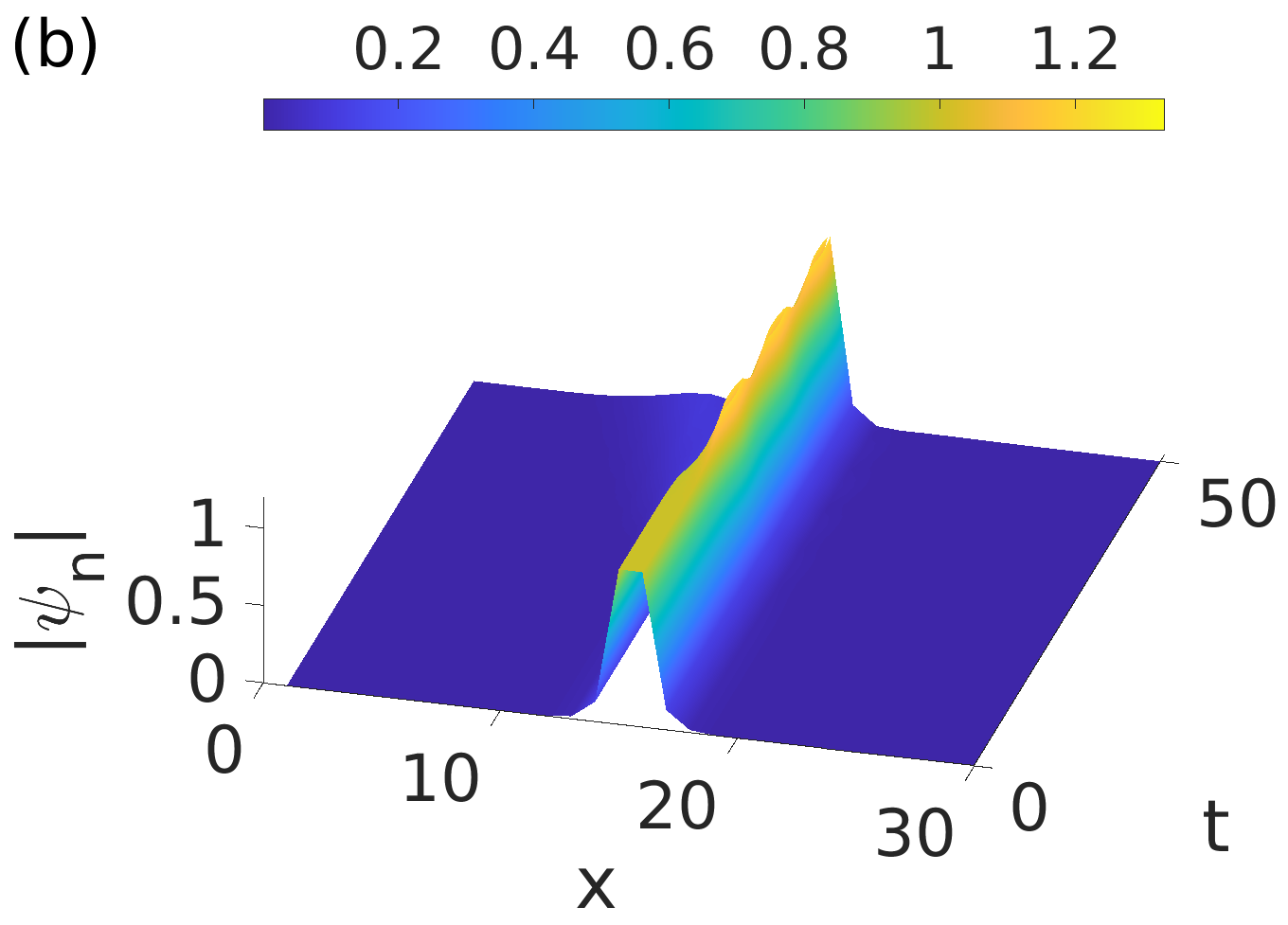}}
  \centerline{ \includegraphics[width=4.1cm]{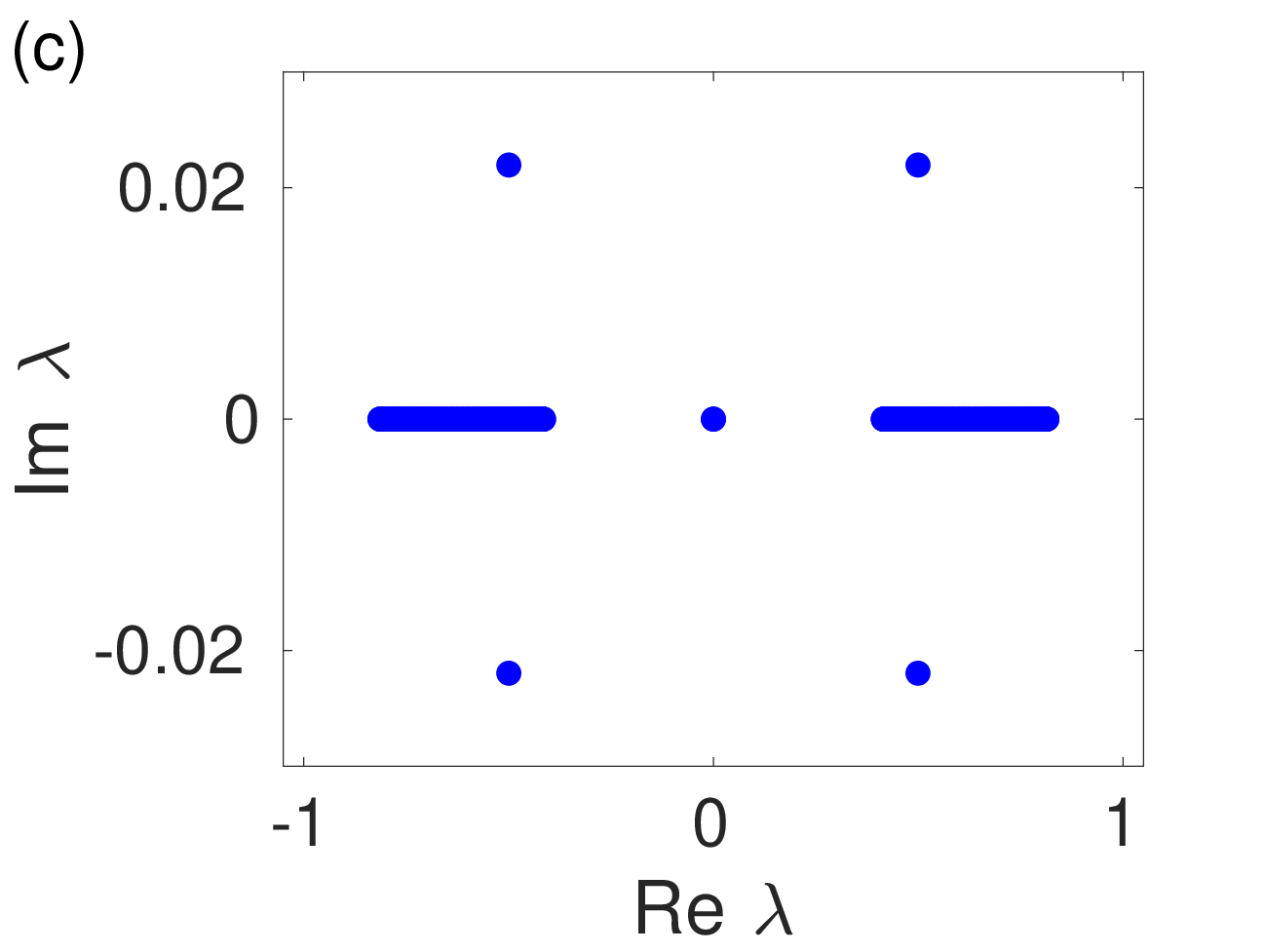}
   \includegraphics[width=4.1cm]{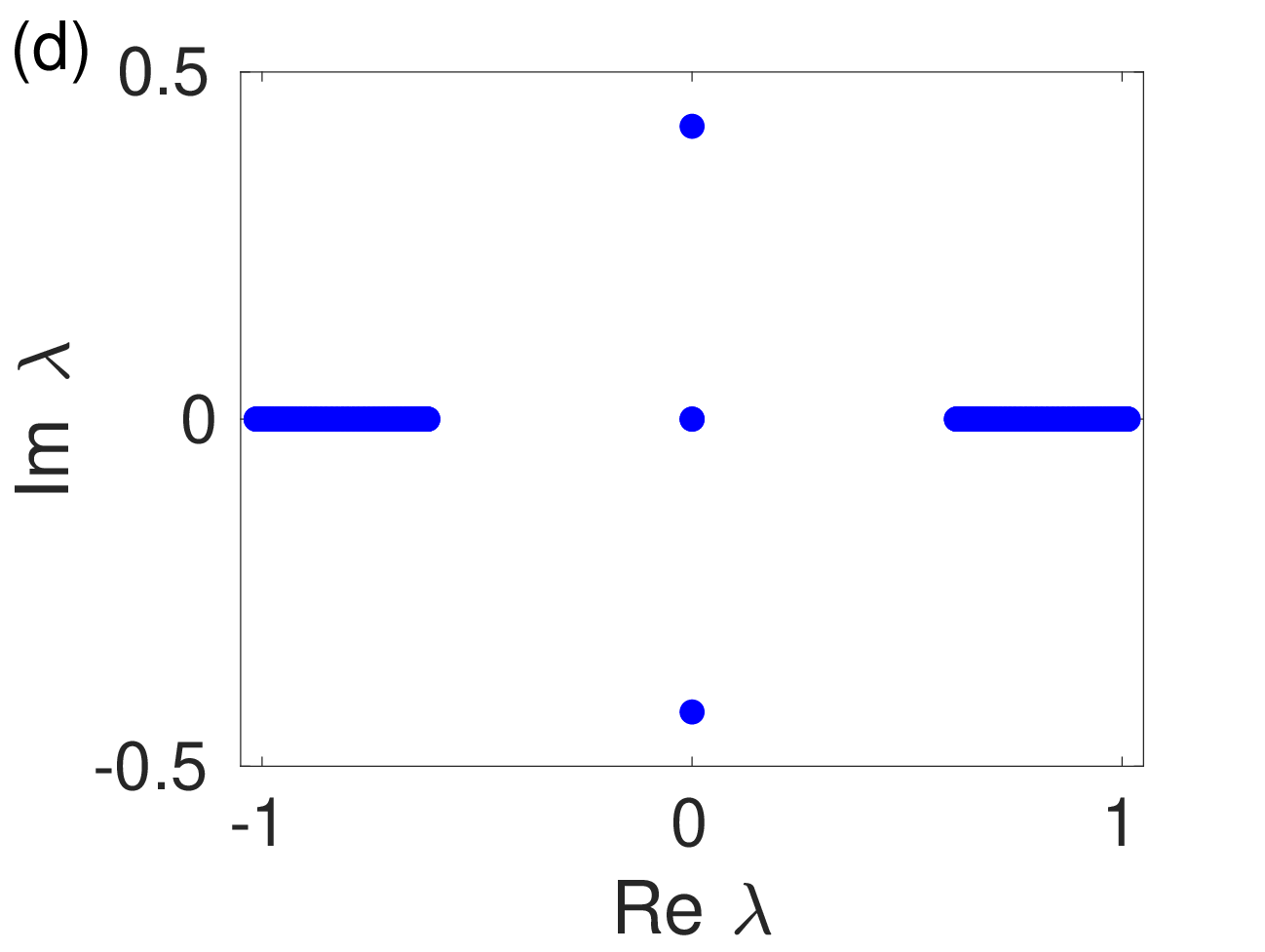}}
\caption{Plots (a) and (b) illustrate the time evolution of strongly localized even modes subjected to small stochastic perturbations applied to the exact initial numerical solutions for $s = -1$ and $s = 1$, respectively. The eigenvalue spectra corresponding to real and imaginary parts are presented in plots (c) and (d). The parameters used in the simulations are $A = 1$, $\kappa = 0.1$, $\gamma = 1$, and $\delta = 0.3$. These figures demonstrate the instability of both solutions.}
\label{fig-6}
\end{figure}

The stability of the obtained solutions was checked by solving the linearized eigenvalue problem, see Eq.~(\ref{LinEignVal}) and also by direct numerical solution of discrete evolution equation Eq.~(\ref{moddnlseQ1D}) with the initial conditions chosen as the exact numerical solutions of the stationary equation obtained, by Newton method and in addition perturbed by small amplitude noise. The results presented in Fig.~\ref{fig-6} show that both symmetric even mode and antisymmetric one are unstable.
Here Fig.~\ref{fig-6} (a) and Fig.~\ref{fig-6} (c) show results for the evolution of perturbed antisymmetric even mode and corresponding result for stability, and Fig.~\ref{fig-6} (b) and Fig.~\ref{fig-6} (d) represent similar results for symmetric mode. 
 One may note from Fig.~\ref{fig-6} (a) that it takes a longer time for the instability of the antisymmetric even mode to be seen since the imaginary part of the eigenvalue is smaller for the given set of parameters.

In the next Sec.~\ref{sec:Odd}, we derive the odd solutions.

\subsection{Odd bright quantum droplets}
\label{sec:Odd}
Similarly, in order to characterize the odd (on-site) mode we use the following ansatz~\cite{Darmanyan1998JETP}:
\begin{eqnarray}
e_n=B(...,0,\beta_2,\beta_1,\beta_0,s\beta_1,s\beta_2,0,...),
\label{ODDen}
\end{eqnarray}
where $|\beta_2|\ll |\beta_1| \ll 1$. We consider the symmetric $s=1$ and antisymmetric $s=-1$ modes separately. 
Applying analogous algebraic operations to those executed for the even mode yields the dispersion relation and the corresponding formula for the secondary amplitude. 

For the $s=1$ symmetric mode, we have
\begin{eqnarray}
&\mu= -\gamma B^2 + \delta B^3 + \cfrac{2 \kappa^2}{-\gamma B^2 +\delta B^3}\,, 
\nonumber \\
&\beta_0=1, \quad \beta_1=-\cfrac{\kappa}{-\gamma B^2 +\delta B^3} \quad 
\beta_2= \beta_1^2. 
\label{discMU2}
\end{eqnarray}
For the $s=-1$ antisymmetric mode, we get
\begin{eqnarray}
&\mu= -\gamma B^2 + \delta B^3 + \cfrac{\kappa^2}{-\gamma B^2 +\delta B^3}\,, 
\nonumber \\
&\beta_0=0, \quad \beta_1=1 \quad \beta_2= -\cfrac{\kappa}{-\gamma B^2 +\delta B^3}. 
\label{discMU2}
\end{eqnarray}
%
\begin{figure}[htbp]
  \centerline{ \includegraphics[width=4.55cm]{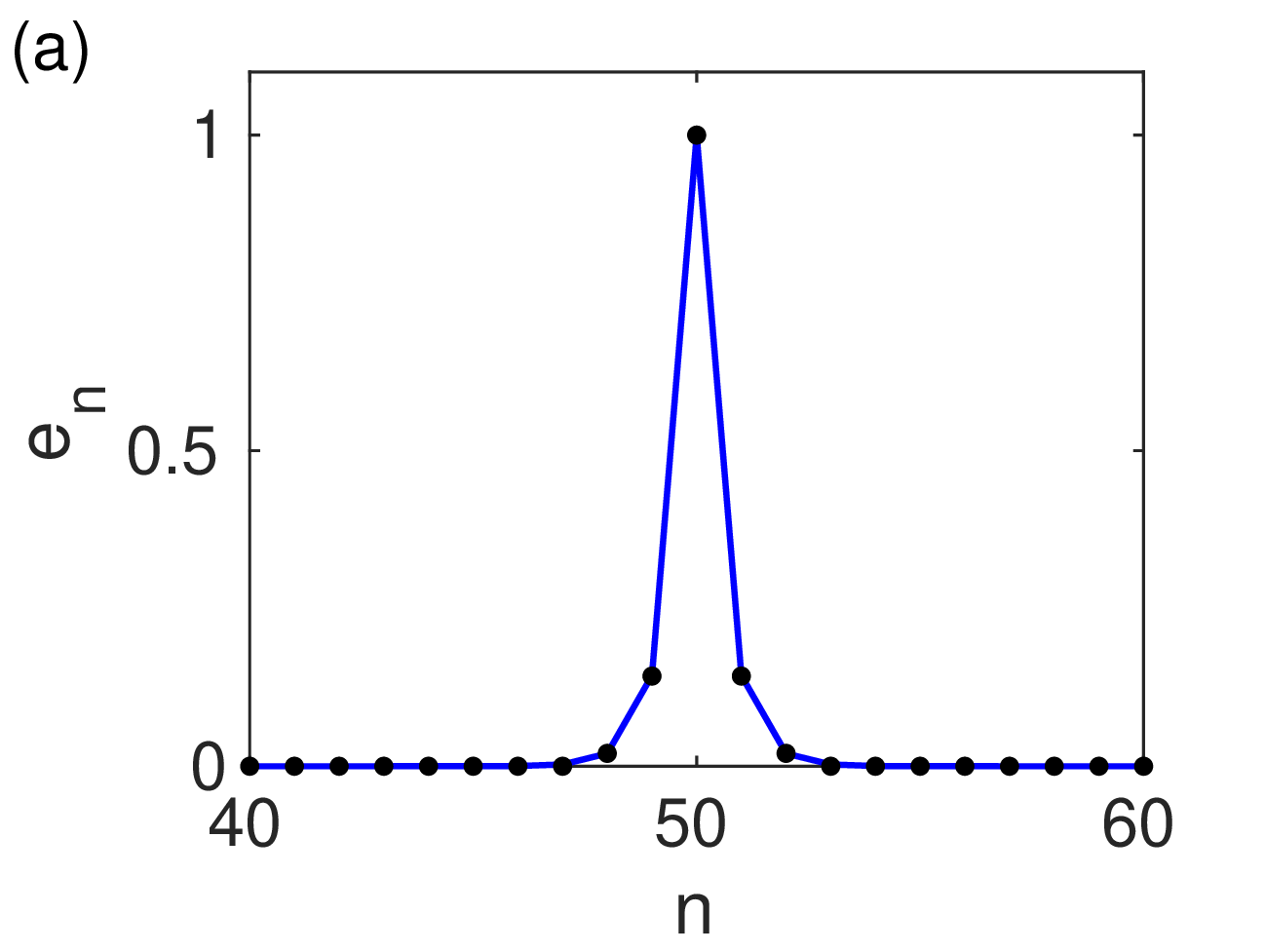} 
  \includegraphics[width=4.5cm]{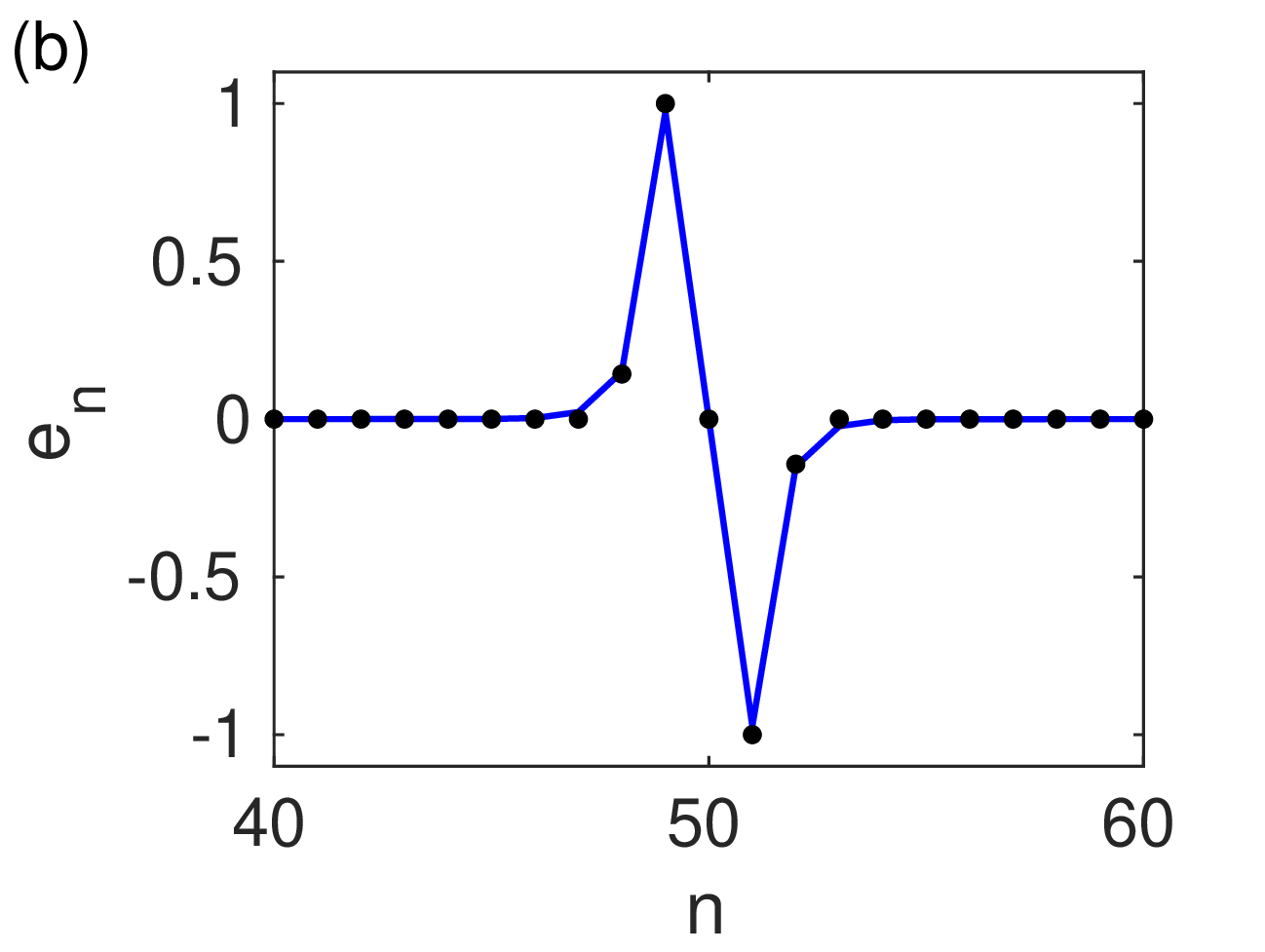}}
\caption{Plots (a) and (b) display the odd strongly localized modes for $s = 1$ and $s = -1$, respectively. The solid lines represent the solutions of Eq.~(\ref{discStatEQ}) obtained using the Newton method, while the points indicate the approximate solutions from the Page method as given by Eq.~(\ref{discMU2}). The parameters used are $A=1$, $\kappa=0.1$, $\gamma=1$ and $\delta=0.3$.}
\label{fig-8}
\end{figure}
%

\begin{figure}[htbp]
  \centerline{ \includegraphics[width=4.55cm]{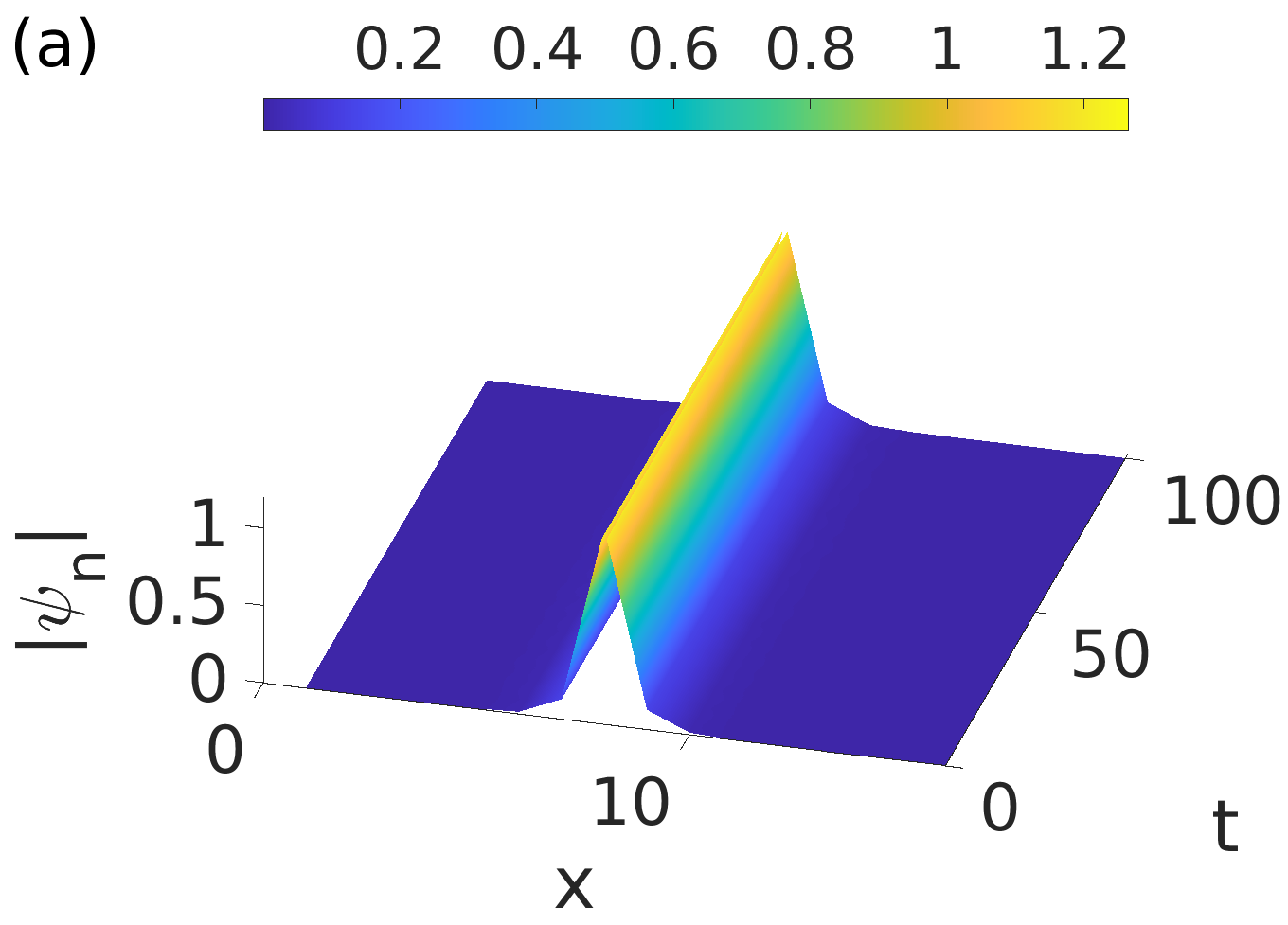} 
  \includegraphics[width=4.5cm]{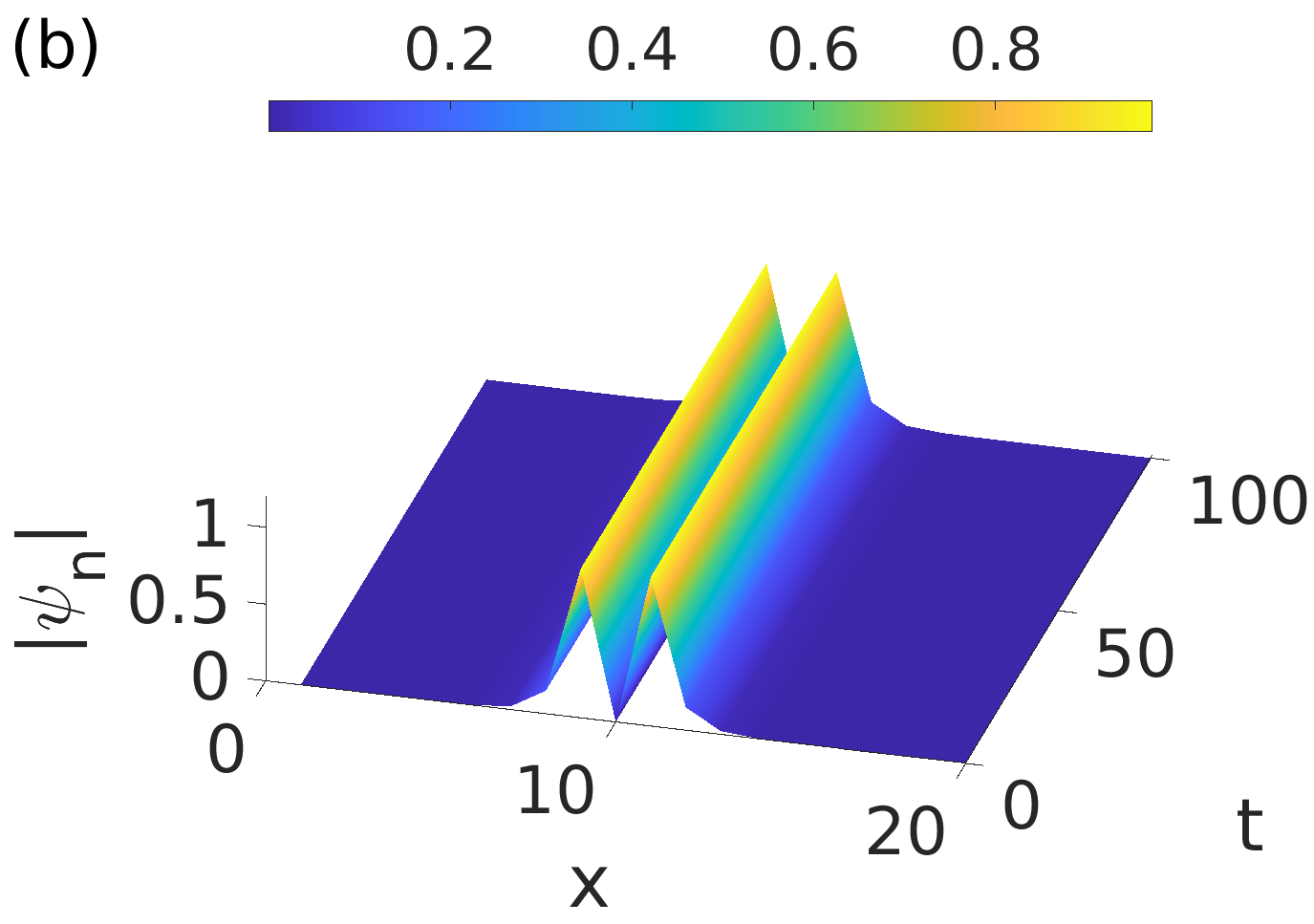}}
   \centerline{ \includegraphics[width=4.55cm]{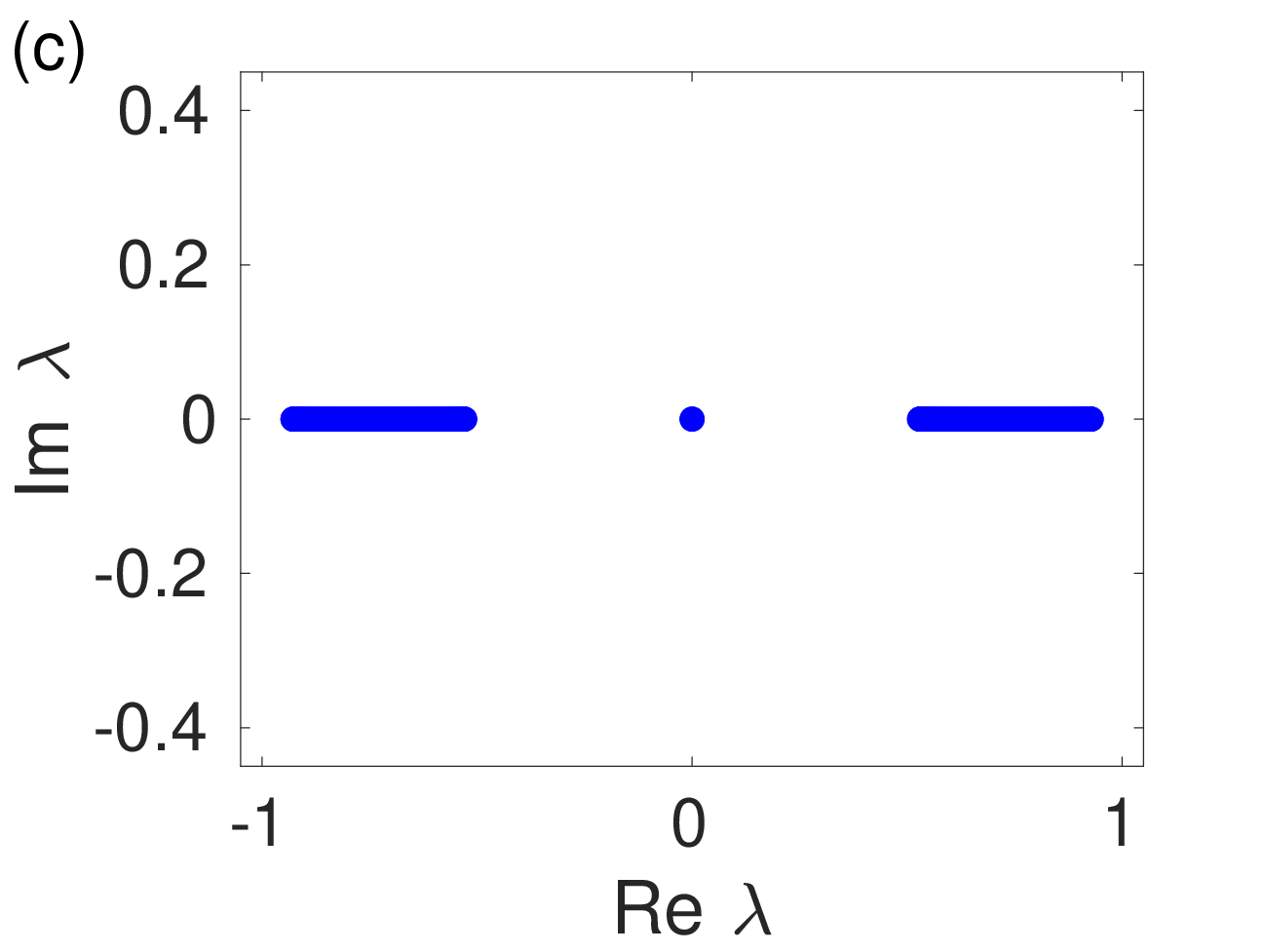}
   \includegraphics[width=4.55cm]{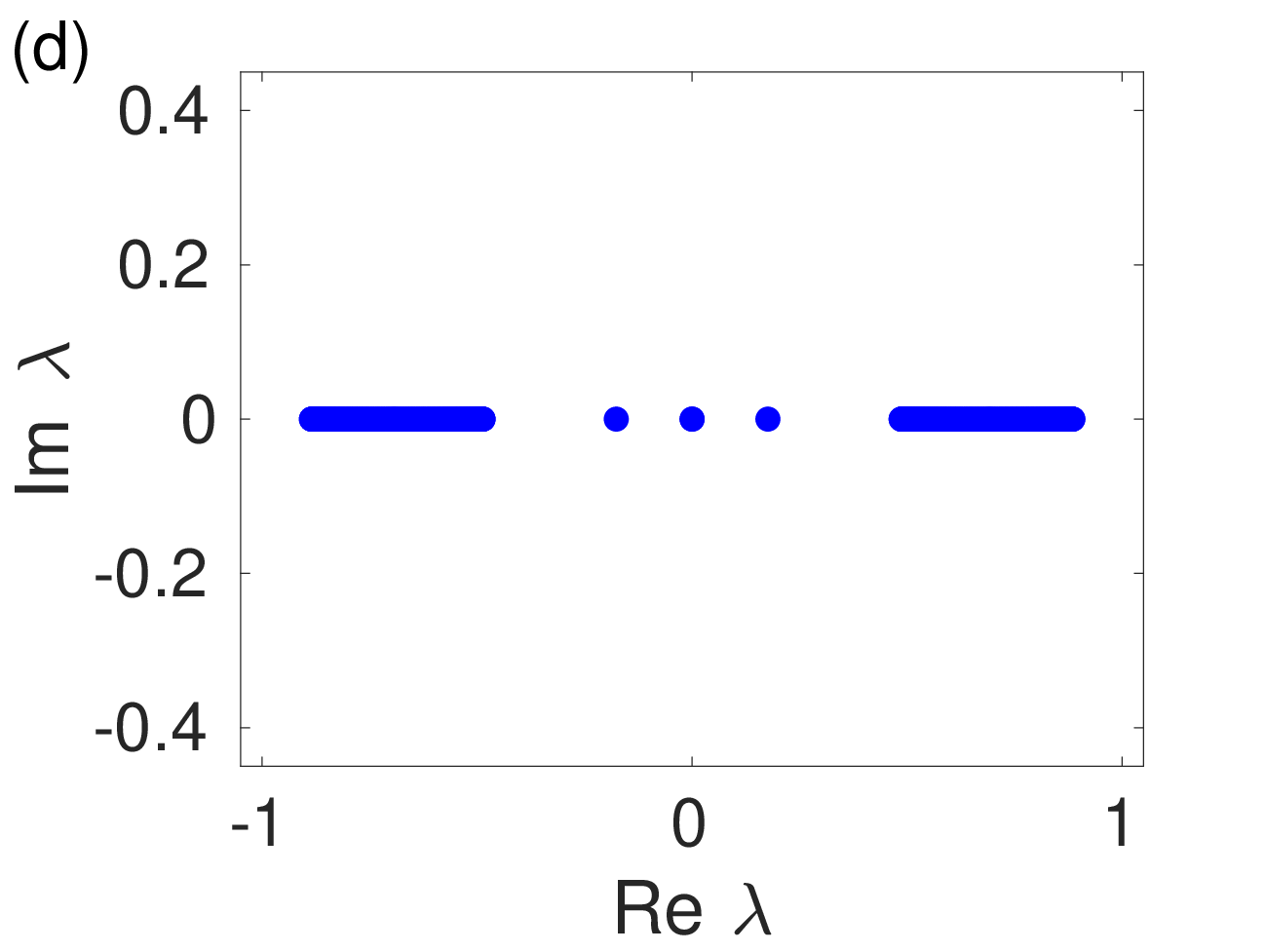}}
\caption{Plots (a) and (b) illustrate the time evolution of strongly localized odd modes subjected to small stochastic perturbations applied to the exact initial numerical solutions for $s = 1$ and $s = -1$, respectively. The corresponding real and imaginary parts of the eigenvalue spectra are presented in plots (c) and (d), with (c) representing $s = 1$ and (d) representing $s = -1$. The parameters used are $A=1$, $\kappa=0.1$, $\gamma=1$ and $\delta=0.3$.}
\label{fig-9}
\end{figure}

In our theoretical approach, we limited the analysis to second-order terms that involve the small parameters $\alpha$ and $\beta$. These approximations provide a reasonably accurate representation of cases of strong localization, a validation of this approach supported by direct numerical simulations.
Let us repeat the numerical procedure, applied in the previous subsection, now for the case of strongly localized odd modes. The results presented in Figs.~\ref{fig-8} and \ref{fig-9} confirm that the approximate Page method can be successfully used for strongly localized odd modes and that both symmetric and antisymmetric modes are stable.
\subsection{Topological and flat-top solitons}
\label{sec:flat-top}
\begin{figure}[htbp]
  \centerline{ \includegraphics[width=4.55cm]{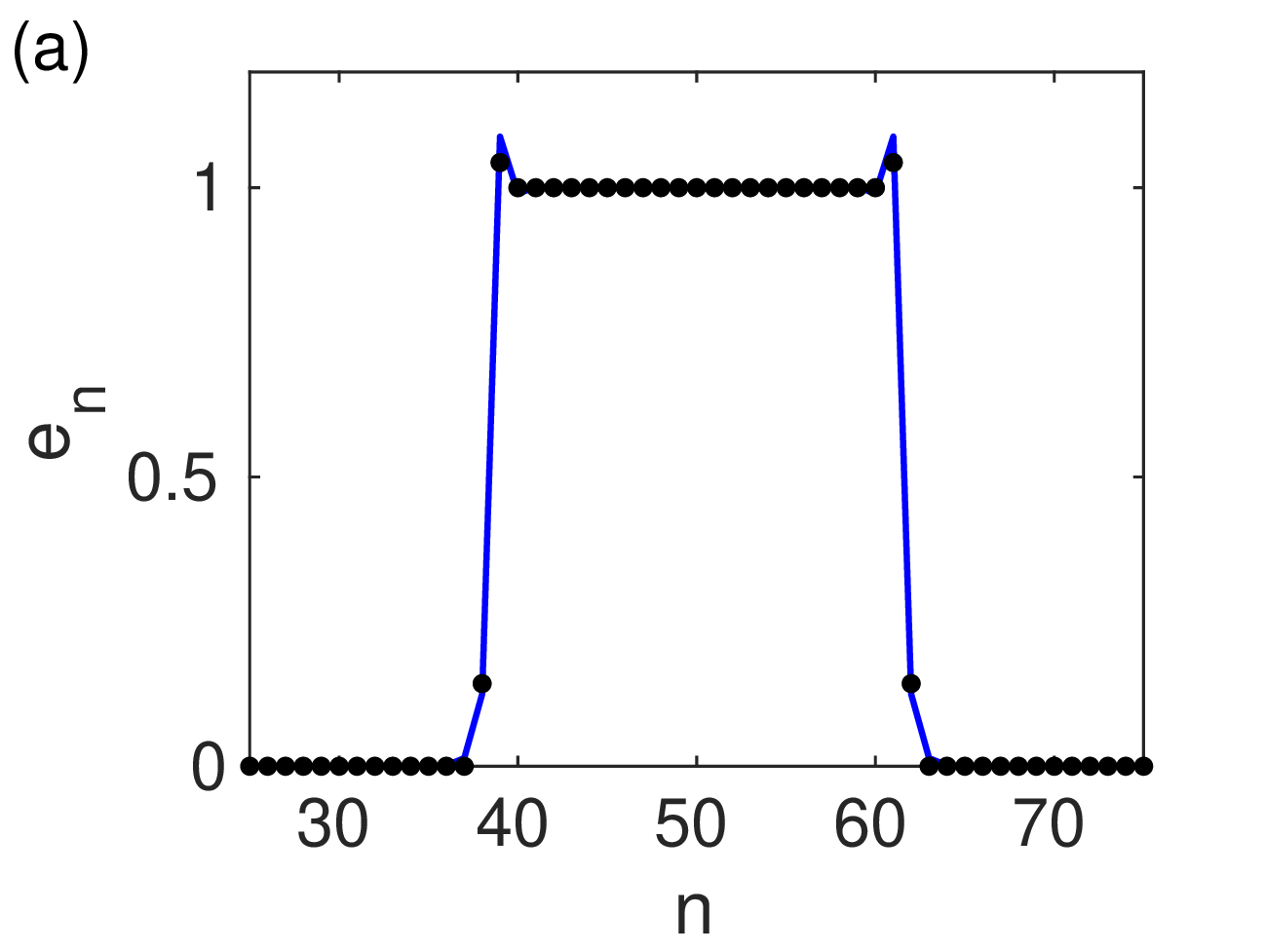}
   \includegraphics[width=4.55cm]{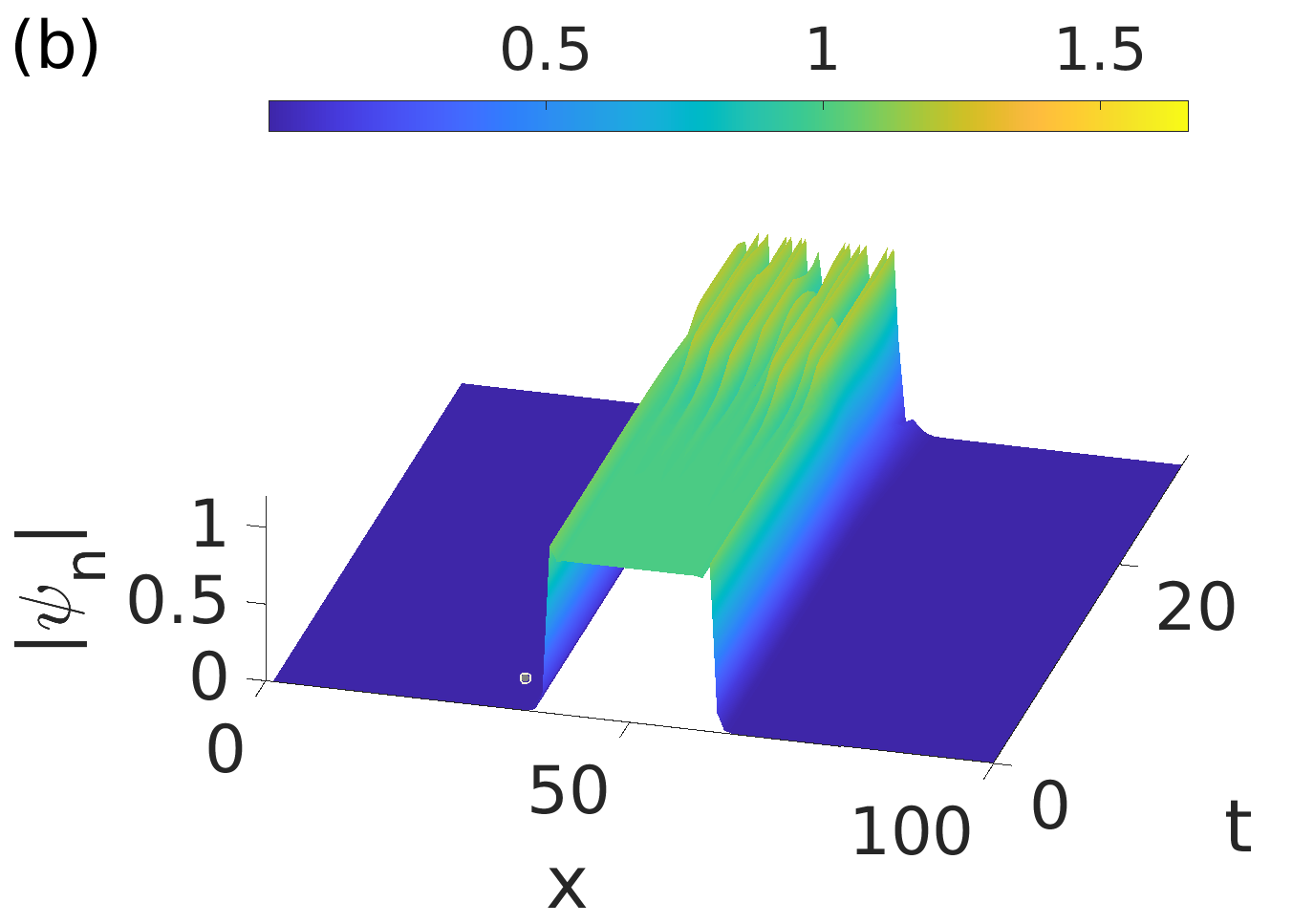}}
\caption{(a) Strongly localized flat-top mode profile: solid line shows Newton method solution of Eq.~(\ref{discStatEQ}), points represent Page method approximation from Eq.~(\ref{amplu1234}). (b) Time evolution under small stochastic perturbation of the exact initial solution. Parameters are $A=1$, $\kappa=0.1$, $\gamma=1$ and $\delta=0.3$.}
\label{fig-11}
\end{figure}

The front profile, represented as $e_n = A(...,0,0,0,...,1,1,1,...)$, exhibits both a zero and a nonzero asymptotic value, with the corresponding dispersion relation taking the following form:
\begin{equation}
\mu=-2 \kappa -\gamma V^2 + \delta V^3\,.
\label{muFT}
\end{equation}
We use
\begin{equation}
e_n=V (...,0,0,0,u_1,u_2,u_3,u_4,1,1,1,...),
\label{FrontAnsatz}
\end{equation}
ansatz function~\cite{Darmanyan1999}, where $V>0$. By introducing the small parameter $\epsilon=\kappa /(-\gamma V^2 + \delta V^3) \ll 1$, which characterizes the ratio between linear coupling and nonlinearity and serves as a measure of the degree of localization. It is known that no front solutions exist when the transition domain is larger; in other words, the front solution is consistently strongly localized. Attempts to find broader fronts yield only solutions with a nonmonotonous transition region~\cite{Darmanyan1999}. By substituting Eq.~(\ref{FrontAnsatz}) into Eq.~(\ref{discStatEQ}) and considering terms up to the first order in $\epsilon$, we get following equations for small amplitudes:
\begin{eqnarray}
& u_1 \approx 0, \qquad u_2 \simeq - \epsilon,
\nonumber \\
&u_3=1-\epsilon \, \cfrac{\gamma V^2 -\delta V^3}{2 \gamma V^2 + \delta V^3}, \qquad u_4 \approx 1. 
\label{amplu1234}
\end{eqnarray}
Flat-top solutions can be considered as a superposition of such two topological ``kink-like'' solutions, see Fig.~\ref{fig-11}. When kinks are close to each other profile has a bell shape, and when they are well-separated profile has a flat-top shape~\cite{Otajonov2019}. We recall that a flat-top profile is a characteristic property of quantum droplets~\cite{Petrov2015}. 

%
\begin{figure}[htbp]
  \centerline{ \includegraphics[width=4.55cm]{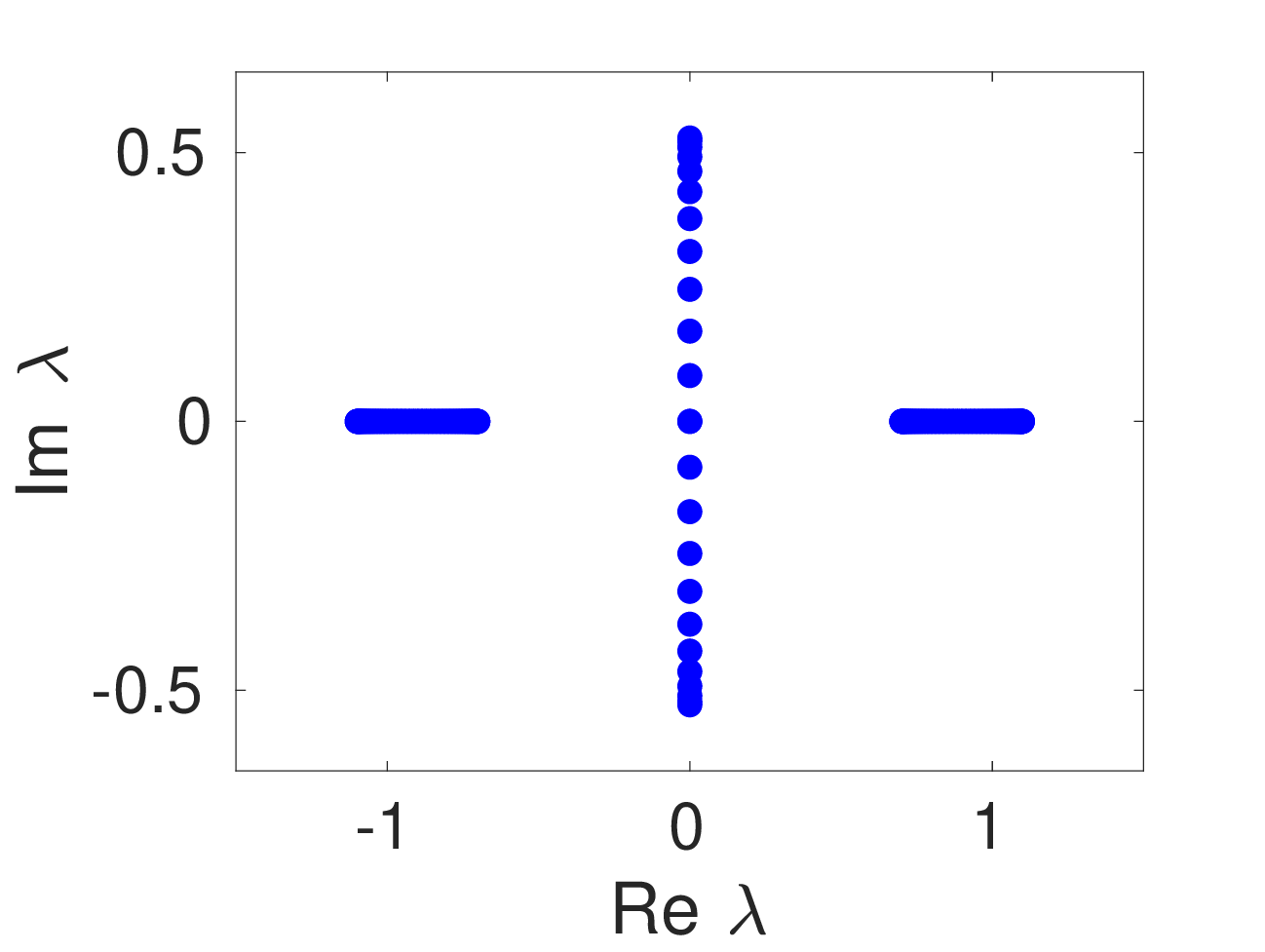}}
\caption{ The real and imaginary parts of the eigenvalue spectrum for the strongly localized flat-top mode in FIG.~\ref{fig-11}.}
\label{fig-12}
\end{figure}

The numerical results for flat top solution are presented in Figs.~\ref{fig-11} and \ref{fig-12} and we proceed with the same steps as in previous subsections. 
Again the solution exists and is well approximated by the results of application of the Page method. The exact numerical solution for flat top discrete mode is unstable as follows from a solution of time-dependent evolution equation Eq.~(\ref{moddnlseQ1D}), also the eigenvalues of linearized equation (Eqs.~(\ref{LinEignVal})) have non zero imaginary parts, see Fig.~\ref{fig-12}.

Similarly, dark flat-top localized mode solutions can be obtained. These solutions are represented by the superposition of two anti-kink and kink profiles. The next subsection \ref{sec:LHY} is devoted to the problem of the existence and stability of strongly localized discrete modes in the case of Lee-Yang-Huang quantum liquid located in a one-dimensional optical lattice~\cite{Jorgensen, Skov}.

\subsection{LHY discrete strongly localized modes }
\label{sec:LHY}

\begin{figure}[htbp]
  \centerline{ \includegraphics[width=4.55cm]{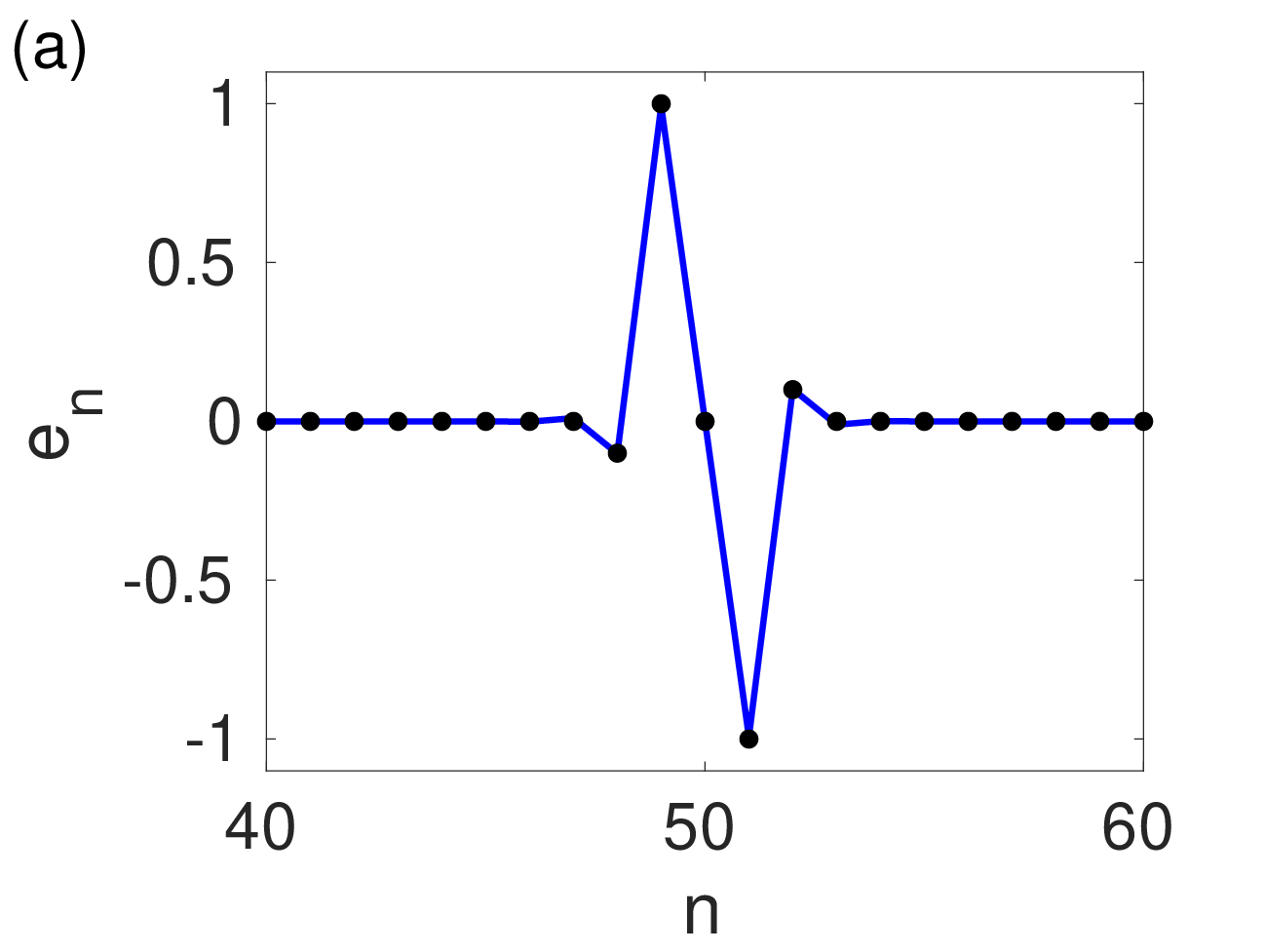} 
  \includegraphics[width=4.5cm]{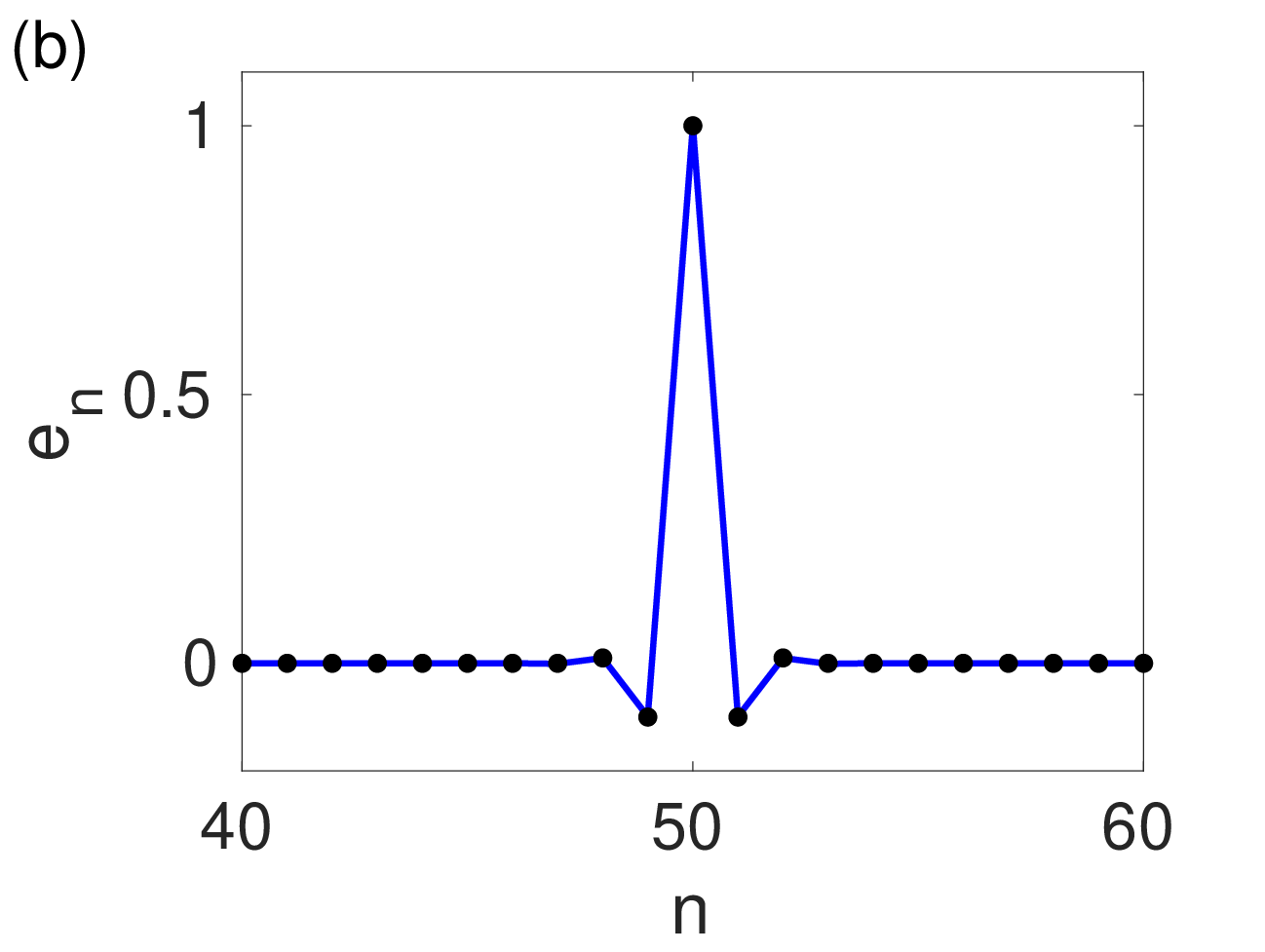}}
\caption{Odd strongly localized modes: (a) $s=-1$, (b) $s=1$. The solid line shows the Newton method solution of Eq.(\ref{discStatEQ}), and points show Page method approximation from Eq.(\ref{discMU}). Other parameters are $A=1$, $\kappa=0.1$, $\gamma=0$ and $\delta=1$.}
\label{fig-13}
\end{figure}

\begin{figure}[htbp]
  \centerline{ \includegraphics[width=4.55cm]{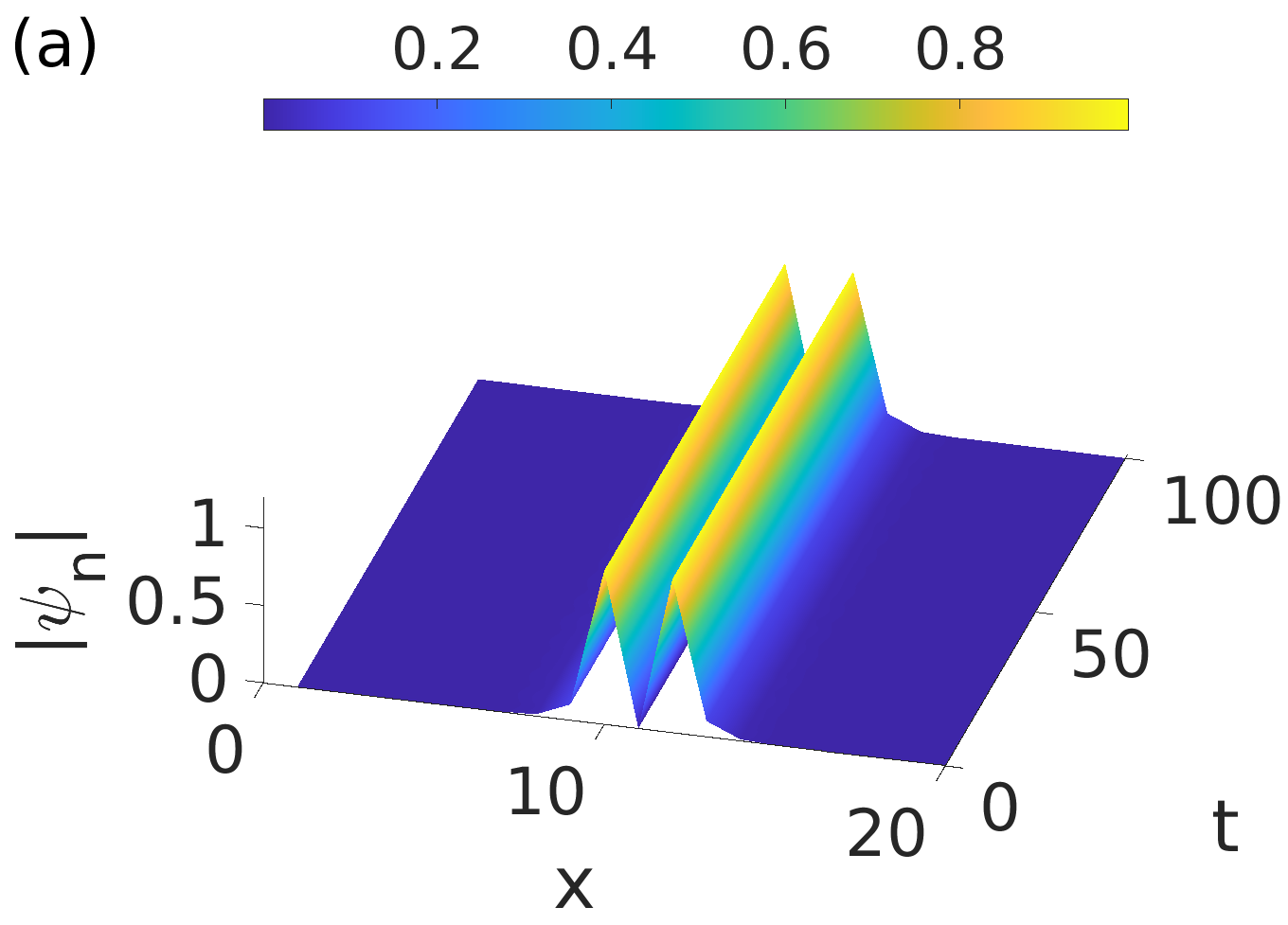} 
  \includegraphics[width=4.5cm]{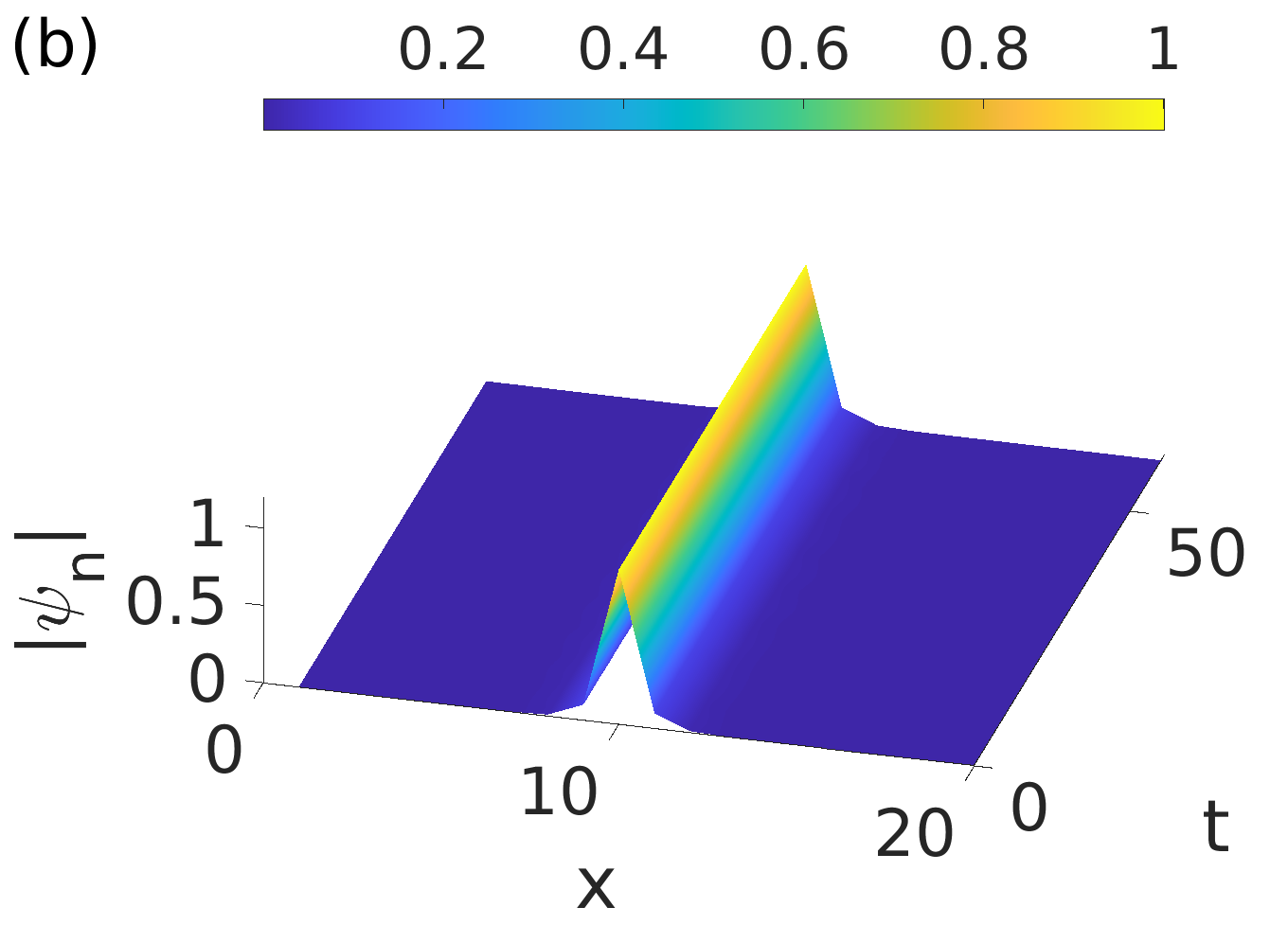}}
  \centerline{ \includegraphics[width=4.55cm]{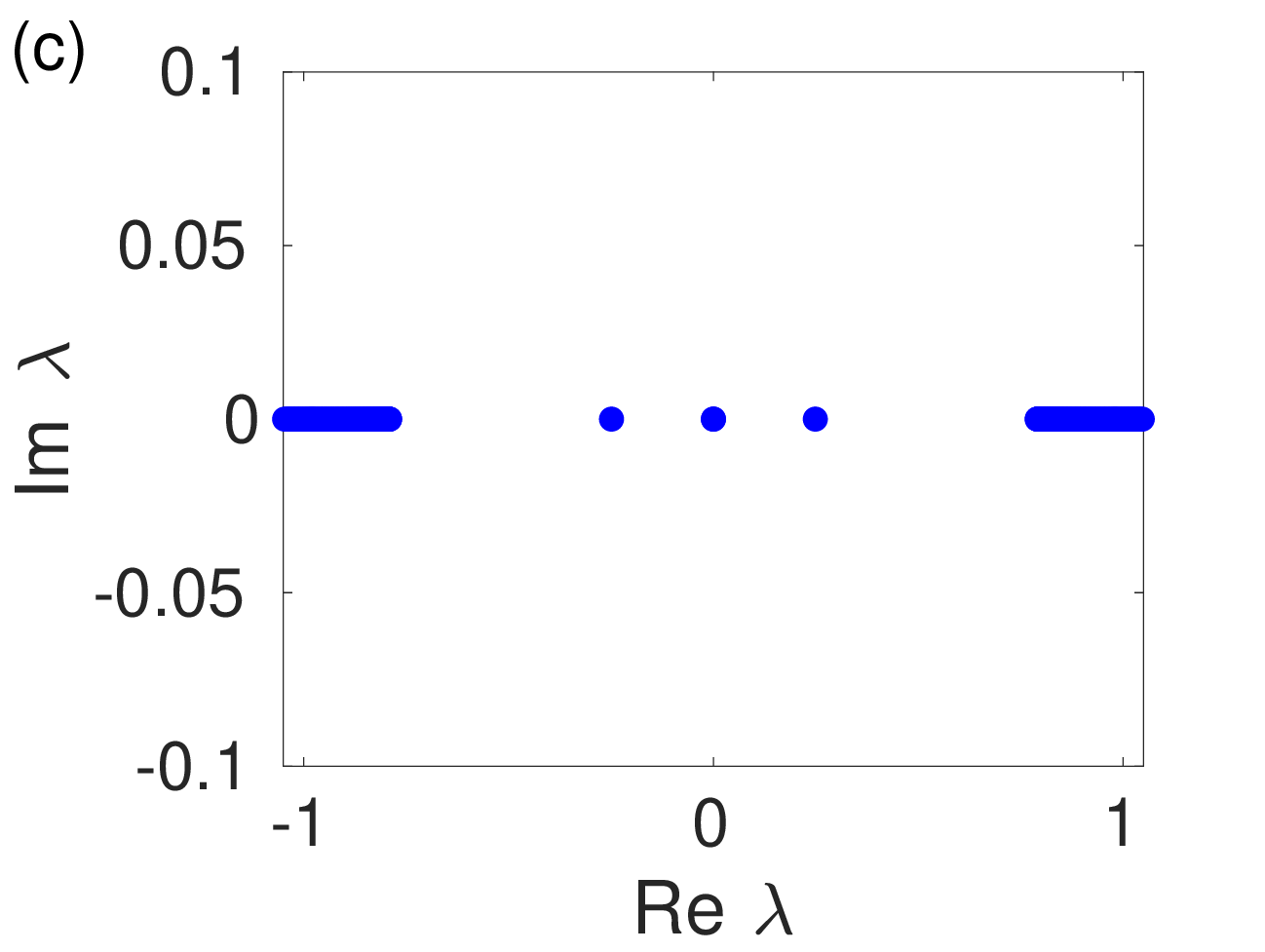}
   \includegraphics[width=4.55cm]{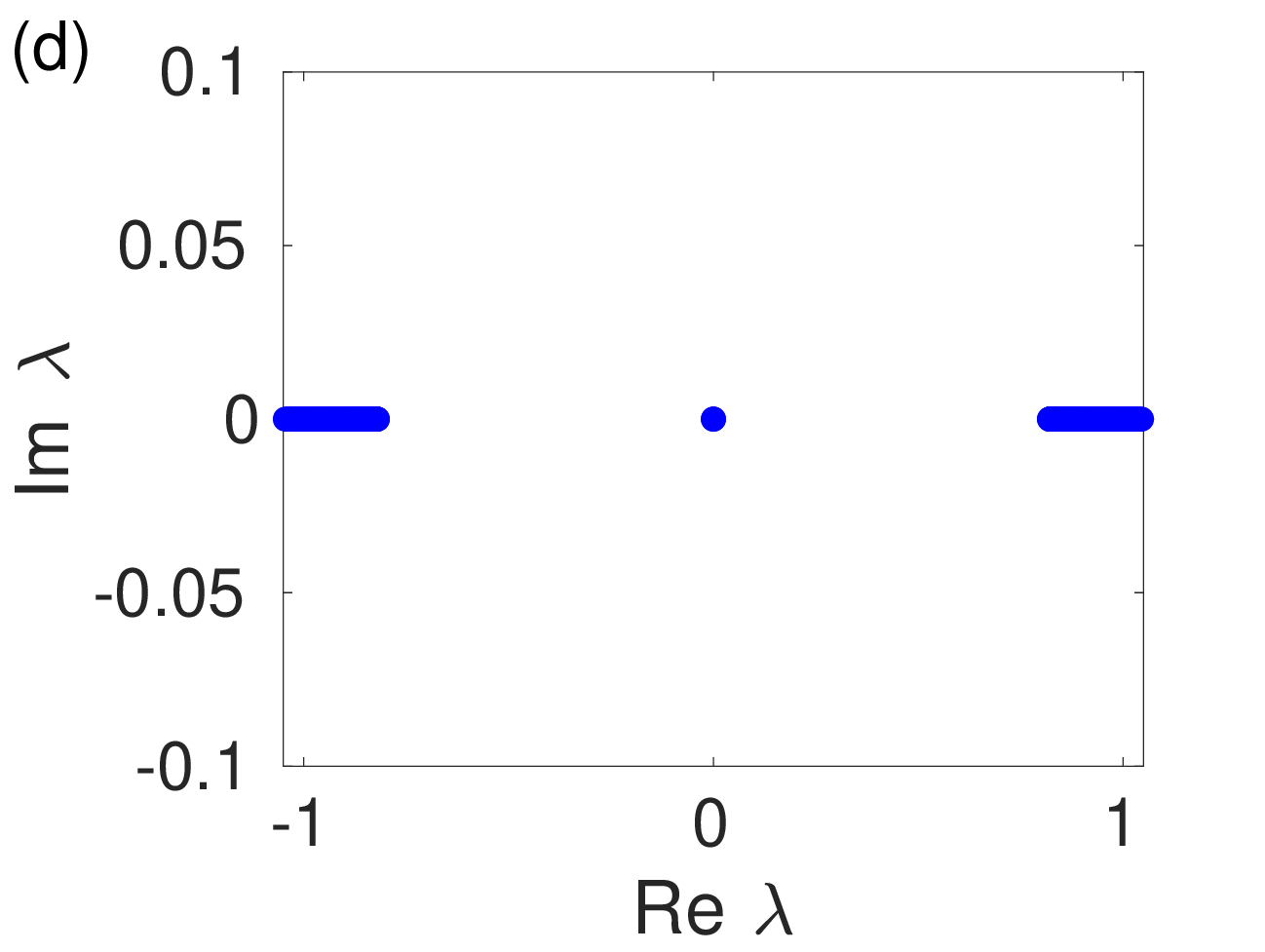}}
\caption{Time evolution of strongly localized odd modes with small stochastic perturbations applied to the initial exact numerical solution for (a)$s = -1$ and (b) $s = 1$. Real and imaginary parts of the eigenvalue spectrum for the corresponding modes are shown in (c) $s = -1$ and (d) $s = 1$. Other parameters are $A=1$, $\kappa=0.1$, $\gamma=0$ and $\delta=1$. 
}
\label{fig-14}
\end{figure}

\begin{figure}[htbp]
  \centerline{ \includegraphics[width=4.55cm]{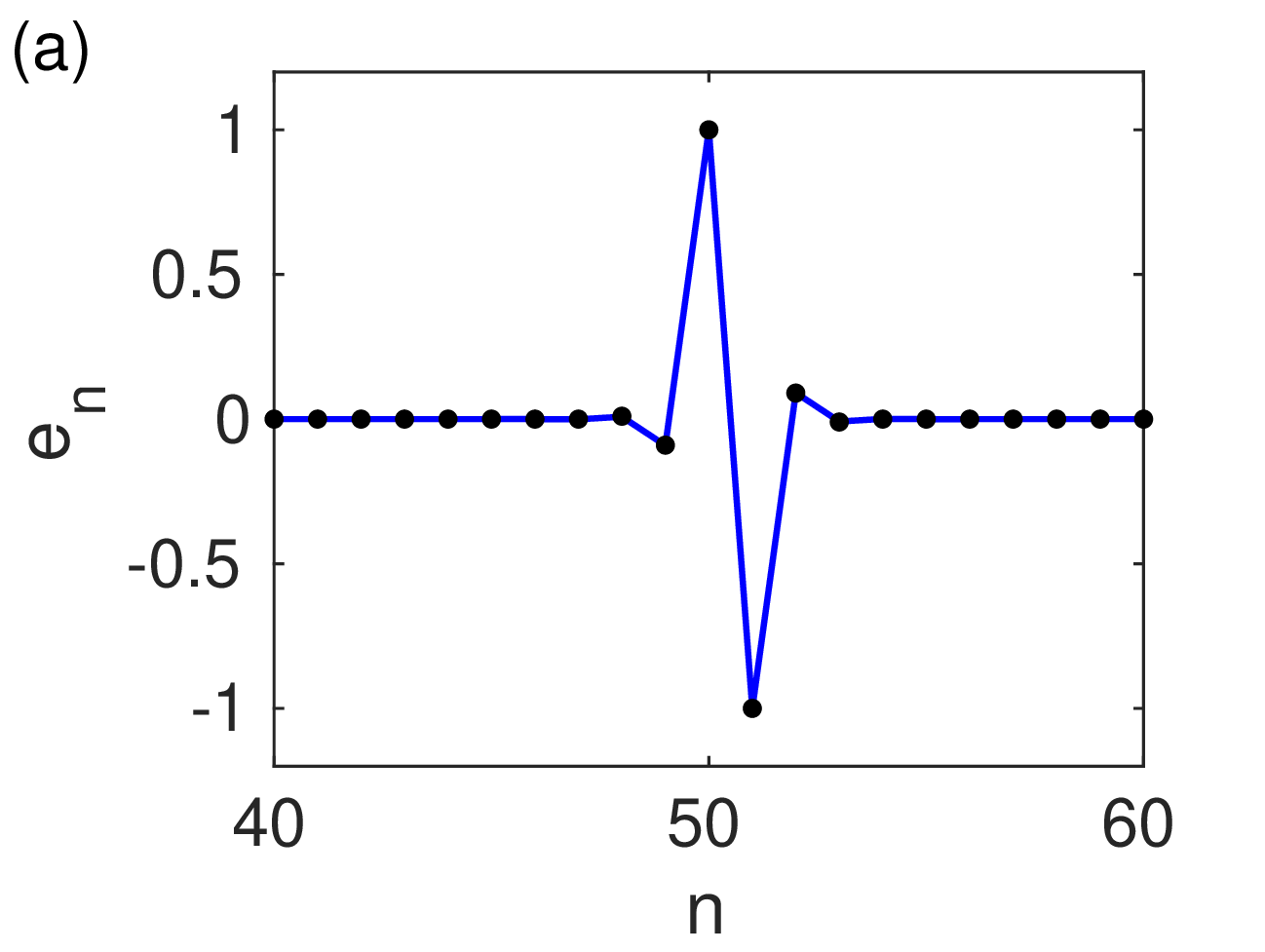} 
  \includegraphics[width=4.5cm]{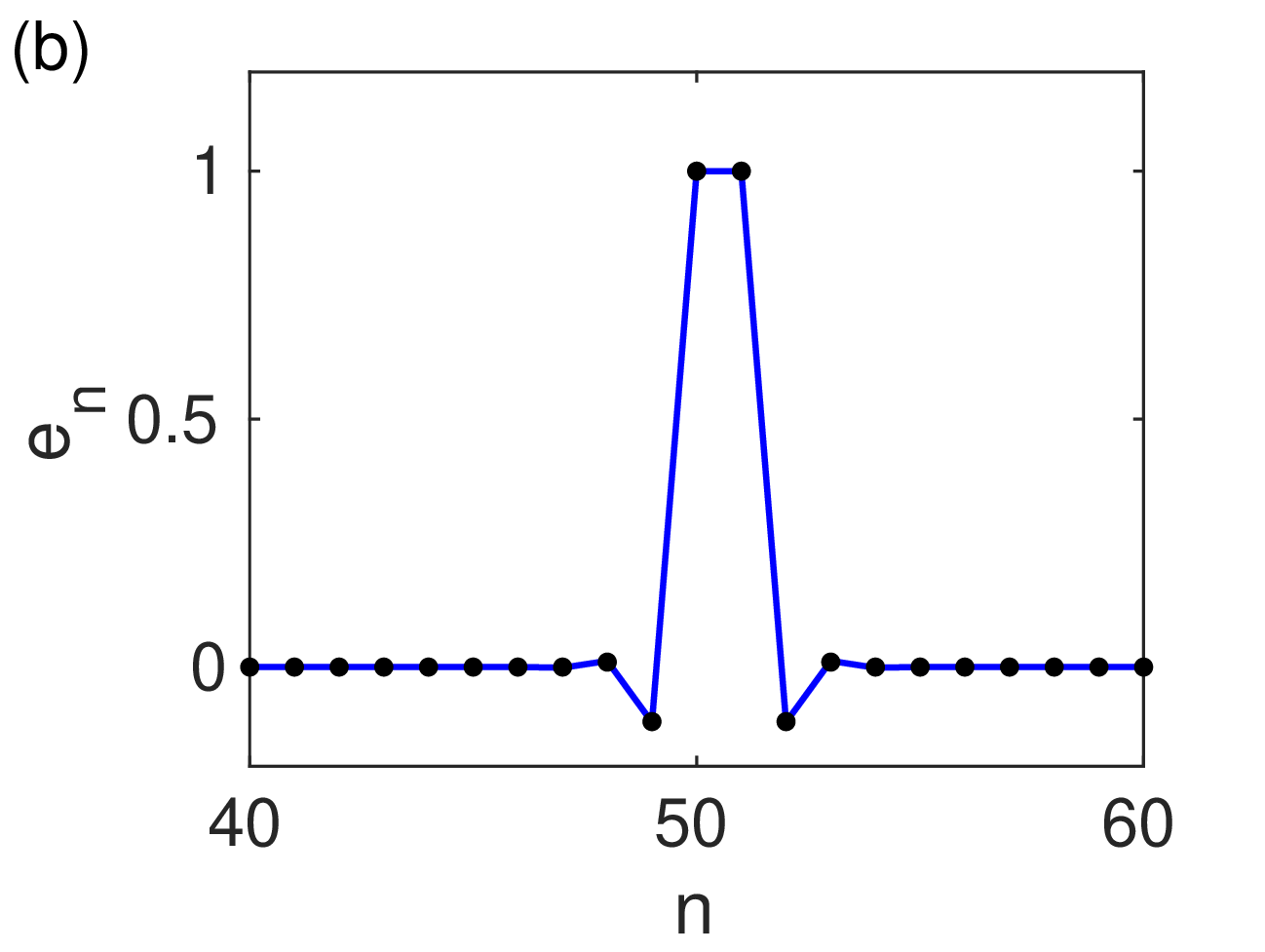}}
\caption{Even strongly localized modes: (a) \(s = -1\), (b) \(s = 1\). The solid line represents the Newton method solution of Eq.~(\ref{discStatEQ}), and points correspond to the Page method approximation from Eq.~(\ref{discMU}). Other parameters are $A=1$, $\kappa=0.1$, $\gamma=0$ and $\delta=1$.}
\label{fig-16}
\end{figure}

\begin{figure}[htbp]
  \centerline{ \includegraphics[width=4.55cm]{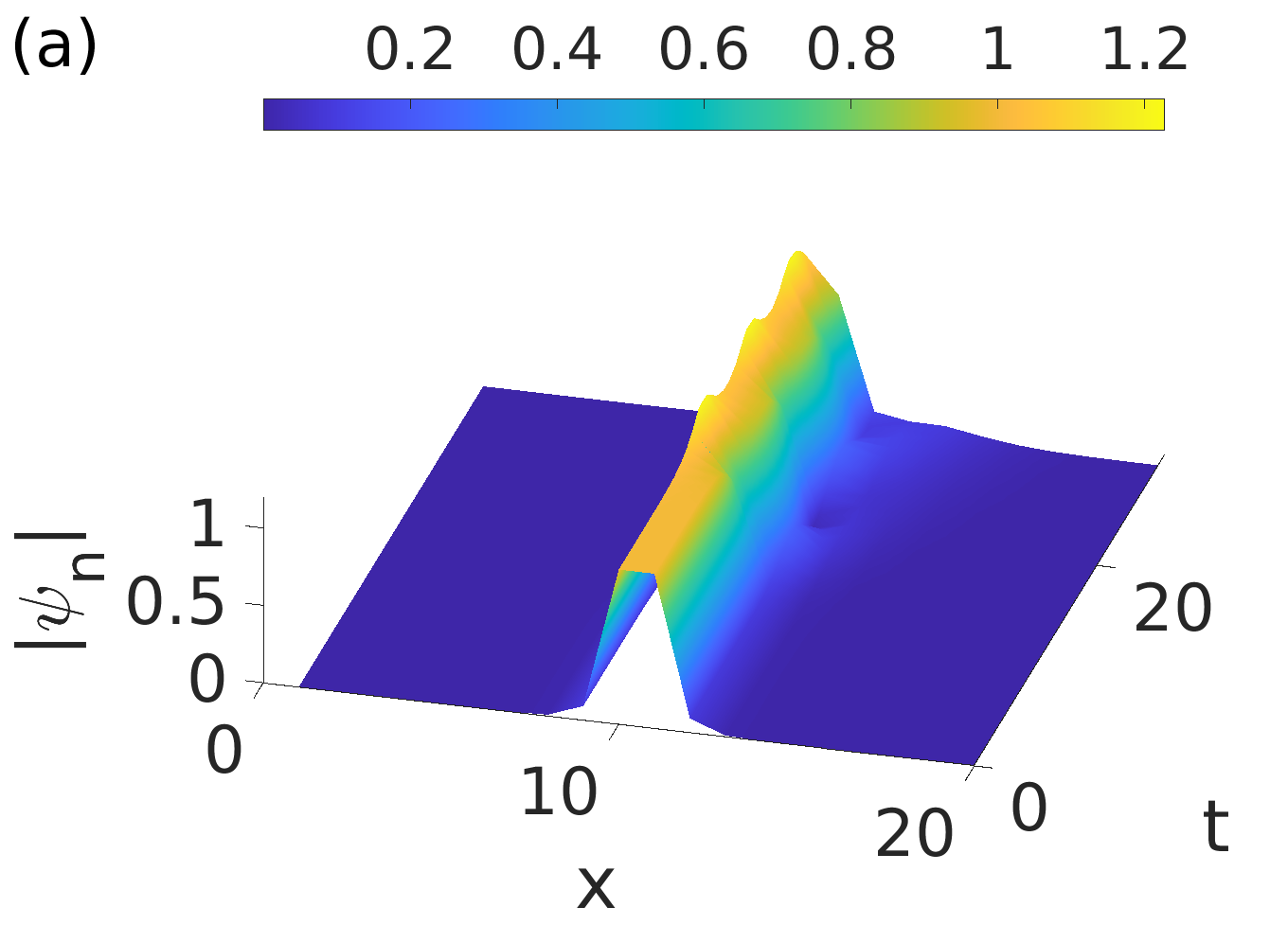} 
  \includegraphics[width=4.5cm]{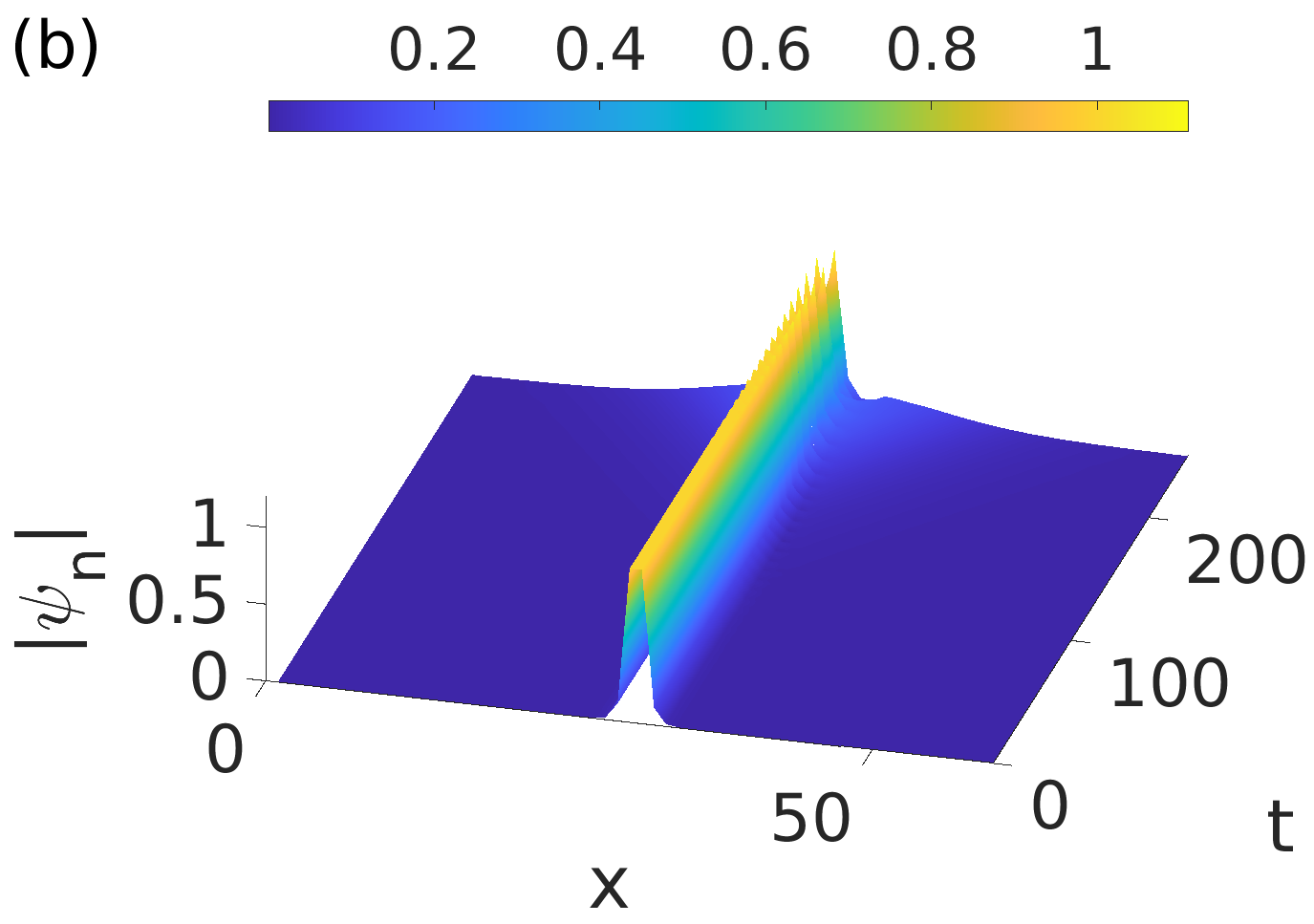}}
  \centerline{ \includegraphics[width=4.55cm]{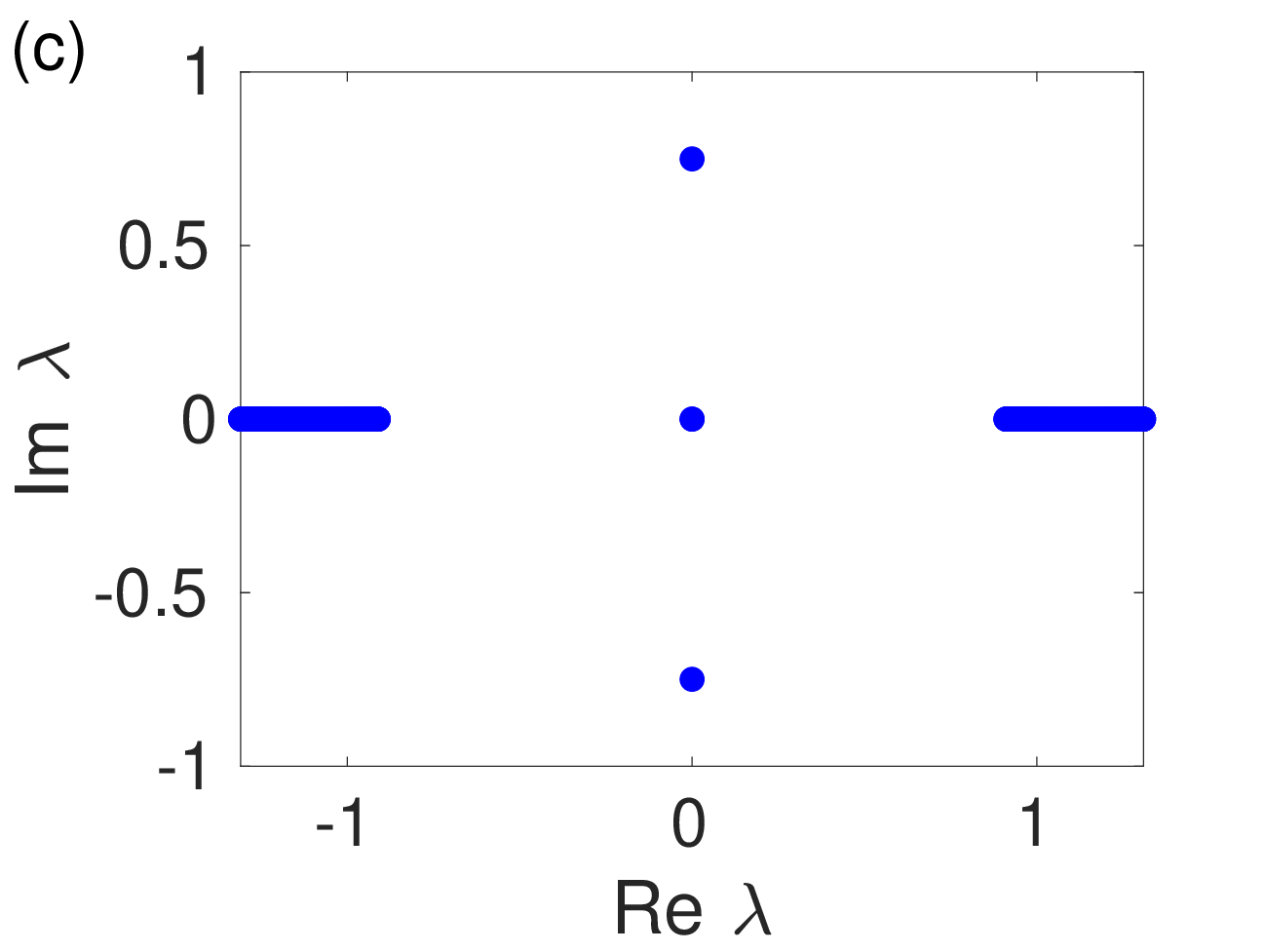}
   \includegraphics[width=4.55cm]{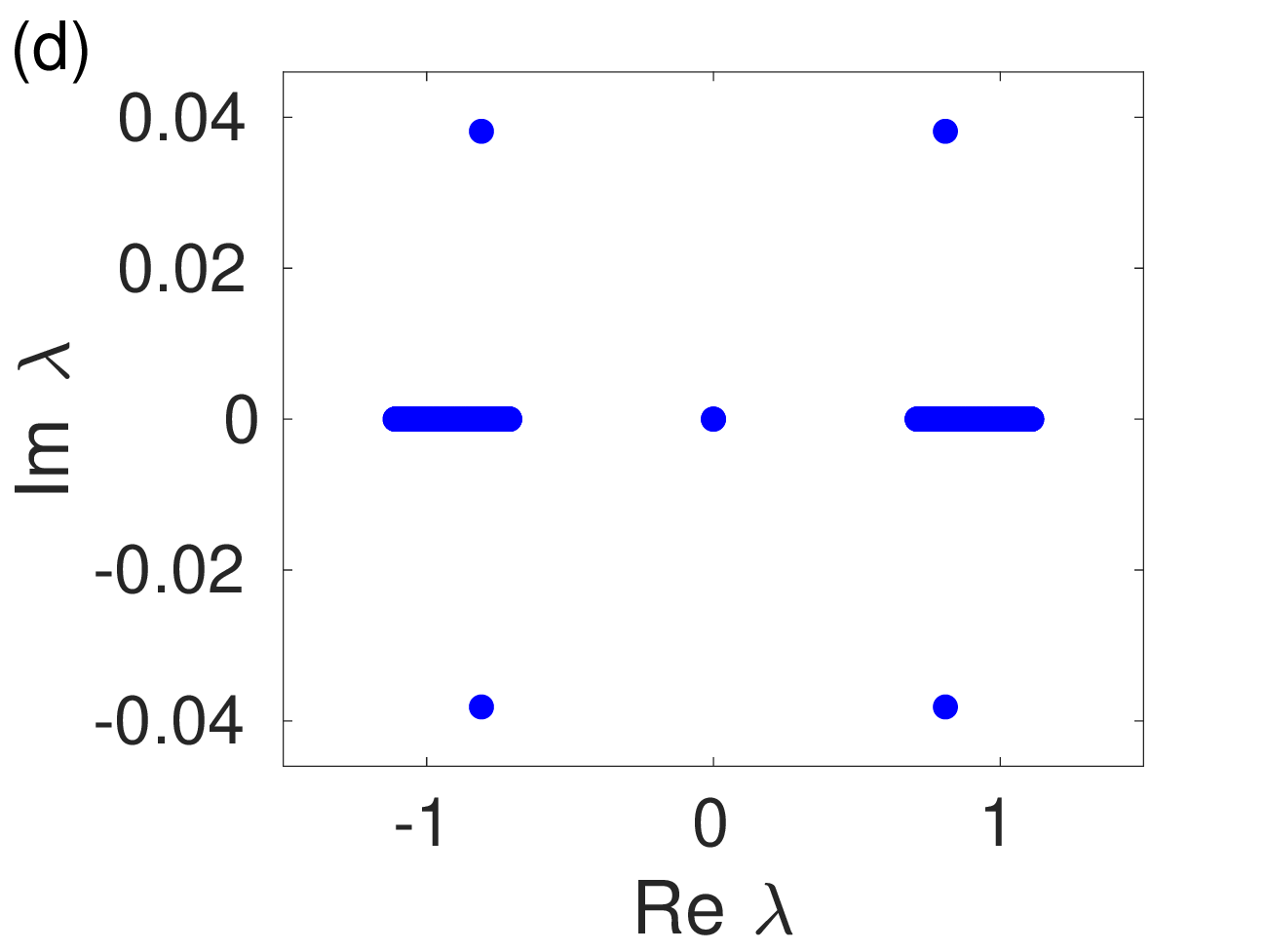}}
\caption{Time evolution of strongly localized even modes under small stochastic perturbations: (a) $s = -1$, (b) $s = 1$. Eigenvalue spectrum (real and imaginary parts) for these modes: (c) $s = -1$, (d) $s = 1$. Other parameters are $A=1$, $\kappa=0.1$, $\gamma=0$ and $\delta=1$.
}
\label{fig-17}
\end{figure}

\begin{figure}[htbp]
  \centerline{ \includegraphics[width=4.55cm]{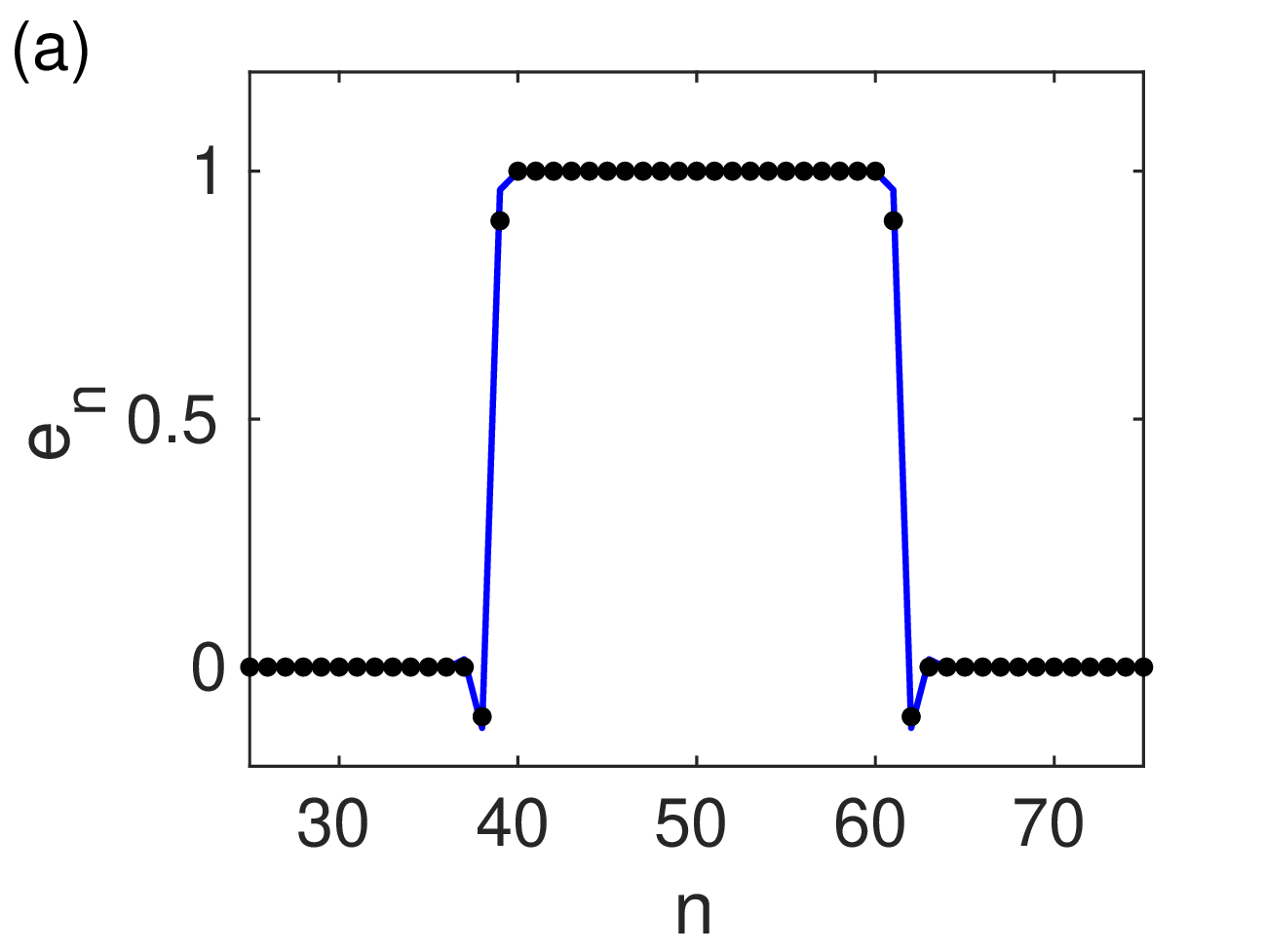}
   \includegraphics[width=4.55cm]{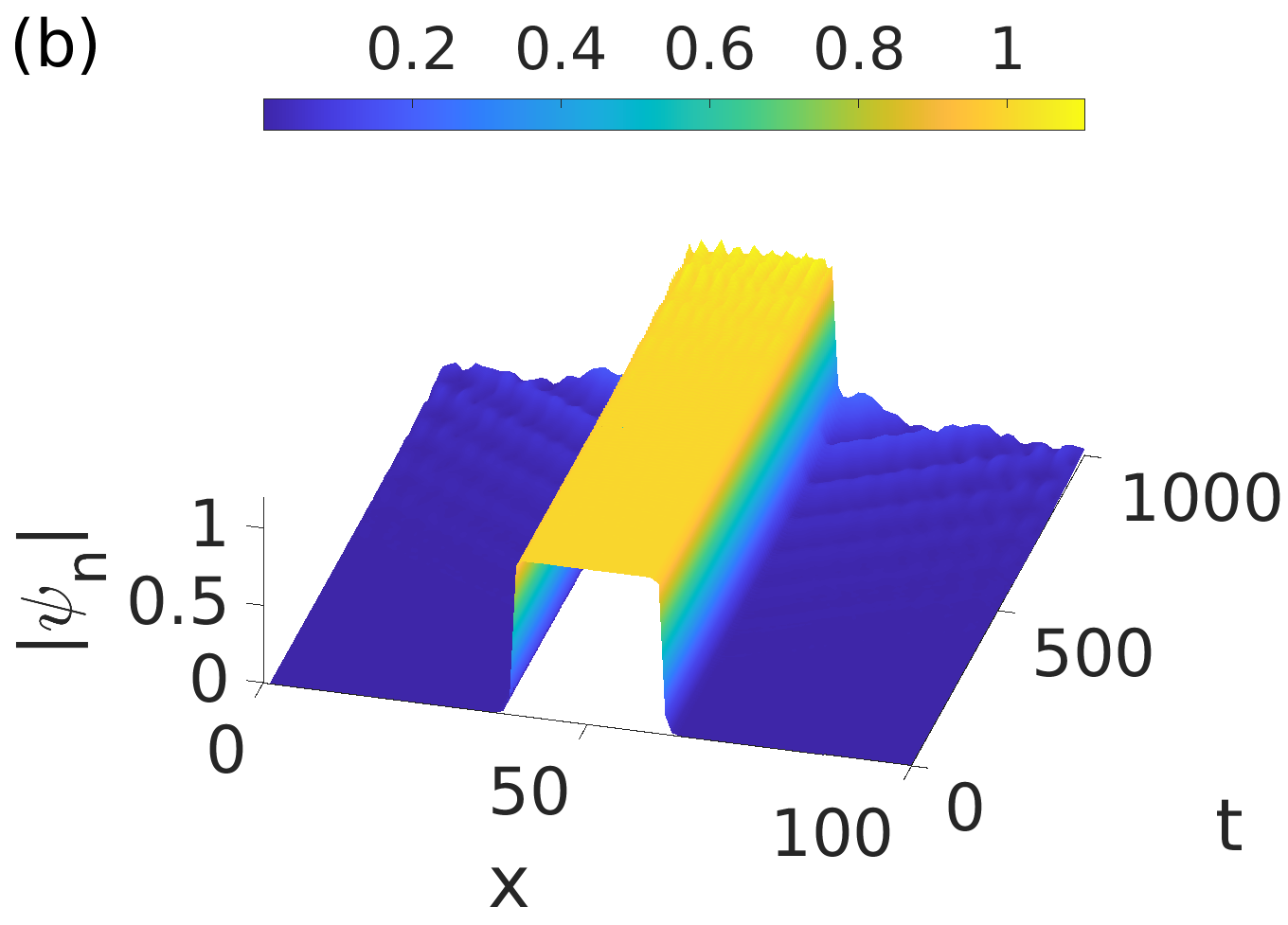}}
\caption{
(a) Strongly localized flat-top mode: solid line shows the Newton method solution of Eq.~(\ref{discStatEQ}), while points indicate the Page method approximation from Eq.~(\ref{amplu1234}). (b) Time evolution of the flat-top mode under small stochastic perturbations applied to the initial exact numerical solution. Parameters are $A=1$, $\kappa=0.1$, $\gamma=0$ and $\delta=1$.}
\label{fig-19}
\end{figure}

\begin{figure}[htbp]
  \centerline{ \includegraphics[width=4.55cm]{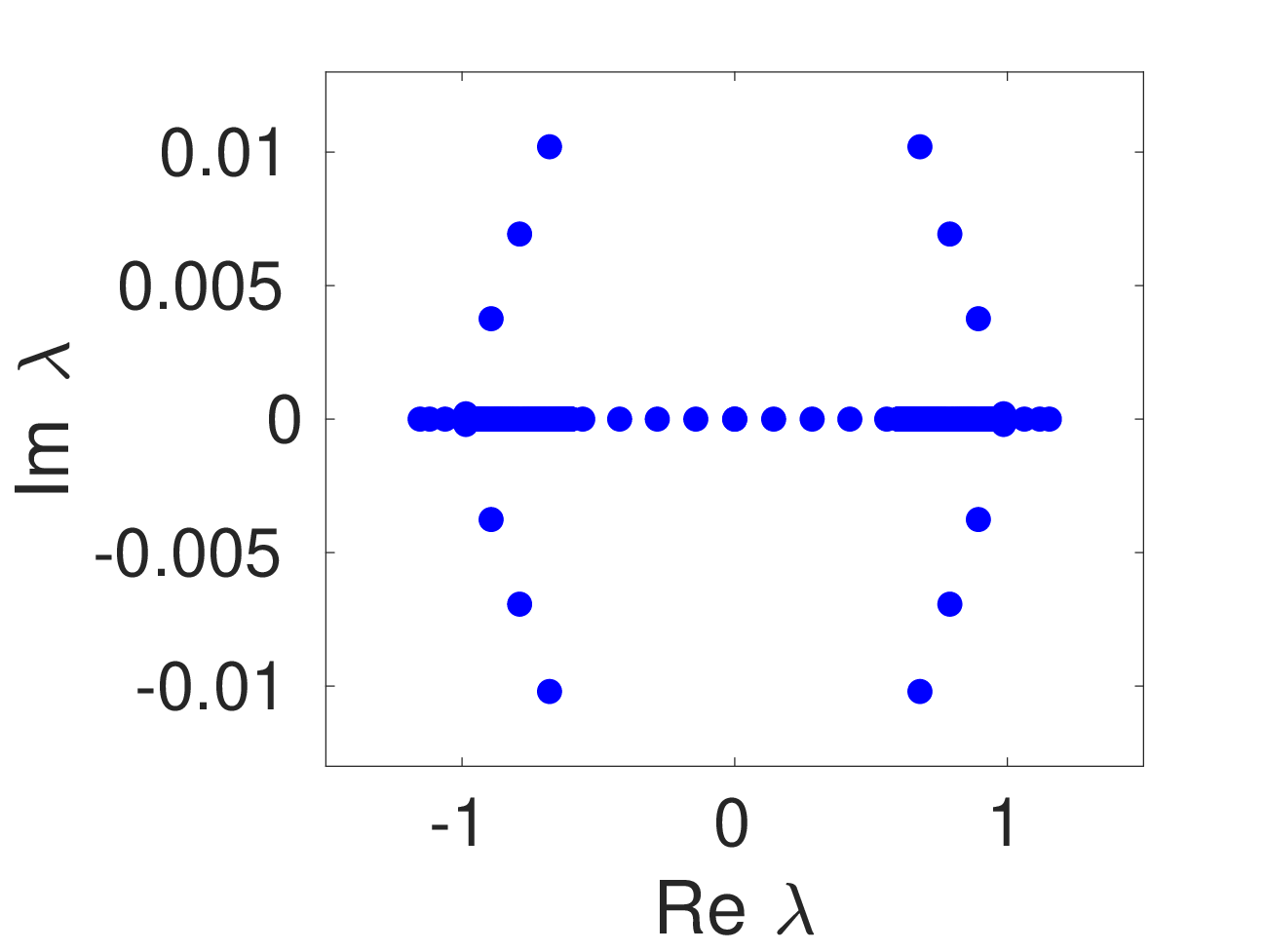}}
\caption{The real and imaginary parts of the eigenvalue spectrum for the strongly localized flat-top mode in Fig.~\ref{fig-19}.}
\label{fig-20}
\end{figure}

 The system can be described by the Eq.~(\ref{moddnlseQ1D}) and Eqs.~(\ref{LinEignVal}) with $\gamma=0$, 
so the two-body mean field interaction is eliminated. 
We apply the following values for parameters $\kappa=0.1$, $\gamma=0$ and $\delta=1$.
 To check the existence and stability the same steps are applied as in previous subsections and the results are presented in Figs.~\ref{fig-13}-\ref{fig-20}.  

%

Figures~\ref{fig-13} and \ref{fig-14} show that LHY strongly localized symmetric and antisymmetric odd modes exist and stable. 
%
%
%

From Figs.~\ref{fig-16} to \ref{fig-20} one may conclude that LHY strongly localized symmetric and antisymmetric modes and flat top modes exist, but they are unstable. For all discussed cases solutions obtained by the Page method are good approximations for numerically exact solutions. The stability and existence conditions for LHY strongly localised modes are qualitatively the same as for the solutions discussed in previous subsections when we assumed that attractive two-body interaction was dominant.  

In the next Sec.~\ref{sec:Variational} we are looking for localized quasi-continuous solutions for discrete quantum droplets.

\section{Quasi-continuous approximation. Variational method}
\label{sec:Variational}

Let us consider first in more detail the case when the coupling constant is large, as an example we take the following values for coefficients $ \kappa=1/2$, $\gamma=\delta=1$, so it is assumed that attractive mean field and repulsive LHY interactions are of the same order and there is a competition between them. Then Eq.~(\ref{dnlseQ1D}) for $\psi_{n,t}$ takes the following form,  
\begin{equation}
i \psi_{n,t}+ \cfrac{1}{2} (\psi_{n+1}+\psi_{n-1}-2 \psi_n) + |\psi_n|^2 \psi_n-|\psi_n|^3\psi_n=0\,.
\label{dnlseQ1D-C1}
\end{equation}
Replacing discrete variable $n$ with continuous variable $x$ and then writing an equation for stationary solutions using transformation $\psi(x,t)= \phi(x) \exp(-i \mu t)$, one gets the equation for steady solutions in continuous approximation as,
\begin{equation}
\mu \phi + \frac{1}{2}\phi_{xx} + \phi^3 - \phi^4 = 0.
\label{dnlseQ1D-C2}
\end{equation} 
Here we drop the norm sign, and in the following will consider only positive solutions.

The Lagrangian density for Eq.~(\ref{dnlseQ1D-C2}) is
\begin{equation}
{\cal L} = \frac{1}{4} \phi_x^2 - \frac{\mu}{2}\phi^2 -
\frac{1}{4}\phi^4 + \frac{1}{5}\phi^5.
\end{equation}

The super-Gaussian ansatz will be used, which is known as a good approximation
for both soliton and droplet shapes ~\cite{Otajonov2024},
\begin{equation}\label{ansatz1}
\phi(x) = A \exp\left[-\frac{1}{2}\left(\frac{x}{a}\right)^{2
m}\right]. 
\end{equation}

Substituting this  ansatz and integrating over $x$ one gets the averaged Lagrangian  

\begin{eqnarray}\label{lagrang1}
& L=\frac{A^2}{8 M a} \Gamma(2-M)-\mu A^2 a M\Gamma(M)-A^4 a\frac{
 M \Gamma(M)}{2^{M+1}}
 \nonumber \\
& +A^5 a(2/5)^{M+1}M \Gamma(M)\,.
\end{eqnarray}

The variational equations are derived from  $\partial
L/\partial A = 0$, \ $\partial L/\partial a=0$, \ $\partial
L/\partial M=0$. 

Also, we apply the normalization condition for $\phi(x)$, where $\Gamma(z)$ is a Gamma function.
\begin{equation}
\qquad N=\int_{-\infty}^{+\infty} |\phi(x)|^2dx = 2A^2a
\Gamma(1+M), \quad M = \frac{1}{2 m}.
\end{equation}
Then we get four equations for variational parameters $A,\;a,\;M$ and $\mu$.
From these equations parameter $a$ can be excluded using the equation for $N$, and the chemical potential can be found explicitly, then we have
  \begin{eqnarray}
 \frac{A^2 \Gamma(M)\Gamma(2-M)}{N^2}-2^{-(M+1)}+\frac{3}{5}(\frac{2}{5})^M A=0\,,
 \nonumber \\
 \frac{1}{2^{M+2}} \left[ 
	2\log(2)-\frac{1}{M}+\Psi(1+M)+\Psi(2-M)\right] +
\nonumber\\
 A(\frac{2}{5})^M \left[\frac{3}{10M}(1-M\Psi(2-M)-M \Psi(1+M)) + 
\right.
\nonumber\\
\left.
	\frac{2}{5}\log(\frac{2}{5})
\right]\,,
\nonumber\\
\mu=-A^2[3(2^{-(M+2)})- 0.7A(2/5)^M].
\label{VarEq1}
\end{eqnarray} 

\begin{figure}[htbp]
  \centerline{ \includegraphics[width=4.55cm]{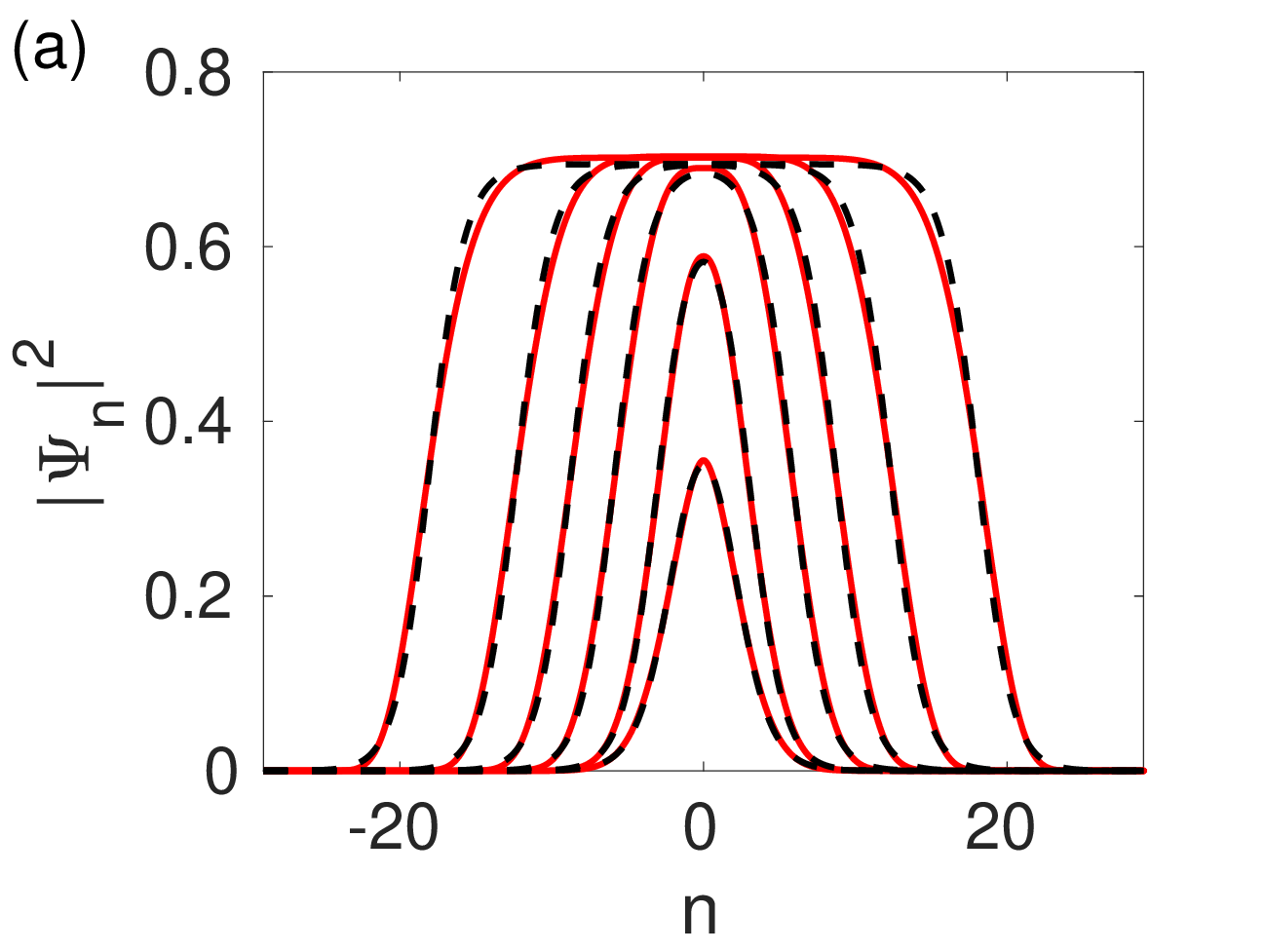}
  \includegraphics[width=4.55cm]{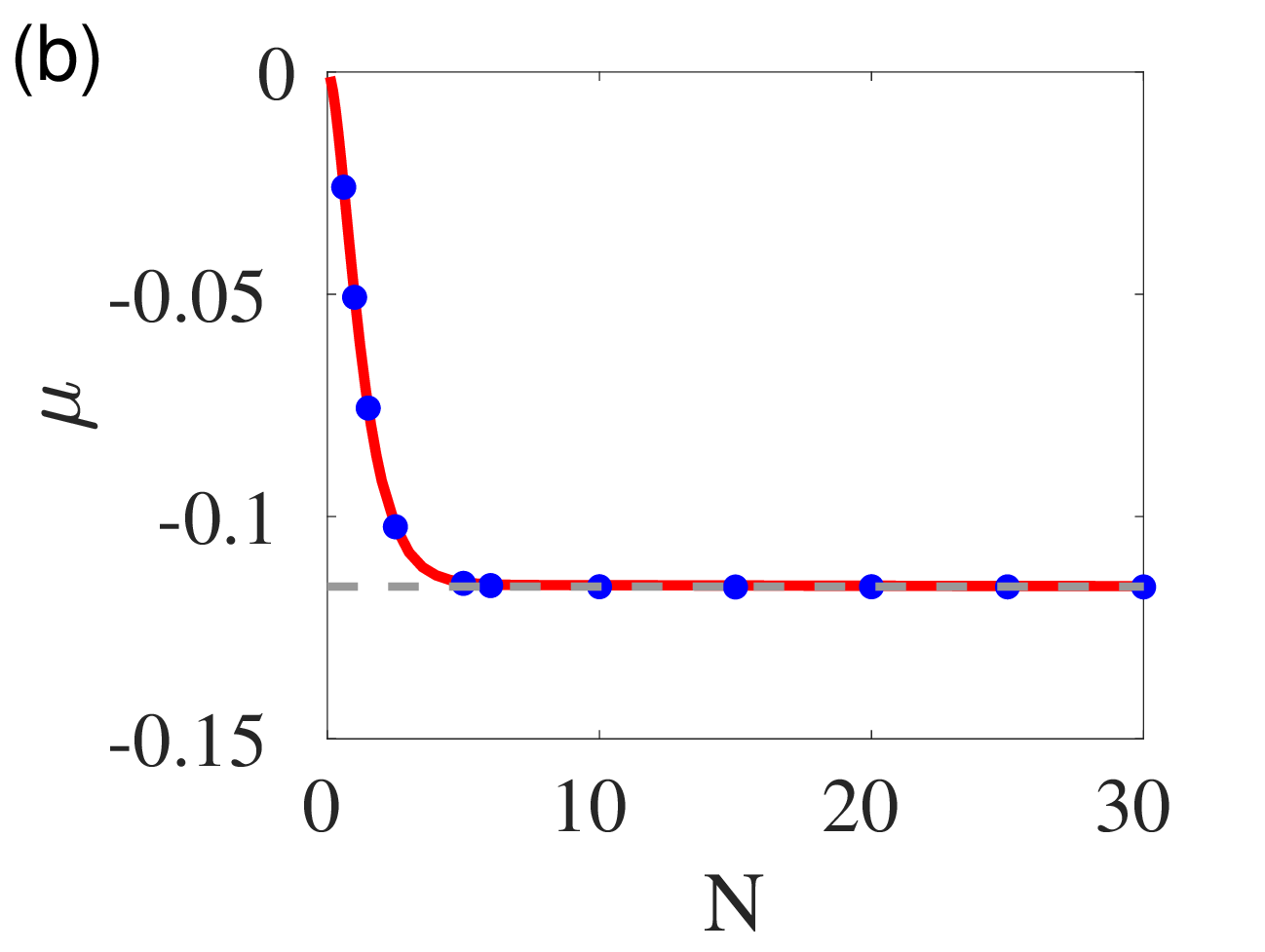}}
\caption{(a) Comparisons of quantum droplet profiles for different values of $N$. The straight lines correspond to VA for quasi-continuous cases while dashed lines represent numerically exact droplet solutions from bottom to top $N=2, 4, 8, 12, 17$ and $25$, respectively. In the quasi-continuous case, panel (b) shows the chemical potential versus norm $N$ for numerically exact droplet solutions (solid lines) and approximate variational solutions (points). The dashed line represents the Thomas-Fermi limit.
Parameters are $\kappa=0.5$, $\gamma=1$ and $\delta=1$.}
\label{fig-21}
\end{figure}

Solving these equations numerically allows one to determine all parameters for fixed norm $N$. 
Thomas-Fermi limit achieved ~\cite{Otajonov2024} when $N$ is large and $M$ is close to zero, in this limit we can get that $A=5/6$ and $\mu=-\frac{1}{6}(\frac{5}{6})^2$. The result for the Thomas-Fermi limit can be derived directly from both continuous and discrete equations.  
Now we proceed with the discussion of discrete quantum droplet localized solutions in quasicontinuous case. Figure~\ref{fig-21}(a) represents the comparisons of VA-predicted solutions with numerically exact solutions for different values of a number of atoms. One may note that for smaller values of $N$, the droplet has a solitonic shape, but for increased values of atom number the shape of the droplet becomes closer to the flat-top, and for the larger $N$ the Thomas-Fermi limit will be achieved. For all values of $N$ the variational super-Gaussian trial function quite well describes solutions, but still obtained solutions are approximate, and when we use them as the initial conditions to solve the time-dependent problem, the unwanted oscillations of the shape appear. To get a numerically exact solution we have applied Nijhof's iteration method~\cite{nijhof2000} adopted for Eq.~(\ref{dnlseQ1D}), and as an initial guess we pick the solution obtained by variational method.

Figure~\ref{fig-21}(b) shows the dependence of chemical potential on the norm for discrete droplets, one can see that for large values of a number of atoms Thomas-Fermi limit for chemical potential is achieved, and the variational method with supergaussian anzatz well describes this behavior of chemical potential. 

\begin{figure}[htbp]
  \centerline{ \includegraphics[width=4.55cm]{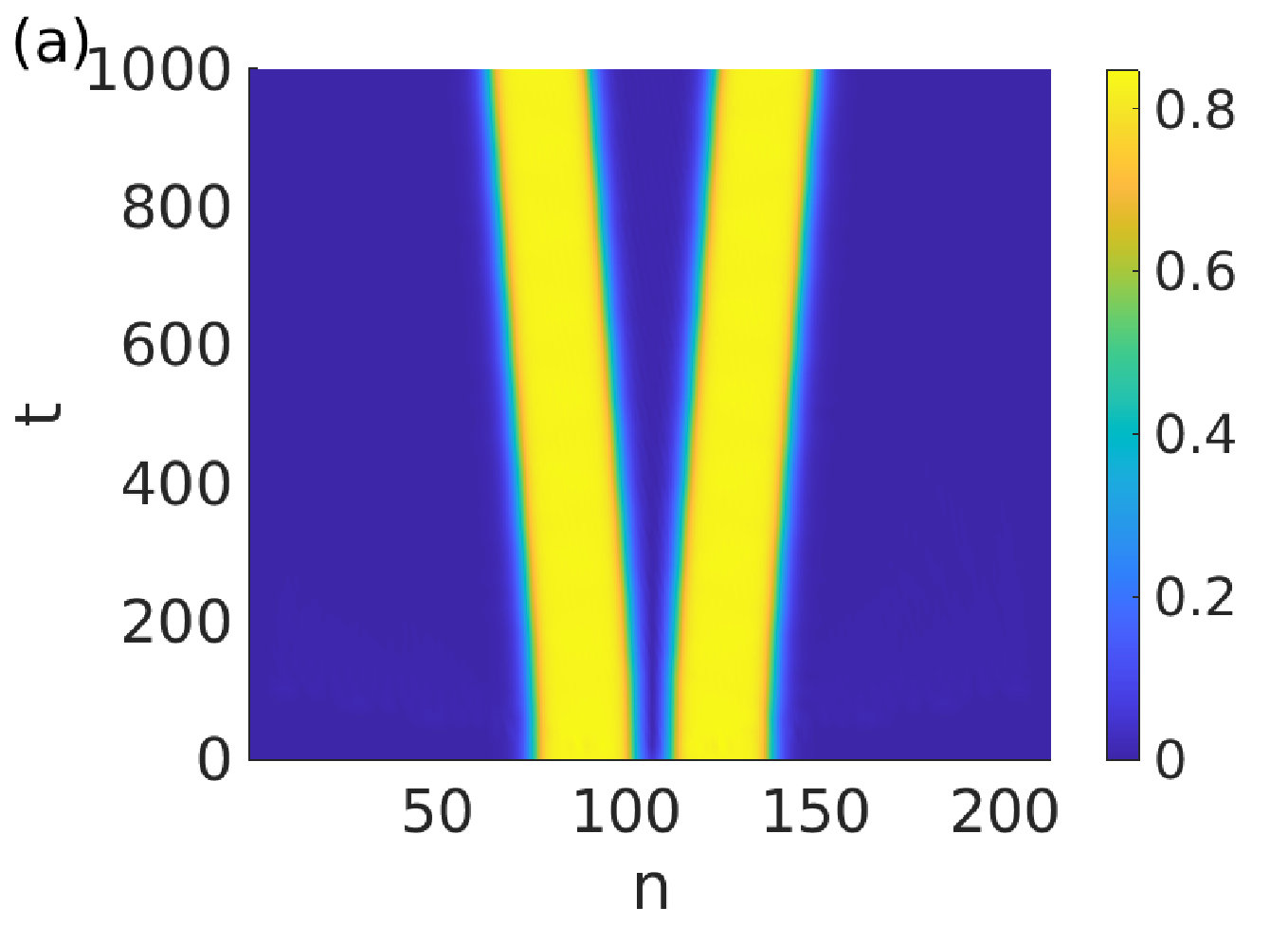}
   \includegraphics[width=4.55cm]{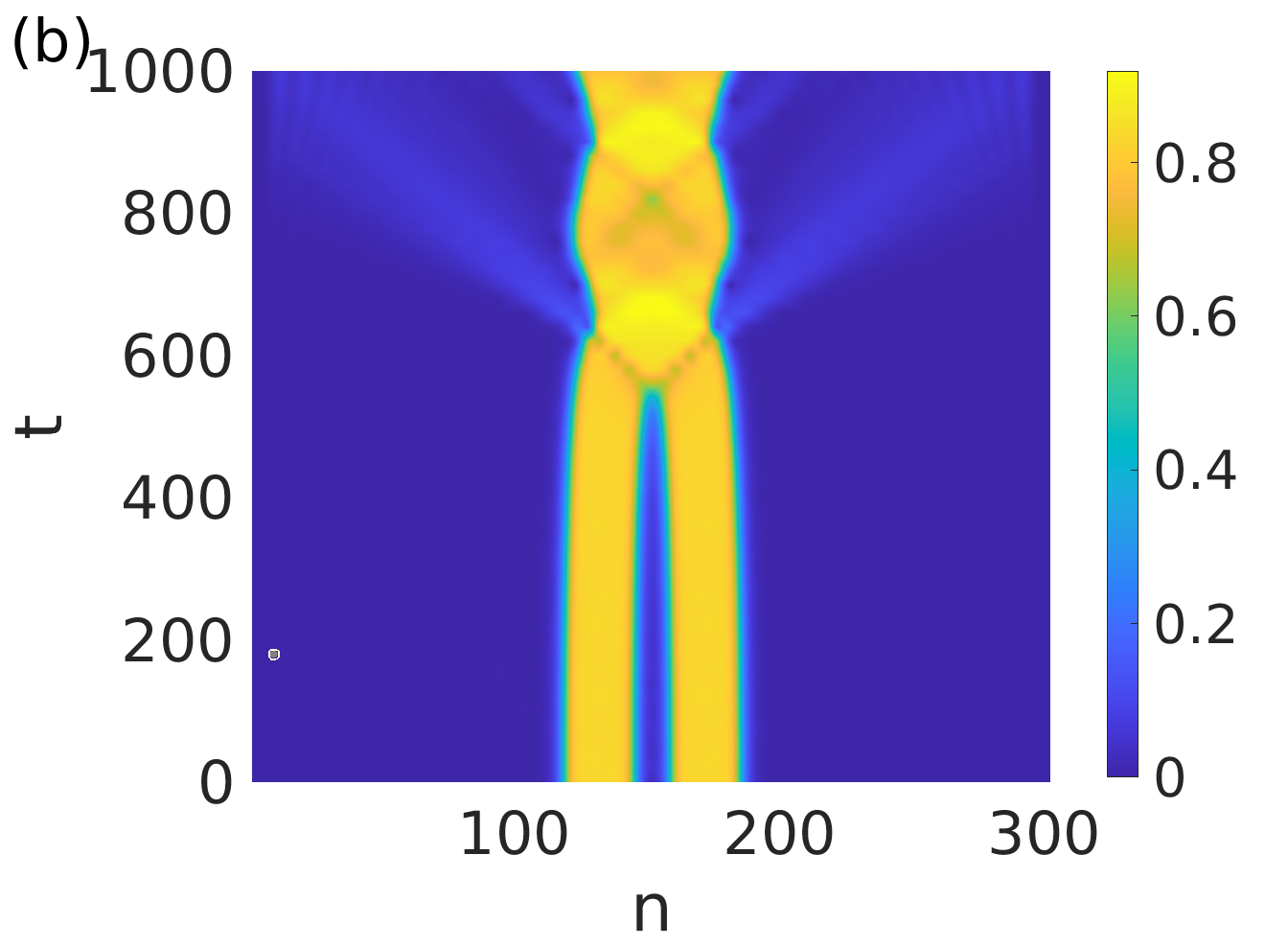}}
\caption{Interaction of flat-top droplets in the quasi-continuous case. Parameters are $N=17$, $\kappa=0.5$, $\gamma=1$ and $\delta=1$. (a) In-phase interaction $\phi = 0$; (b) Out-of-phase interaction $\phi = \pi$.}
\label{fig-22}
\end{figure}

\begin{figure}[htbp]
  \centerline{ \includegraphics[width=4.55cm]{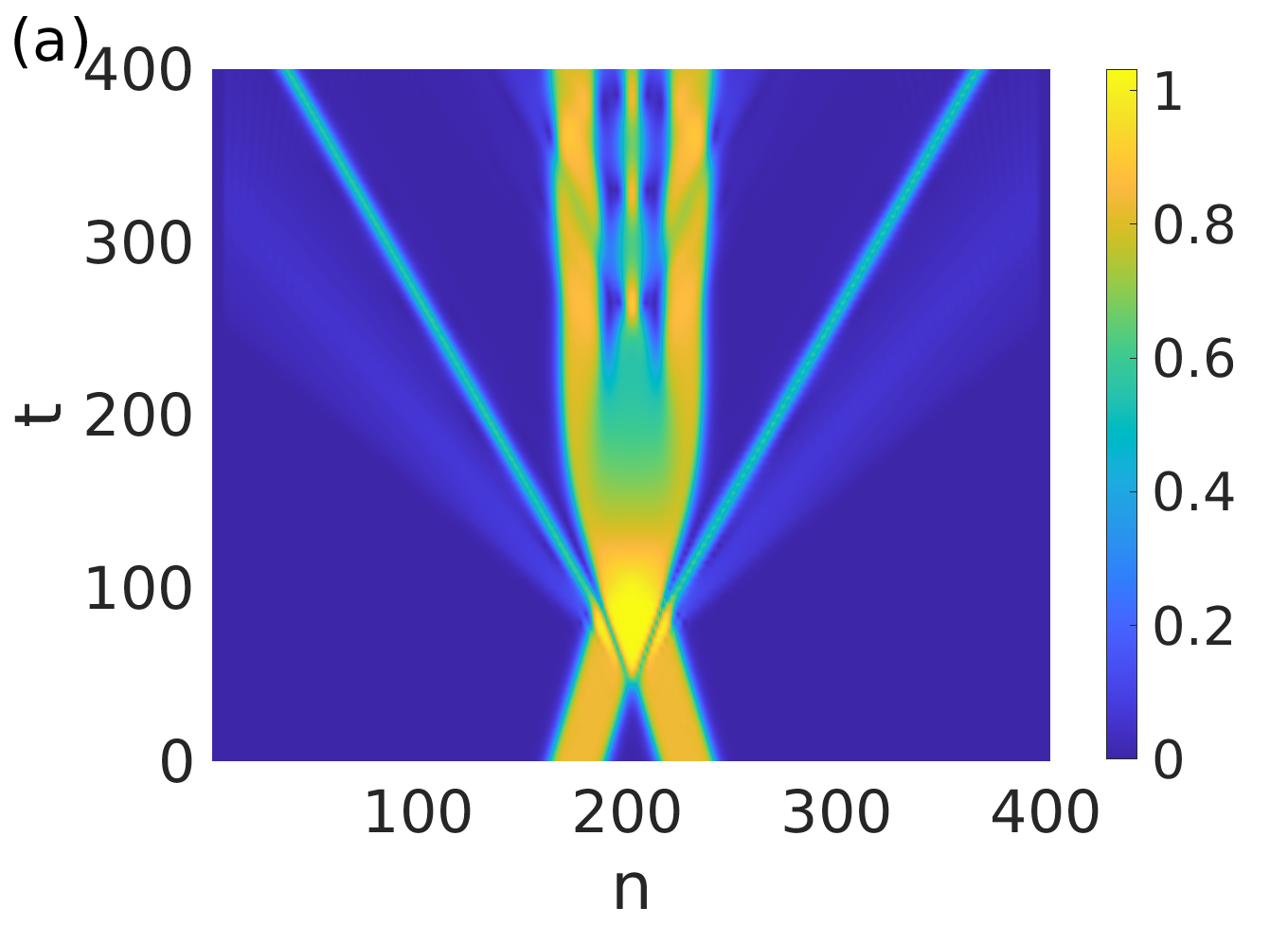}
   \includegraphics[width=4.55cm]{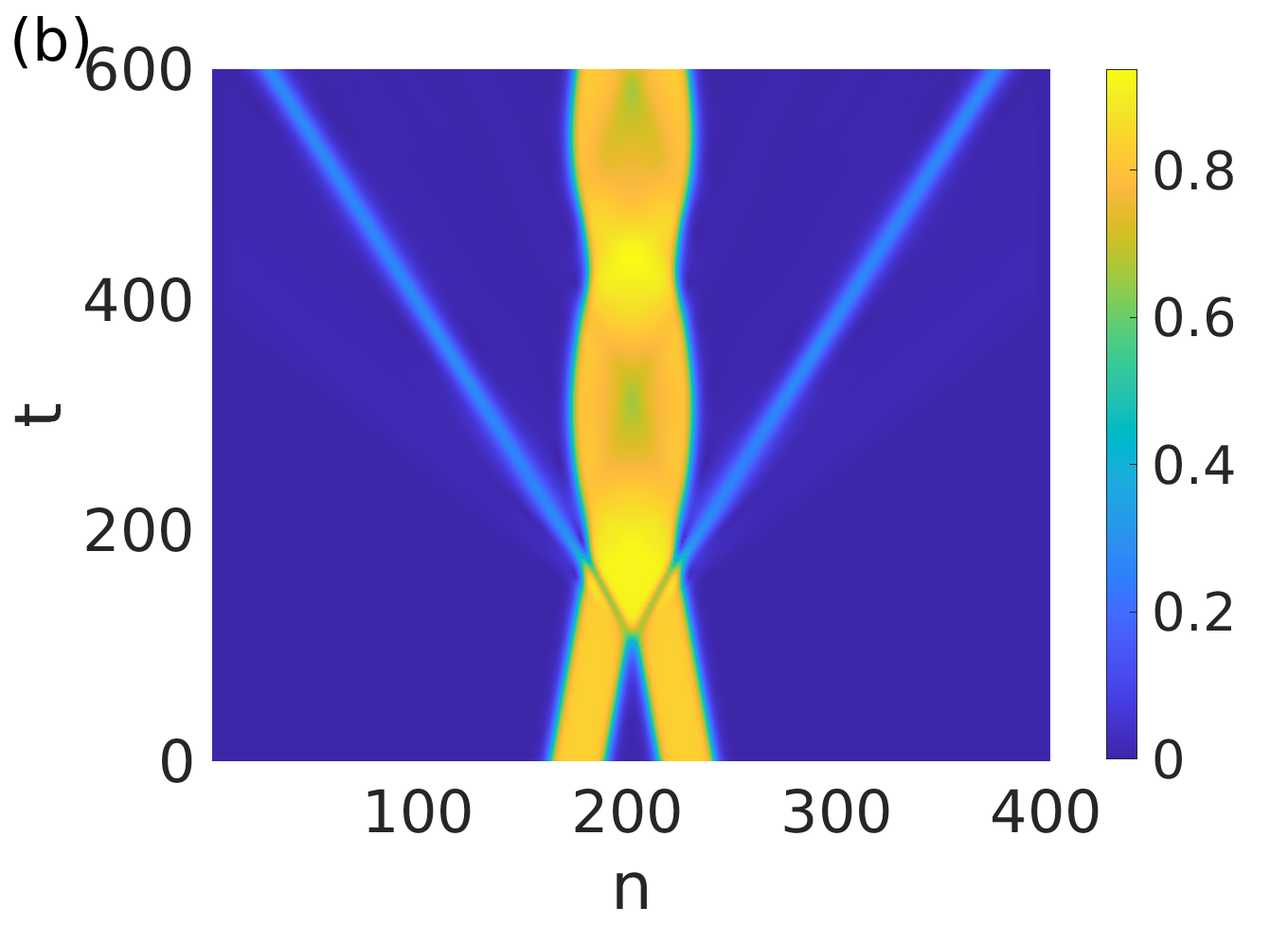}}
\caption{Interaction of flat top droplets in quasi-continuous case. Parameters are $N=17$, 
$\kappa=0.5$, $\gamma=1$ and $\delta=1$. (a) $v_1=0.5, v_2=-0.5$. (b) $v_1=0.2, v_2=-0.2$}
\label{fig-23}
\end{figure}

Using obtained solutions we can study an interaction of discrete droplets. Let us take as an initial condition two discrete droplets 
\begin{eqnarray}\label{VarEq1}
& \psi_{n,0}=\phi_1 (n-n_0)*\exp(i(v_1/2)(n-n_0)+\phi_1)+
 \nonumber \\
& \phi_1 (n+n_0)*\exp(i(v_2/2)(n+n_0)+\phi_2)
\end{eqnarray} 
located on some distance $2n_0$ from each other and apply an initial kick $v_1,\;v_2 $, so they will start to move. The results of numerical simulations of Eq.~(\ref{dnlseQ1D}) with these initial conditions are presented in Figs.~\ref{fig-22}-\ref{fig-23}.

Figure~\ref{fig-22} (a) illustrates the evolution of two flat-top droplets initially separated by a distance of $2n_0 = 10$, with equal phases and no initial momentum. Due to their interaction, the droplets repel each other and move freely in opposite directions with equal velocities. In Fig.~\ref{fig-22} (b), a similar initial configuration is shown, but with a phase difference of $\pi$ between the droplets. In this case, the droplets attract each other, eventually merging to form a larger droplet.

Figure~\ref{fig-23} shows the evolution of two flat-top discrete droplets initially separated by a distance of $2n_0 = 50$, with equal phases and initial kicks in opposite directions. The outcome of their collision depends on the relative speed. For higher relative speeds, as shown in Fig.\ref{fig-23} (a), three droplets form: two flat-top droplets with increased widths moving in opposite directions, and one stationary solitonic localized mode. For lower relative speeds, as shown in Fig.\ref{fig-22} (b), the droplets merge to form a larger droplet. Additionally, small amplitude localized wave packets, moving rapidly in opposite directions, are observed.

\begin{figure}[htbp]
  \centerline{ \includegraphics[width=4.55cm]{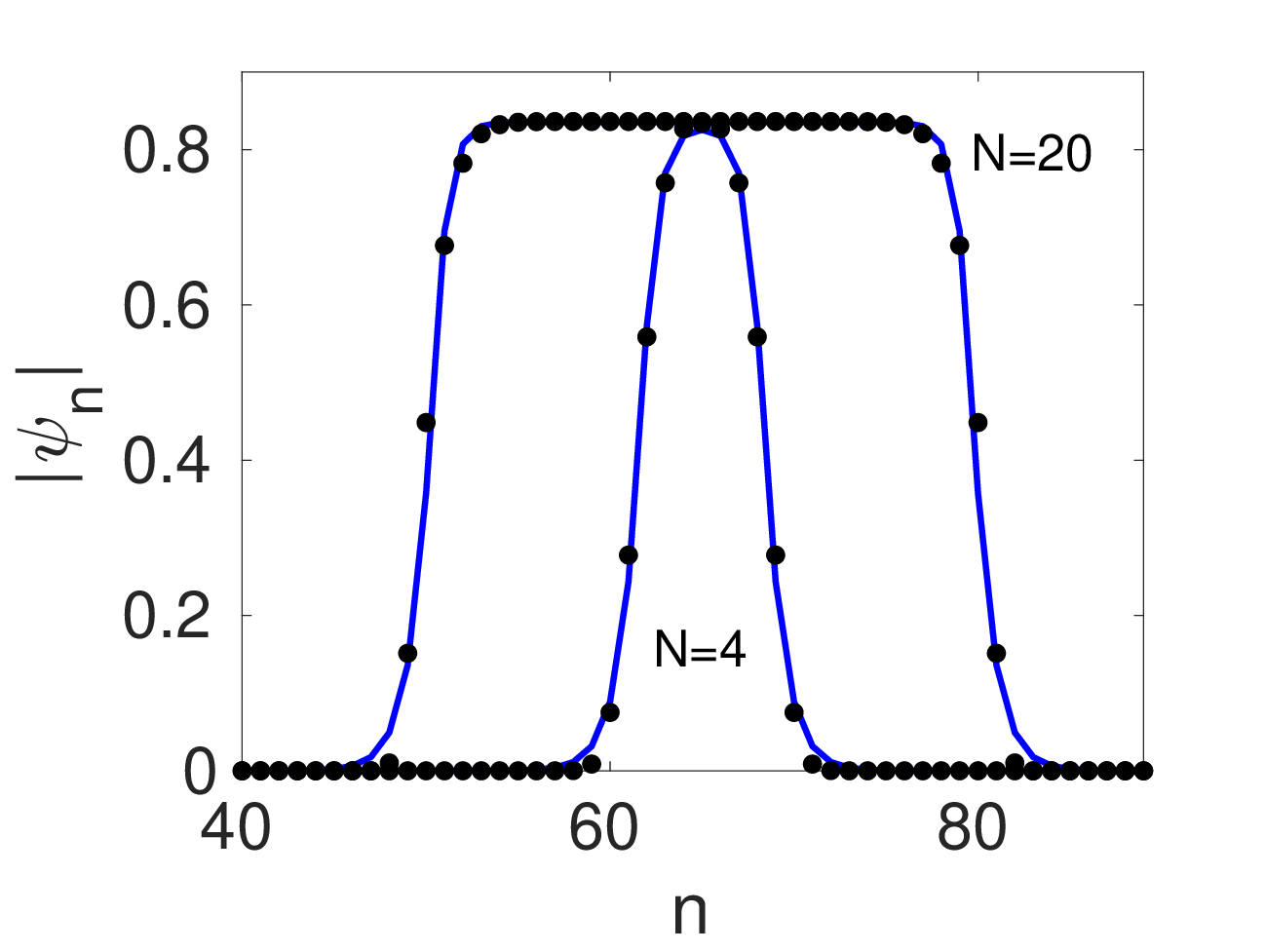}}
\caption{Numerically exact droplet solutions (solid lines) and approximate variational solutions (points) in the quasi-continuous case. For a small number of atoms ($N = 4$), the solution is bell-shaped, whereas for a larger number of atoms ($N = 20$), it is flat-top shaped. 
Parameters are $\kappa=0.2$, $\gamma=1$ and  $\delta=1$.}
\label{fig-24}
\end{figure}

\begin{figure}[htbp]
  \centerline{ \includegraphics[width=4.55cm]{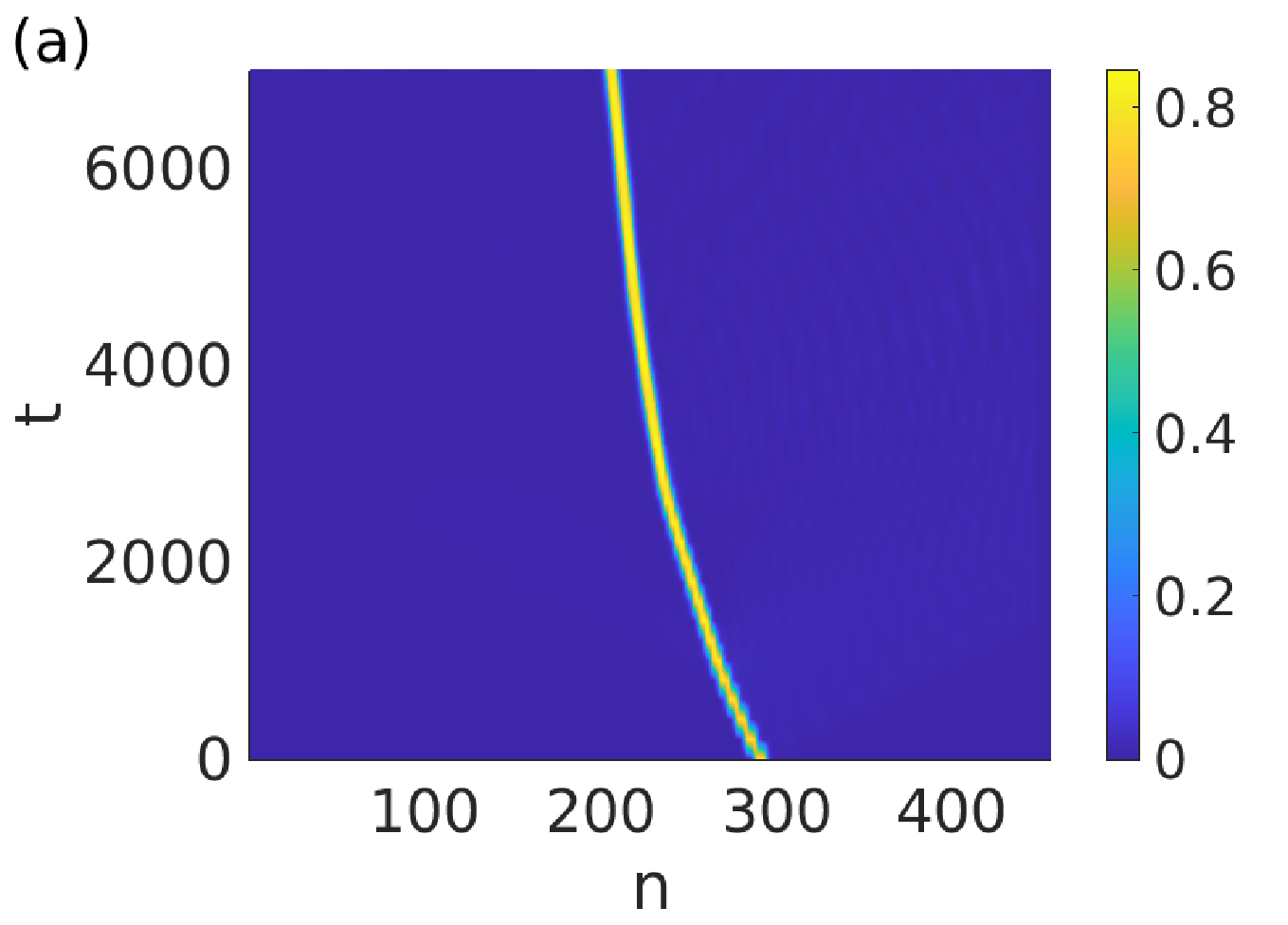}
   \includegraphics[width=4.55cm]{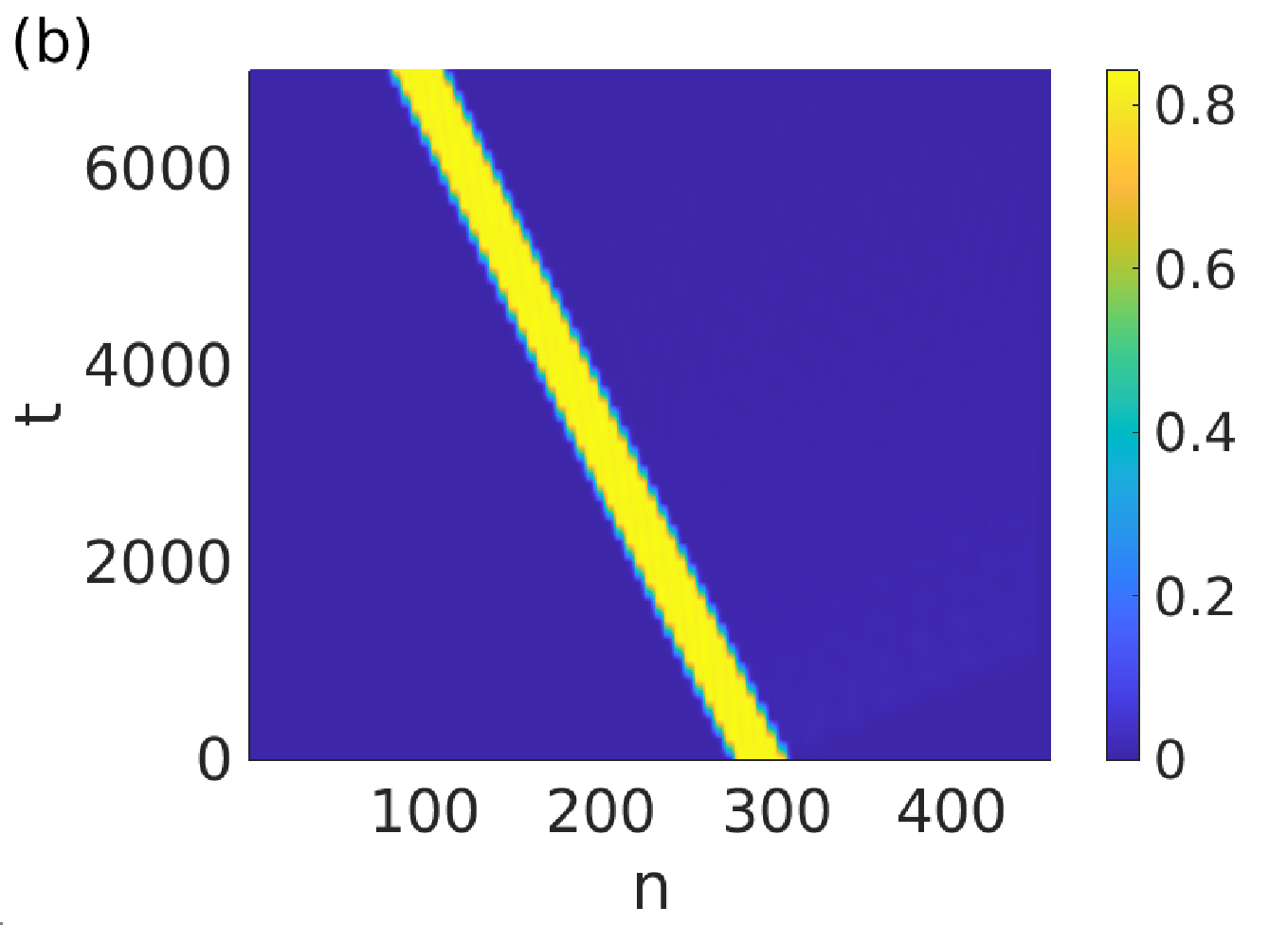}}
    \centerline{ \includegraphics[width=4.55cm]{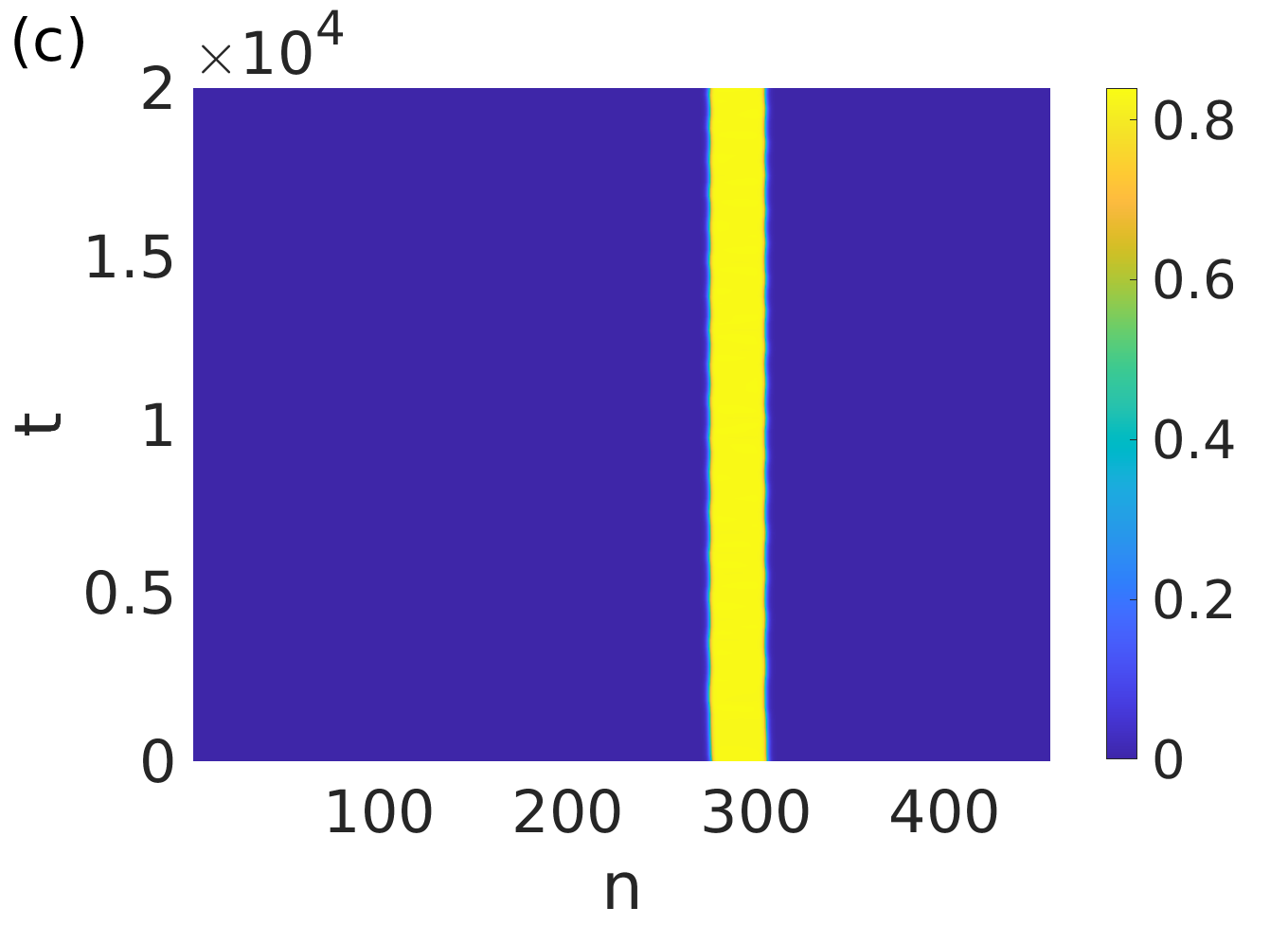}
    \includegraphics[width=4.55cm]{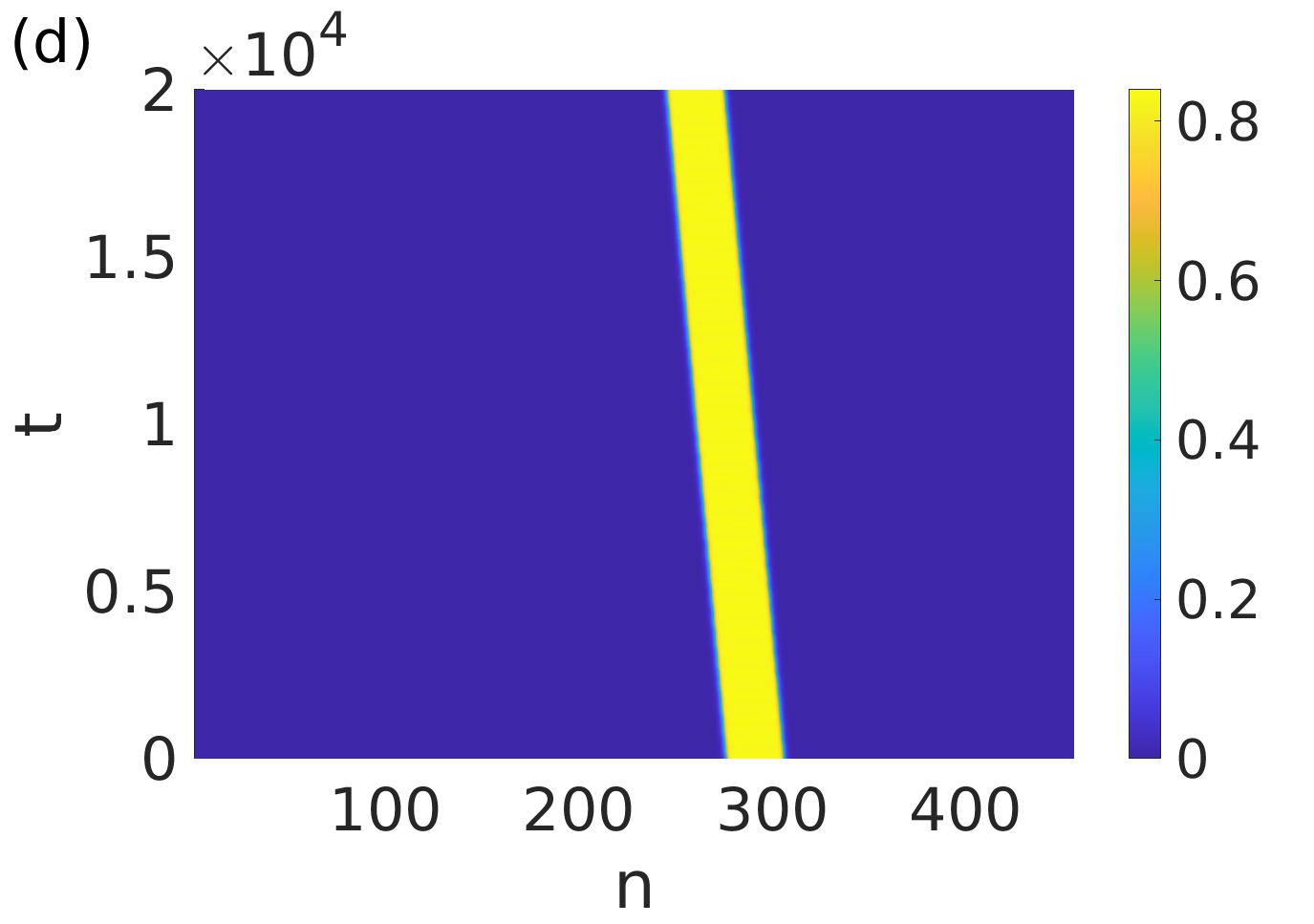}}
\caption{Time evolution of flat-top droplets in the quasi-continuous case:
(a) $(N, v)=(4,0.3)$; (b) $(N, v)=(20, 0.65)$. (c) $(N, v)=(20, 0.02)$; (d) $(N, v)=(20, 0.025)$.
Other parameters are $\kappa=0.2$, $\gamma=1$ and $\delta=1$. }
\label{fig-25}
\end{figure}

Next, we examine the existence and dynamics of quasi-continuous discrete quantum droplets for arbitrary values of $\kappa$, based on Eq.~(\ref{dnlseQ1D}). The equation for stationary solutions in the continuous approximation takes the following form:
\begin{equation}
\mu \phi + \frac{\kappa}{2}\phi_{xx} + \phi^3 - \phi^4 = 0.
\label{dnlseQ1D-C3}
\end{equation}
This equation can be reduced to the  Eq.~(\ref{dnlseQ1D-C2}) with $\kappa=1$, by redefining the spatial variable $x' = x/\sqrt{\kappa}$. From that one can deduce that if we redefine the width $a' = \sqrt{\kappa}a $ and norm $N' =  \sqrt{\kappa}N $, the results obtained from the application of super-Gaussian ansatz with $\kappa=1$ can be used for the case with arbitrary $\kappa$. We are particularly interested in the tight bound case with small hopping rates, so in Figs.~\ref{fig-24}-\ref{fig-25} the results for $\kappa=0.2$ are presented. 
%

Figure~\ref{fig-24} presents the comparison of results obtained for soliton and flat top shapes by variational method and by exact numerical solution. One can see that the variational method quite well describes the static solution even for strongly localized modes.
In Fig.~\ref{fig-25} (a), we present the numerical results for the evolution of a strongly localized soliton-like mode with an initial kick. One can observe that this mode preserves its identity for a long time, but its velocity gradually decreases. The decrease in velocity can be qualitatively deduced from the figure by noting that the rate of change of the droplet's position over time decreases compared to its initial rate. In contrast, as shown in Fig.~\ref{fig-25} (b), a strongly localized mode with a flat-top shape and a larger $N$ preserves both its shape and velocity. 
In Fig.~\ref{fig-25} (c) and (d) the results of the evolution of a flat-top discrete droplet are presented when a small kick is applied. From these figures, one can deduce that even a small increase in the initial kick may change the dynamics of the droplet from localized at an initial position to moving with some nonzero velocity.
These results can be qualitatively explained by the effect of the Peierls-Nabarro potential barrier
on localized modes~\cite{Malomed2013}.            

\section{Conclusions}
\label{sec:conc}

We have examined the conditions for MI and the existence, stability, and interaction dynamics of discrete breathers in a elongated Bose-Einstein condensate loaded in a deep optical lattice, considering quantum fluctuations. These are described by the discrete Gross-Pitaevskii equation with cubic-quartic nonlinearity. We derived analytical equations for the MI growth rate and identified the instability regions of nonlinear plane-wave solutions across different parameter spaces. 
Under similar initial conditions, such as equal coupling constants and plane wave amplitudes, cubic nonlinearity has a stronger effect on MI compared to quartic nonlinearity. In certain regions of the parameter space, the combined effect of cubic and quartic nonlinearity can induce MI, whereas neither cubic nor quartic nonlinearity alone can cause instability.
The Page method is employed to derive analytical approximate solutions for various types of strongly localized discrete breathers, including flat-top discrete quantum droplets. Numerical simulations demonstrate that strongly localized even modes and flat-top modes are unstable, while odd modes are stable.

It is also shown that strongly localized discrete modes persist even in the absence of attractive mean-field nonlinearity, as in the Lee-Huang-Yang fluid case. The existence of these modes is only due to the effective nonlinearity induced by quantum fluctuations. This is in contrast to the cubic-quintic discrete nonlinear Schrödinger equation, where the cubic (Kerr) nonlinearity is considered the primary contributing factor \cite{Malomed2013, Abdullaev2007, Gonzales2006}.

A variational method using a super-Gaussian ansatz is applied to study localized modes within a quasi-continuous approximation. It is shown that this variational approach provides a good approximation for both soliton-like and flat-top discrete modes, regardless of whether the linear coupling is strong or weak, as confirmed by numerical simulations.

The evolution and interaction of the resulting discrete quantum droplets are further studied numerically. In the weak coupling regime, the interaction of flat-top discrete droplets closely resembles the behavior of continuous quantum droplets~\cite{Otajonov2024}: two discrete droplets attract when the phase difference is zero and repel when the phase difference is $\pi$. The dynamics of collisions between two discrete droplets strongly depend on their initial velocities. At low velocities, the droplets merge to form larger ones, while at higher velocities, the collision may result in several smaller droplets.

In the strong coupling regime, the discrete nature of the system significantly influences the evolution of moving droplets. For flat-top droplets with large particle numbers $N$, discreteness only affects mobility for very small initial impulses. In contrast, soliton-like droplets with small $N$ may experience deceleration due to Peierls-Nabarro barriers.

\section*{Acknowledgements}

This work has been funded from the State budget of the Republic of Uzbekistan.


\newpage
\appendix
\section{Effective one-dimensional extended Gross-Pitaevskii equation with Lee-Huang-Yang correction term in an elongated trap}
\label{appenA}
Let us derive the effective one-dimensional extended GP equation with the LHY term describing BEC in the elongated cigar-type trap. The approach is to consider $l_{\perp}$ as the variational variable, i.e. calculate it with the aid of the variational approach~\cite{Salasnich1, Blackie1, Blackie2}. 

The energy is~\cite{Jorgensen}:
\begin{equation}
E= \int d^3r[\frac{\hbar^2}{2m_0}|\nabla\Psi|^2 +V(r)\Psi +\frac{\delta g}{2}|\Psi|^4 + \frac{2}{5}\gamma_{QF}|\Psi|^5],
\end{equation}
where 
$$
V(r)=\frac{1}{2}m_0 \omega_z^2 + \frac{1}{2}m_0 \omega_{\perp}^2, \ \gamma_{QF}=\frac{4m_0^{3/2}}{3\pi^2\hbar^3}g^{5/2}.
$$
Looking for the solution in the form
\begin{equation}
\Psi(r,t)=R(x,y)\psi(z,t),
\end{equation}
with
\begin{equation}
R(x,y)=\frac{1}{\sqrt{\pi}l_{\perp}}e^{-\frac{x^2+y^2}{2l_{\perp}^2}},
\end{equation}
and integrating over the transverse coordinates, we obtain for the energy:
\begin{eqnarray}
E_{\mathrm{1D}}& =& \int dx [\frac{\hbar^2}{2m_0}(|\psi_z|^2) + \frac{|\psi|^2}{l_{\perp}^2}) +
\nonumber\\ 
&& (\frac{m_0 \omega_z^2}{2}x^2 + \frac{m_0\omega_{\perp}^2l_{\perp}^2}{2})|\psi|^2 +
\nonumber\\  
&& \frac{\delta g}{4\pi l_{\perp}^2}|\psi|^4 + \frac{4\gamma_{QF}}{25\pi^{3/2}l_{\perp}^3}|\psi|^5].
\end{eqnarray}
Performing minimization on $l_{\perp}$ 
$$
\frac{\delta E}{\delta l_{\perp}}=0,
$$
and by taking $l_{\perp}=l_0\sigma, \, l_0=\sqrt{\hbar/m_0 \omega_{\perp}}$, we find:
\begin{equation}\label{sig1}
\sigma^5 -(1+ \frac{\delta a}{2 \pi a}\xi )\sigma -\frac{64 a}{25\pi^{5/2} l_0}\xi^{3/2}=0.
\end{equation}
where $\xi=4\pi a|\psi|^2$, and $n_{\mathrm{1D}}=|\psi|^2$ 

Let us determine the parameters of our model under realistic experimental conditions. We consider $^{39}\mathrm{K}$ atoms in different spin states with mass $m_0 = 6.49 \times 10^{-26} \,\mathrm{kg}$. Both intra-species scattering lengths are set to $ a_{11} = a_{22} = a = 500\,a_0$, where $a_0$ is the Bohr radius. The residual scattering length is taken within the interval $\delta a \in (0.02 - 0.05) a$.
We choose a transverse trap frequency of $\omega_\perp = 2\pi \times 200 \,\mathrm{Hz}$, which provides tight confinement and allows the condensate dynamics to be treated within an effective-1D approximation. The corresponding harmonic oscillator length is given by $l_0 = \sqrt{\frac{\hbar}{m_0 \omega_\perp}}=1.137$ $\mu$m. The parameter $\xi$ is defined according to the characteristic scales of our model (see after Eq.~(\ref{dlessGPE1D})). For $|\tilde{\gamma}| = |\tilde{\delta}| = 1$, the values of $\xi$ fall within the range $\xi \approx 0.31\text{--}1.96$. From Eq.~(\ref{sig1}), we then obtain $ \sigma \approx 1$ for this interval, indicating only a small renormalization. This confirms the applicability of the proposed model for higher values of $\xi$.


\begin{thebibliography}{99}

\bibitem{Petrov2015} D. S. Petrov, Quantum mechanical stabilization of a
collapsing Bose-Bose mixture, Phys. Rev. Lett. {\bf 115}, 155302 (2015).

\bibitem{Barbut2019} I. Ferrier-Barbut, Ultradilute quantum droplets, Phys. Today {\bf 72}(4), 46 (2019).

\bibitem{Luo2021} Z. H. Luo, W. Pang, B. Liu, Y. Y. Li, and B. A. Malomed, A new form of liquid matter:
Quantum droplets, Frontiers of Physics \textbf{16}, 32201 (2021).

\bibitem{LHY} T. D. Lee, K. Huang, and C. N. Yang, Eigenvalues and eigenfunctions of a Bose system of hard spheres and its low-temperature properties, Phys. Rev. \textbf{106}, 1135 (1957).

\bibitem{Barbut2016} I. Ferrier-Barbut, H. Kadau, M. Schmitt, M. Wenzel, and T.
Pfau, Observation of quantum droplets in a strongly dipolar Bose Gas, Phys.
Rev. Lett. {\bf 116}, 215301 (2016).

\bibitem{Cheiney2018} P. Cheiney, C. R. Cabrera, J. Sanz, B. Naylor, L. Tanzi, and
L. Tarruell, Bright soliton to quantum droplet transition in a mixture of
Bose-Einstein condensates, Phys. Rev. Lett. 120 (2018) 135301.

\bibitem{D'Errico2019} C. D'Errico, A. Burchianti, M. Prevedelli, L. Salasnich,
F. Ancilotto, M. Modugno, F. Minardi, and C. Fort,
Observation of quantum droplets in a heteronuclear bosonic mixture,
Phys. Rev. Research {\bf 1}, 033155 (2019).

\bibitem{Petrov2016} D. S. Petrov and G. E. Astrakharchik, Ultradilute low-dimensional liquids,
Phys. Rev. Lett. \textbf{117}, 100401 (2016).

\bibitem{Otajonov2019} Sh. R. Otajonov, E. N. Tsoy, and F. Kh. Abdullaev,
Stationary and dynamical properties of one-dimensional quantum droplets,
Phys. Lett. A, {\bf 383}, 125980 (2019).

\bibitem{Otajonov2020} Sh. R. Otajonov, E. N. Tsoy, and F. Kh. Abdullaev,
Variational approximation for two-dimensional quantum droplets, Phys. Rev. E \textbf{102}, 062217 (2020).

\bibitem{Otajonov2022_1} Sh. R. Otajonov,
Quantum droplets in three-dimensional Bose-Einstein condensates,
J. Phys. B: At. Mol. Opt. Phys. {\bf 55}, 085001 (2022).

\bibitem{Otajonov2024} Sh. R. Otajonov, B. A. Umarov, and F. Kh. Abdullaev,
Dynamics of quasi-one-dimensional quantum droplets in Bose–Bose mixtures,
Chaos, Solitons and Fractals, {\bf 186}, 115212 (2024).

\bibitem{Debnath2021} A. Debnath, A. Khan, Investigation of quantum droplets: An analytical approach, Ann. Phys. (Berlin) {\bf 533}, 2000549 (2021).

\bibitem{Debnath2022} A. Debnath, A. Khan, S. Basu, Dropleton-soliton crossover mediated via trap modulation, Physics Letters A {\bf 439}, 128137 (2022). 

\bibitem{Liu2023} B. Liu, X. Cai, X. Qin, X. Jiang, J. Xie, B. A. Malomed, and Y. Li, Ring-shaped quantum droplets with hidden vorticity in a radially periodic potential, Phys. Rev. E {\bf 108}, 044210 (2023).

\bibitem{Zhang2019} X. Zhang, X. Xu, Y. Zheng, Z. Chen, B. Liu, C. Huang, B. A. Malomed, and Y. Li,
Semidiscrete quantum droplets and vortices, Phys. Rev. Lett. \textbf{123}, 133901 (2019).

\bibitem{Zhao2021} F. Zhao, Z. Yan, X. Cai, C. Li, G. Chen, H. He, B. Liu, Y. Li, Discrete quantum droplets in one-
dimensional optical lattices, Chaos, Solitons and Fractals \textbf{152}, 111313 (2021).

\bibitem{Katsimiga2023} G. S. Katsimiga, I. G. Mistakidis, B. A. Malomed, D. J. Frantzeskakis, R. Carretero-Gonzalez, P. G. Kevrekidis, Interactions and dynamics of one-dimensional droplets bubbles and kinks. Condens Matter \textbf{8}, 67 (2023).
 
\bibitem{Debnath2023} A. Debnath, A. Khan, B. A Malomed, Interaction of one-dimensional quantum
droplets with potential wells and barriers, Commun. Nonlinear. Sci. Numer. Simul. \textbf{126}, 107457 (2023).
 
\bibitem{Khan2024} S. S. Adusumalli, K. Senapati, S. Singh, A. Khan, Quantum liquid in lower dimensions: From the perspective of surface tension. Phys. Lett. A, \textbf{516} 129638 (2024).
 
\bibitem{Debnath2023a} A. Debnath, A.Khan, P. K. Panigrahi, Dynamics of bright soliton under cubic-quartic interactions in quasi-one-dimensional geometry, Eur. Phys. J. Plus. \textbf{138}, 954 (2023).

\bibitem{Sakkaf2024} U. Al Khawaja, M. O. D. Alotaibi , and L. Al Sakkaf, Elastic and inelastic scattering of flat-top solitons, Phys. Rev. E \textbf{110}, 044215 (2024).

\bibitem{Benjamin1967} T. B. Benjamin and J. E. Feir, The disintegration of wave trains on deep water.
1. Theory, J. Fluid Mech. {\bf 27}, 417 (1967).

\bibitem{Kivshar1992} Y.S. Kivshar and M. Peyrard, Modulational instabilities in discrete lattices, Phys. Rev. A \textbf{46}, 3198 (1992).

\bibitem{Kevrekidis2009} P. G. Kevrekidis, The Discrete Nonlinear Schr\"odinger Equation: Mathematical Analysis, Numerical Computations, and Physical Perspectives (Springer: Berlin and Heidelberg, 2009).

\bibitem{Sukhorukov2003} A. A. Sukhorukov, Y. S. Kivshar, H. S. Eisenberg and Y. Silberberg, Spatial optical solitons in waveguide arrays, IEEE Journal of Quantum Electronics, \textbf{39}, 31-50 (2003).

\bibitem{Mingaleev2000} S. F. Mingaleev, Yu. S. Kivshar, and R. A. Sammut, Long-range interaction and nonlinear localized modes in photonic crystal waveguides, Phys. Rev. E \textbf{62}, 5777 (2000).

\bibitem{Trombettoni2001} A. Trombettoni and A. Smerzi, Discrete Solitons and Breathers with Dilute Bose-Einstein Condensates, Phys. Rev. Lett.  \textbf{86}, 2353 (2001). 

\bibitem{Abdullaev2001} F. Kh. Abdullaev, B. B. Baizakov, S. A. Darmanyan, V. V. Konotop, and M. Salerno, Nonlinear excitations in arrays of Bose-Einstein condensates, Phys. Rev. A \textbf{64}, 043606 (2001).

\bibitem{Morsch2006} O. Morsch and M. Oberthaler, Dynamics of Bose-Einstein condensates in optical lattices, Rev. Mod. Phys. \textbf{78}, 179 (2006).

\bibitem{Flach2008} S. Flach and A. V. Gorbach, Discrete breathers--advances in theory and applications, Phys. Rep. \textbf{467}, 1 (2008).

\bibitem{Edler2017} D. Edler, C. Mishra, F. Wächtler, R. Nath, S. Sinha, and L. Santos, Quantum fluctuations in quasi-one-dimensional dipolar bose-einstein condensates, Phys. Rev. Lett. {\bf 119}, 050403 (2017).

\bibitem{Zin2018} P. Zi\'n, M. Pylak, T. Wasak, M. Gajda, and Z. Idziaszek, Quantum Bose-Bose droplets at a dimensional crossover, Phys. Rev. A{\bf  98}, 051603 (2018).

\bibitem{Salasnich1} L. Salasnich, A. Parola, and L. Reatto, Effective wave equations for the dynamics of cigar-shaped and disk-shaped Bose condensates, Phys. Rev. A {\bf 65}, 043614 (2002).

\bibitem{Blackie1} P. Blakie, D. Baillie and S. Pal, Variational theory for the ground state and collective excitations of an elongated dipolar condensate, Commun. Theor. Phys., {\bf 72}, 085501 (2020).

\bibitem{Blackie2} Blakie, P., Baillie, D., Chomaz, L. and Ferlaino, F. Supersolidity in an elongated dipolar condensate. Phys. Rev. Res. {\bf 2}, 043318 (2020).

\bibitem{Bloch_discr} G. Natale, Th. Bland, S. Gschwendtner, L. Lafforgue, D. S. Gr\"un, A. Patscheider, M. J. Mark  and F. Ferlaino, Bloch oscillations and matter-wave localization of a dipolar quantum gas in a one-dimensional lattice, Comm. Phys. {\bf  5},  227 (2022).

\bibitem{Malomed1} E. Shamriz, Z. Chen , and Boris A. Malomed, Suppression of the quasi-two-dimensional quantum collapse in the attraction field by the Lee-Huang-Yang effect, Phys. Rev. A {\bf 101}, 063628 (2020).

\bibitem{Malomed2} L. Dong, M. Fan, and B. A. Malomed, Stable higher-order vortex quantum droplets in an annular potential, Chaos, Solitons \& Fractals, {\bf 179}, 114472 (2024).

\bibitem{China1}
Z. Lin, X. Xu, Z. Chen, Z. Yan, Z. Mai, and B. Liu, Two-dimensional vortex quantum droplets get thick, Commun. in Nonlin. Sci. and Num. Sim. {\bf 93}, 105536 (2021).

\bibitem{Edmonds2020}
Edmonds, M., Bland, T. and  Parker, N. Quantum droplets of quasi-one-dimensional dipolar Bose–Einstein condensates. J. Phys. Comm. {\bf 4}, 125008 (2020).

\bibitem{Palo2023} S. D. Palo, E. Orignac, R. Citro, and L. Salasnich, Effect of Transverse Confinement on a Quasi-One-Dimensional Dipolar Bose Gas, Condens. Matter {\bf 8}, 26 (2023).

\bibitem{Alfimov} G. L. Alfimov, P. G. Kevrekidis, V. V. Konotop, and M. Salerno, Wannier functions analysis of the nonlinear Schr\"odinger equation with a periodic potential, Phys. Rev. E {\bf 66}, 046608 (2002).

\bibitem{Lewenstein2012} M. Lewenstein, A. Sanpera, V. Ahuﬁnger, Ultracold atoms in optical lattices, Oxford university Press, (2012).

\bibitem{Celi2014} A. Celi, P. Massignan, J. Ruseckas, N. Goldman, I. B. Spielman, G. Juzeliunas, and M. Lewenstein, Synthetic gauge fields in synthetic dimensions, Phys. Rev. Lett. \textbf{112}, 043001 (2014).

\bibitem{Page} J. B. Page,  Asymptotic solutions for localized vibrational modes in strongly anharmonic periodic
systems, Phys. Rev. B {\bf 41}, 7835 (1990).

\bibitem{Darmanyan1998JETP} S. Darmanyan, A. Kobyakov and F. Lederer, Stability of strongly localized excitations in discrete media with cubic nonlinearity, J. Exp. Theor. Phys. \textbf{86}, 682-686 (1998).

\bibitem{Darmanyan1999} S. Darmanyan, A. Kobyakov, F. Lederer and L. V\'azques, Discrete fronts and quasirectangular solitons, Phys. Rev. B \textbf{59}, 5994 (1999).

\bibitem{Jorgensen} N. B. Jørgensen, G. M. Bruun, and J. J. Arlt, Dilute fluid governed by quantum fluctuations, 
Phys. Rev. Lett. {\bf 121}, 173403 (2018).
 
\bibitem{Skov} T. G. Skov, M. G. Skou, N. B. Jørgensen, and J. J. Arlt, Observation of a Lee-Huang-Yang fluid, 
Phys. Rev. Lett. {\bf 126}, 230404 ( 2021).

\bibitem{nijhof2000}
J. H. B. Nijhof, W. Forysiak, and N. J. Doran, The averaging method
for finding exactly periodic dispersion-managed solitons, IEEE J.
Sel. Top. Quantum Electron. 6, 330 (2000).

\bibitem{Malomed2013} C. Mejıa-Cortes, R. A. Vicencio and B. A. Malomed,  Mobility of solitons in one-dimensional lattices with the cubic-quintic nonlinearity, Phys. Rev. E  \textbf{88}, 052901 (2013).

\bibitem{Abdullaev2007} F. Kh. Abdullaev, A. Bouketir, A. Messikh, B. A. Umarov, Modulational instability and discrete breathers in the discrete cubic–quintic nonlinear Schrödinger equation, Physica D {\bf 232}, 54 (2007).

\bibitem{Gonzales2006} R. Carretero-Gonzales, J. D. Talley, C. Chong and B. A. Malomed, Multistable solitons in the cubic-quintic discrete nonlinear Schrodinger equation, Physica D, {\bf 216}, 77-89 (2006).


\end{thebibliography}
\end{document}